\documentclass[11pt]{article}

\usepackage[legalpaper, margin=1in]{geometry}
\usepackage{graphics}
\usepackage{caption}
\usepackage{subcaption}
\usepackage{amsmath}
\usepackage{amssymb}
\usepackage{pdfpages} 
\usepackage{authblk}
\usepackage{pdflscape}
\usepackage{multirow}
\usepackage{hyperref}
\usepackage{rotating} 
\usepackage{array}
\usepackage{comment}

\usepackage[utf8]{inputenc}
\usepackage[T1]{fontenc}

\title{Registration of pre-surgical MRI and whole-mount histopathology images in prostate cancer patients with radical prostatectomy via RAPSODI}
\date{}

\author[a]{Mirabela Rusu\thanks{To whom correspondence should be addressed. E-mail: mirabela.rusu@stanford.edu}}
\author[b]{Christian A. Kunder}
\author[c]{Nikola C. Teslovich}
\author[d]{Jeffrey B. Wang}
\author[e]{Rewa R. Sood}
\author[a]{Wei Shao}
\author[c]{Leo C. Chen}
\author[b]{Robert West}
\author[c]{Richard Fan}
\author[a]{Pejman Ghanouni}
\author[c]{James D. Brooks}
\author[a,c]{Geoffrey A. Sonn}

\affil[a]{Department of Radiology, Stanford University, 300 Pasteur Drive Stanford, 94305, CA, USA;}
\affil[b]{Department of Pathology, Stanford University, 300 Pasteur Drive Stanford, 94305, CA, USA;}
\affil[c]{Department of Urology, Stanford University, 300 Pasteur Drive Stanford, 94305, CA, USA;}
\affil[d]{School of Medicine, Stanford University, 350 Serra Mall, Stanford, 94305, CA, USA;}
\affil[e]{Department of Electrical Engineering, Stanford University, 350 Serra Mall, Stanford, 94305, CA, USA;}

\begin{document}

\maketitle

\begin{abstract}
Magnetic resonance imaging (MRI) has great potential to improve prostate cancer diagnosis. 
It can spare men with a normal exam from undergoing invasive biopsy while making biopsies more accurate in men with lesions suspicious for cancer. 
Yet, the subtle differences between cancer and confounding conditions, render the interpretation of MRI challenging. 
The tissue collected from patients that undergo pre-surgical MRI and radical prostatectomy provides a unique opportunity to correlate histopathology images of the entire prostate with MRI in order to accurately map the extent of prostate cancer onto MRI. Such mapping will help improve existing MRI interpretation schemes, e.g. PIRADS, and will facilitate the development of quantitative image analysis methods to assess the imaging characteristics of prostate cancer on MRI. Here, we introduce the RAPSODI (\underline{RA}diology \underline{P}athology \underline{S}patial \underline{O}pen-Source multi-\underline{D}imensional \underline{I}ntegration) framework for the registration of radiology and pathology images. RAPSODI relies on a three-step procedure that first reconstructs in three dimensions (3D) the resected tissue using the serial whole-mount histopathology slices, then registers corresponding histopathology and MRI slices, and finally maps the cancer outlines from the histopathology slices onto MRI. We tested RAPSODI in a phantom study where we simulated various conditions, e.g., tissue specimen rotation upon mounting on glass slides, tissue shrinkage during fixation, or imperfect slice-to-slice correspondences between histopathology and MRI images. Our experiments showed that RAPSODI can reliably correct for rotations within $\pm15^{\circ}$ and shrinkage up to 10\%. We also evaluated RAPSODI in 89 patients from two institutions that underwent radical prostatectomy, yielding 543 histopathology slices that were registered to corresponding T2 weighted MRI slices. We found a Dice similarity coefficient of 0.98$ \pm $0.01 for the prostate, prostate boundary Hausdorff distance of 1.71$ \pm $0.48 mm, a urethra deviation of 2.91$ \pm $1.25 mm, and a landmark deviation of 2.88$ \pm $0.70 mm between registered histopathology images and MRI. Our robust framework successfully mapped the extent of disease from histopathology slices onto MRI and created ground truth labels for characterizing prostate cancer on MRI. Our open-source RAPSODI platform is available as a 3D Slicer plugin or as a stand-alone program and can be downloaded from \url{https://github.com/pimed/Slicer-RadPathFusion}. 
\end{abstract}

Keywords: radiology-pathology registration $|$ prostate cancer $|$ whole-mount histopathology $|$ radical prostatectomy $|$ magnetic resonance imaging

\section{Introduction}

Despite advances in diagnosis and treatment, prostate cancer remains the second leading cause of cancer death in American men \cite{siegel_cancer_2019}.
Overdiagnosis of low-grade cancers that do not require treatment and the underdiagnosis of aggressive cancers are still a concern \cite{futterer_can_2015}, even after the changes in the recommendation of prostate biopsy for elevated Prostate Specific Antigen (PSA).
Magnetic Resonance Imaging (MRI) can help address all of these problems \cite{ahmed_diagnostic_2017}.
When MRI is normal, up to 50\% of men can safely avoid prostate biopsy, thereby reducing overdiagnosis of low-grade cancer and infectious complications of biopsy. However, this is only true when MRI is interpreted by world-leading experts \cite{van_der_leest_head--head_2019}. 
In practice, lack of widespread expertise and alarming levels of inter-reader variation greatly reduce the potential of MRI to revolutionize prostate cancer diagnosis \cite{sonn_prostate_2017}. 
Both false negatives and false positives, even when using the recommended PIRADS reporting system \cite{weinreb_pi-rads_2016}, are very common and the vast majority of men who undergo MRI still undergo biopsy. Finally, MRI has yet to supplant biopsy which is still required to confirm the presence and aggressiveness of prostate cancer \cite{barentsz_synopsis_2016}. 

In men diagnosed with prostate cancer on biopsy, radical prostatectomy remains the most common treatment \cite{cooperberg_trends_2015}. 
The resected prostate provides a unique opportunity to correlate pre-surgical MRI with digitized histopathology images and map the exact extent of cancer from histopathology images onto MRI. Developing a large dataset of prostatectomy cases where cancer and Gleason grade is accurately mapped on MRI has two potentially transformative applications. First is helping to improve existing MRI interpretation schemes that are still affected by many false positive and false negative findings. Second, it may facilitate the development of machine learning methods to identify prostate cancer on MRI by accurately labeling of cancer for model training and validation.

\begin{table}[t]
\begin{tabular}{|m{0.8in}|m{0.5in}|m{1.70in}|m{1.2in}|m{0.6in}|m{0.8in}|}
\hline
Publication & Subject \# & Approach & Additional Input & Dice Coef. & Landmark Error (mm) \\ \hline
Park 2008 \cite{park_registration_2008} & 2& 3D reconstruction + affine and TPS registration & block face picture, ex vivo MRI & NA & 3-3.74  \\ \hline
Chappelow 2011\cite{chappelow_elastic_2011} & 25 & Feature Based Mutual Information + BSpline & - & NA & NA\\\hline
Ward 2012 \cite{ward_prostate:_2012} & 13 & 2D Affine + TPS Registration &  Strand-shaped fiducials, Ex vivo MRI  & NA & 1.1 \\ \hline
Kalavagunta 2014 \cite{kalavagunta_registration_2015} &  35 & Local affine registration & Internal landmarks, 3D Printed Molds & 0.99 & 1.54$\pm$0.64 \\ \hline
Reynolds 2015 \cite{reynolds_development_2015} &6& 2D TPS registration + deformable registration & Control Points, ex vivo MRI, sectioning box &  0.93 & 3.3 \\ \hline
Li 2017 \cite{li_co-registration_2017}  & 19 & Multi-Scale Representation + deformable registration & - & 0.96$\pm$0.01 &2.96$\pm$0.76 \\ \hline
Losnegard 2018 \cite{losnegard_intensity-based_2018} & 12 & 3D histopathology reconstruction, 3D affine and deformable registration & - &  0.94 & 5.4 \\ \hline
Wu 2019 \cite{wu_system_2019} & 17 & 2D Rigid, TPS Registration (automatic landmarks) & ex vivo MRI, 3D printed molds & 0.87$\pm$0.04&2.0$\pm$0.5\\ \hline
Rusu 2019 \cite{rusu_framework_2019} & 15 & 3D histopathology reconstruction, 2D Affine+Deformable & 3D printed Molds &0.94$\pm0.02$& 1.11$\pm$0.34 \\ \hline
\end{tabular}
\caption{Summary of previous approaches (not exhaustive). We excluded publications with <2 subjects \cite{priester_system_2014}, only synthetic data \cite{priester_registration_2019}, or manually intensive approaches \cite{costa_improved_2017}. All summarized methods require as input the in vivo pre-surgical T2 weighted MRI, digitized serial histopathology images, and the segmentation of the prostate on MRI and histopathology images; Additional input requirements are listed here; Abbreviations: TPS - Thin Plate Spline; NA - Not available}
\label{tab:prior_work}
\end{table}

Although numerous approaches for the radiology-pathology registration in the prostate have been introduced (see section ``Prior Work``), these approaches have not been widely adopted and have not been carefully tested by scientists outside the developer teams. 
Recent publications using histopathology images as reference to improve MRI and automatically detect cancer \cite{penzias_identifying_2018, hurrell_optimized_2018,sumathipala_prostate_2018,cao_joint_2019, reynolds_voxel-wise_2019} still use manual approaches to align the histopathology to MRI images, which are known to be labor-intensive and subjective. 
The reduced adoption of previous methods is due to the challenges associated with managing and registering the histopathology and Magnetic Resonance (MR) images, the lack of open source release of existing methods, and the time constraints associated with running these methods. 

Specifically, the registration of histopathology images and prostate MRI has the following challenges. Histologic processing of the resected tissue causes artifacts, e.g., deformations, shrinkage, and tissue ripping. Some of these artifacts (e.g., deformation and shrinking) can be corrected through registration, while others (e.g. tissue ripping) are challenging to correct and may result in discarding slices when such artifacts are major. Furthermore, our method and many others \cite{kalavagunta_registration_2015, reynolds_development_2015, wu_system_2019} assume slice-to-slice correspondence between histopathology and MRI images, which can be improved through the use of customized 3D printed molds based on pre-operative MRI \cite{turkbey_multiparametric_2011}. However, this approach requires a change in clinical protocol that is not present in the vast majority of institutions performing radical prostatectomy. Finally, the acquired data is different between the histopathology images and MRI. Histopathology images provide a discontinuous serial stack of 4$\mu m$ high-resolution colored images with a pixel size of 0.0005 mm separated by roughly 4 mm spaces, while MRI has a typical resolution of 0.4$\times$0.4$\times$4 mm$^3$.

Here, we introduce the RAPSODI (\underline{RA}diology \underline{P}athology \underline{S}patial \underline{O}pen-Source multi-\underline{D}imensional \underline{I}ntegration) framework for the registration of histopathology slices and pre-operative MRI. RAPSODI includes a dictionary-based data management system, a memory-efficient registration methodology and a Graphical User Interface Plugin to 3D Slicer \cite{fedorov_3d_2012}. Our registration approach relies on the 3D reconstruction of the histopathology specimen to create a digital representation of the tissue before gross sectioning. Next, RAPSODI registers corresponding histopathology and MRI slices.  
Finally, the optimized transforms are applied to the cancer regions outlined on the histopathology images to project those labels onto the pre-operative MRI. 

We evaluated our methodology using a digital phantom study where we simulated various conditions resulting from the histologic preparation of the excised tissue, e.g., rotation of the tissue when mounting on the glass slide or shrinking of the tissue. Moreover, we tested RAPSODI in 89 prostate cancer patients that underwent radical prostatectomy from two institutions, ours and a public cohort \cite{madabhushi_fused_2016}. RAPSODI is open-source and can be downloaded from \url{https://github.com/pimed/Slicer-RadPathFusion}, while the phantom data is available at \url{https://github.com/pimed/rad-path-phantom}.

\subsection{Prior Work and Our Contribution}
Although numerous automated approaches for the registration of radiology and histopathology images have been developed, manual approaches are still employed, even in recent publications \cite{penzias_identifying_2018, hurrell_optimized_2018,sumathipala_prostate_2018,cao_joint_2019, reynolds_voxel-wise_2019}. 
Some manual or semi-automatic approaches utilize landmark-based registration approaches, either alone \cite{penzias_identifying_2018, hurrell_optimized_2018,reynolds_voxel-wise_2019} or in combination with automated registration steps \cite{reynolds_development_2015, hurrell_optimized_2018}. These approaches are labor-intensive and require the human operator to possess expertise in both MRI and histopathology, and necessitate identification of corresponding landmarks on both modalities. Other such approaches \cite{turkbey_multiparametric_2011, sumathipala_prostate_2018} employ cognitive alignment in which a radiologist with the help of a pathologist directly outlines the cancer region on MRI considering the histopathology images as reference. Such methods are tedious to apply and may be prone to underestimating the dimensions of the lesion \cite{priester_magnetic_2017} while MRI invisible lesions are hard if not impossible to outline and thereby they are often omitted from follow-up analysis. 
A few approaches use interactive image transformations \cite{costa_improved_2017}, in which a user indicate scaling, rotations and translations to be applied to the images. Such approaches are also tedious to utilize and require extensive knowledge in both radiology and pathology of the prostate. 

The automated registration of histopathology images with pre-surgical prostate MRI has been performed in proof-of-concept studies, which usually only include a small number of subjects, often < 20 (\tablename~\ref{tab:prior_work}).
Most approaches assume a slice-to-slice correspondence between the histopathology images and T2 weighted (T2w) MRI slices.
Some partial correspondence commonly results from the gross sectioning of the prostate in histologic preparation which is done perpendicular to the urethra. Yet, more advanced methods have been introduced to enforce such correspondences. For example, three dimensional (3D) printed patient-specific molds \cite{turkbey_multiparametric_2011} have been used \cite{kalavagunta_registration_2015, reynolds_development_2015, wu_system_2019} to help preserve the correspondences during tissue sectioning. Some studies additionally included blockface picture \cite{park_registration_2008}, ex vivo MRI \cite{reynolds_development_2015, wu_system_2019, park_registration_2008, ward_prostate:_2012} or external fiducials \cite{ward_prostate:_2012} to help improve the accuracy of the registration. Yet, these approaches required modifications of the clinical protocols usually resulting in only a small number of subjects to be recruited for such research studies.  

{\color{red}
Once correspondences between the histopathology images and T2w MRI are identified, their registration can still be challenging, partially due to the artifacts induced by the tissue preparation. 
Textural features \cite{chappelow_elastic_2011, li_co-registration_2017} have been proposed, yet they may be cumbersome to use due to the high-dimensional scoring function optimization and the choice of textural features. Other approaches rely solely on image intensity to drive the deformable alignment \cite{losnegard_intensity-based_2018, rusu_framework_2019}, but require accurate affine alignment prior to the deformable registration.

Previous work in the lung \cite{rusu_framework_2015, rusu_co-registration_2017}, breast \cite{rusu_rad-path_2019} or prostate \cite{losnegard_intensity-based_2018, rusu_framework_2019}, has relied on approaches that reconstruct the sequential histopathology slices and created a 3D volume representing the histopathology specimen prior to sectioning, which facilitates the spatial registration with the 3D volumetric MRI and alleviates the need for slice correspondences. 
However, these methods are prone to overfitting the histopathology reconstruction due to the large number of degrees of freedom and may suffer from partial volume effects due to the missing data associated with thick MRI slices and the histopathology slice spacing. }

Our approach makes the following contributions: 1) Our registration methodology combines a 3D reconstruction of the histopathology specimen with 2D affine and deformable registration of corresponding histopathology and MRI slices and was optimized for an accurate alignment, 2) Our approach was tested in a digital phantom where the ground truth is known as well as in the largest cohort considered to date in a radiology-pathology registration study, and 3) To the best of our knowledge, we are the first to release the source code for the registration of histopathology and radiology images in the prostate, {\color{red}which is essential to test the reproducibility and robustness of the approach while allowing the wide adaption.}

\section*{Methods}

\subsection*{Notations}
Let $\mathcal{M}$ be the T2 weighted (T2w) MRI image, $\mathcal{M}:\mathcal{R}^3 \rightarrow \mathcal{R}$ and has a matrix size $K \times L \times M$, where $K$, $L$ and $M$ represent the width, height and number of slices of the axial T2w MRI.
Let $\mathcal{H}$ be the stack of histopathology slices, $\mathcal{H}:\mathcal{R}^3\rightarrow\mathcal{R}^3$, obtained by stacking 2D histopathology slices, $\mathcal{H}_i:\mathcal{R}^2\rightarrow\mathcal{R}^3$. The histopathology images are colored images, with Red, Green and Blue channels. The volume $\mathcal{H}$ has $W\times H\times D$ voxels with three components corresponding to the Red, Green  and Blue channels, where $D \in [3,9]$ represents the number of slices, while $W$ and $H$ are the width and height of the histopathology images. The index $i$, is used to indicate either an axial slice within the MRI volume or an image in the histopathology stack.  $\mathcal{M}^{Pr}$ and $\mathcal{H}^{Pr}$ represent the prostate segmentation on MRI and histopathology images, respectively.

\begin{table}[t]
    \centering
    \begin{tabular}{|m{1.0in}|m{1.10in}|m{1.00in}|m{1.10in}|m{1.00in}|}

        \hline
        & \multicolumn{2}{l}{{Cohort C1: Internal}} & \multicolumn{2}{|l|}{Cohort C2: Public \cite{madabhushi_fused_2016}} \\ \hline
        
        Variable  & MRI & Pathology & MRI & Pathology  \\ \hline                
        Manufacturer: Coil type & GE: Surface & - & Siemens: Endorectal & -  \\ \hline 
        Sequence/Data Type & T2w & whole-mount & T2w & pseudo-whole mount  \\ \hline 
        Acquisition  & TR: [3.9, 6.3]; & H\&E & TR: [3.7, 7.0]; &H\&E \\ 
        Characteristics & TE: [122, 130] & & TE: [107] &  \\ \hline
        Number of Patients/Slices & 74/1994 & 74/478 & 16/430 & 16/65 \\ \hline
        Matrix Size & $K,L\in[256,512]$ $M\in[24,43]$ & W,H $\in$ [1663,7556] & $K,L = 320$, $M \in [21,31]$ & W,H $\in$ [2368,6324] \\ \hline
        Pixel Spacing (mm) & $\in$ [0.27,0.94] & $\in$ \{0.0081,0.0162\} & $\in$ [0.41,0.43] & 0.0072*\\ \hline
        Distance Between Slices (mm) & $\in$ [3,5.2] & Same as MRI via 3D printed mold & 4  & Free hand\\ \hline
        Annotations & Prostate, Anatomic Landmarks & Prostate, Anatomic Landmarks, Cancer  & Prostate, Cancer, Urethra & Prostate, Urethra, Cancer\\
        \hline
    \end{tabular}
    \caption{Data Summary: Abbreviations: T2-weighted MRI (T2w), Hematoxylin \& Eosin (H\&E), Relaxation Time (TR) in seconds, Echo Time (TE) in milliseconds; MRI Matrix Size: $K\times L \times M$, Histology Matrix Size: W$\times$H, * estimated, pseudo-whole mount: quadrants were stitched together  }
    \label{tab:data}
\end{table}

\subsection*{Data Description}

Our IRB approved study includes $N_1=73$ subjects from Stanford Hospital (Cohort C1) and $N_2=16$ patients from the "Prostate Fused MRI Pathology" collection, The Cancer Imaging Archive \cite{madabhushi_fused_2016} (Cohort C2) (\tablename~\ref{tab:data}).  A subset of 15 subjects from cohort C1 was previously utilized in \cite{rusu_framework_2019}. The subjects in C1 had an MRI acquired between 2016-2019 prior to the radical prostatectomy, and the excised prostate was submitted for histologic preparation to generate whole-mount sections. {\color{red} The subjects in C2 (description available at \url{https://wiki.cancerimagingarchive.net/display/Public/Prostate+Fused-MRI-Pathology}), underwent radical prostatectomy and had the prostate sectioned in quadrants before being submitted for histology processing.}

\textit{MRI:} The MRI exams for the patients in cohort C1 were acquired using 3 Tesla scanners (MR750, GE Healthcare, Waukesha, WI) with an external 32-channel body array coil without an endorectal coil. The imaging protocol included T2 weighted MRI (T2w), diffusion weighted imaging (DWI) and derived Apparent Diffusion Coefficient (ADC), and dynamic contrast-enhanced imaging sequences. For the patients in Cohort C2 were acquired on a 3T scanner (Siemens) using an endorectal coil. The public repository provides T2w MRI and dynamic contrast-enhanced MRI for these patients. In this study, we utilized the Axial T2w MRI, which is acquired using a 2D Spin Echo protocol (\tablename~\ref{tab:data})

\textit{Histopathology:} The cohort C1 patients, following resection, the prostate was fixed in formalin, sectioned using a patient-specific 3D printed mold built based on the pre-surgical MRI to maintain the correspondence of histopathology slices and T2w images, and embedded in paraffin. This process is now part of our clinical standard of care. The histopathology for the patients in cohort C2, was cut without the use of 3D printed molds, but seeking gross sectioning perpendicular to the urethra. 
Mounting of the 5$\mu m$ thick tissue on the glass slide can result in a rotation of the histopathology slice as well as mounting with either aligned or misaligned left-right orientation. To account for this variability, an expert indicated the gross rotation angle and whether the slice requires left-right flipping. 
The whole-mount slices in C1 and quadrants sections in C2 were stained using Hematoxylin \& Eosin (H\&E) and were digitized at 20x magnification (pixel size 0.5 $\mu m$). 
Pseudo-whole mounts were generated for the images in C2, by stitching adjacent quadrants as described in \cite{singanamalli_identifying_2016}.

\textit{Labels:} Our expert radiologist (PG) outlined the prostate on MRI, $\mathcal{M}^{Pr}$, while our expert pathologist (CK) outlined the prostate, $\mathcal{H}^{Pr}$, and the cancer on the high-resolution scanned images of the histopathology specimen. 
Two hundred fifty-seven matching anatomic landmarks were picked on both histopathology and radiology images for a subset of 12 subjects, targeting 3 landmarks for each corresponding pair of histopathology and MRI slices. Examples of anatomic landmarks include benign prostate hyperplasia nodules, ejaculatory ducts, predominant glands, etc. The urethra was outlined on 22 studies in cohort C1 and the 16 studies in cohort C2. Outlining the urethra on MRI is relatively straightforward at the apex of the prostate, yet it becomes challenging towards the base. Thereby, often urethra annotations are available on the MRI from the mid-gland to apex, but lacking between the mid-gland and the prostate base. Slice correspondences between the MRI and histopathology were identified by an expert urologist and a radiology-pathology registration expert and validated by a multi-disciplinary team of radiologists, pathologists and urologists.

\subsection*{Radiology - Pathology Registration}

Our approach is summarized in \figurename~\ref{fig:flowchart} and described below:

\begin{figure}[t]
\includegraphics[width=\linewidth]{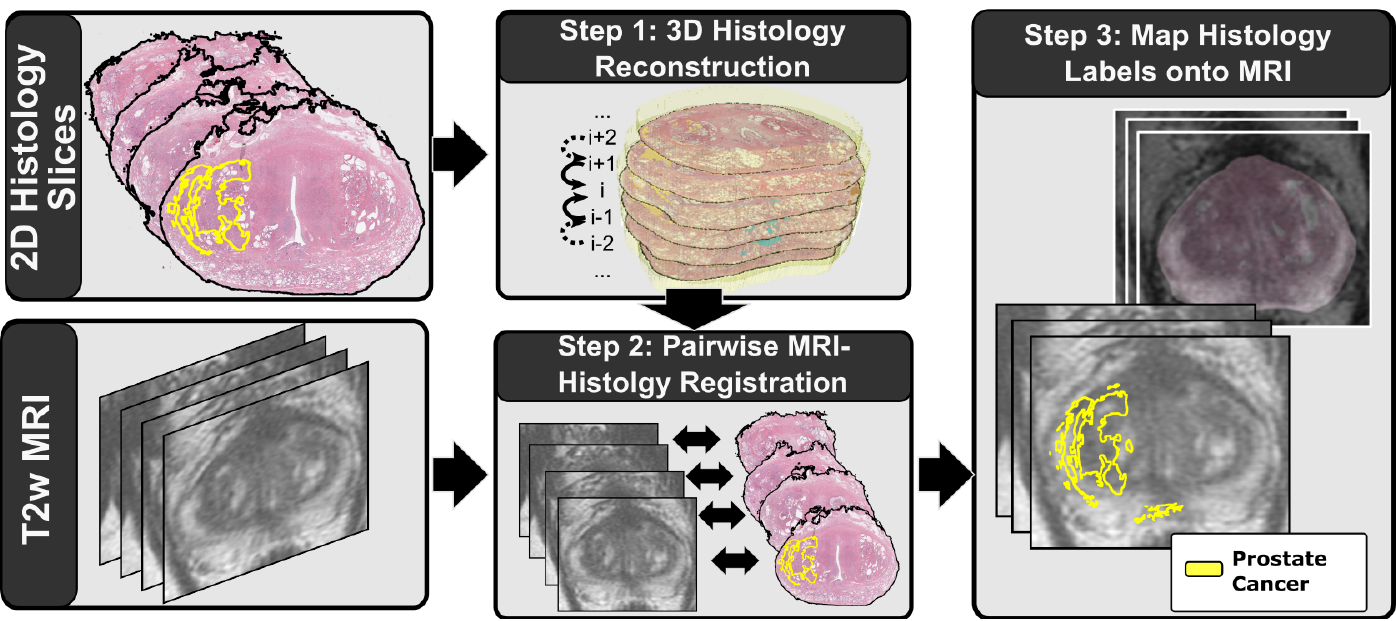}
\caption{Summary of our approach. First, we align the serial histopathology slices relative to each other to reconstruct the 3D histopathology volume. Second, we register the histopathology slices relative to the T2w MRI using rigid, affine and deformable transforms. Finally, we map the extent of cancer from the histopathology images onto the radiology images.}
\label{fig:flowchart}
\end{figure}

\begin{enumerate}
	\item[Step 0)] \textbf{Pre-Processing:} We applied the prostate masks, $\mathcal{M}^{Pr}$ and $\mathcal{H}^{Pr}$ onto $\mathcal{M}$ and $\mathcal{H}$, respectively, to exclude the structures outside the prostate from image registration. The gross rotation angles or left-right flipping was applied as well.
    \item[Step 1)] \textbf{3D Histopathology Reconstruction:} We registered $\mathcal{H}_i$ relative each other, to ensure their 3D consistency within the 3D reconstruction, $\mathcal{H}$. We selected the middle slice $\mathcal{H}_i, i = \frac{M}{2}$ as the first fixed image and registered $\mathcal{H}_{i-1}$ to $\mathcal{H}_i$, $\mathcal{H}_{i-2}$ with $\mathcal{H}_{i-1}$, etc, as well as $\mathcal{H}_{i+1}$ to $\mathcal{H}_{i}$, $\mathcal{H}_{i+2}$ with $\mathcal{H}_{i+1}$, etc. With the exception of the middle slice, all histopathology images will have a corresponding rigid transform following the registration with the adjacent slice $\mathcal{T}^H_{Rig}$. 
    \item[Step 2)] \textbf{2D Registration:} We registered $\mathcal{M}_i$ with $\mathcal{H}_i$, for $\forall i \in [1, D]$, by optimizing rigid $\mathcal{T}^M_{Rig}$, affine $\mathcal{T}^M_{Aff}$ and deformable $\mathcal{T}^M_{Def}$ transforms using gradient-based approaches. The rigid and affine registrations only use the prostate masks during the optimization and applied Sum of Square Differences as scoring function. The deformable registration used Free-Form Deformations \cite{rueckert_nonrigid_1999} to optimize the Mattes Mutual Information computed based on the image intensities. {\color{red}We used a multi-resolution pyramid with three layers (with shrinking factors 16, 8, and 4 respectively, and a smoothing sigma of 4, 2, and 1 respectively). The affine transform optimization was done using a gradient descent optimizer with a learning rate of 0.01 and 250-500 iterations per resolution layer, while, the deformable registration employed a LBFGSB optimizer with 10-50 iterations per resolution layer.}
    \item[Step 3)] \textbf{Mapping Cancer onto MRI:} A composite transform of  $\mathcal{T}^H_{Rig}$, $\mathcal{T}^M_{Rig}$,  $\mathcal{T}^M_{Aff}$ and $\mathcal{T}^M_{Def}$ is applied to deform the histopathology image as well as the cancer label and anatomic landmarks into the coordinates of the T2w MRI.

\end{enumerate}

\begin{figure*}[h]
\includegraphics[width=\linewidth]{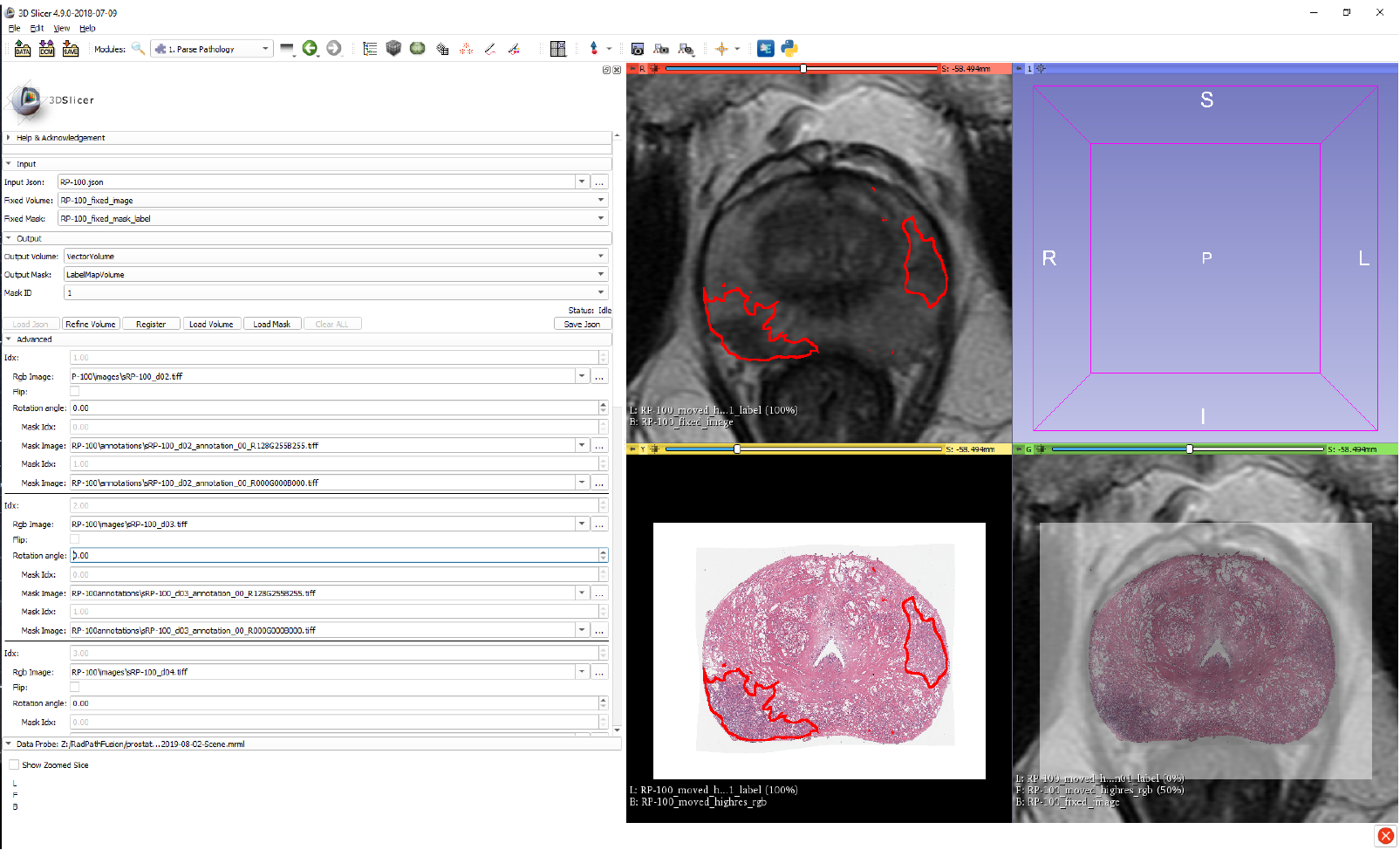}
\caption{Slicer Interface}
\label{fig:slicer}
\end{figure*}

Our approach was developed using the Insight Toolkit (ITK) \cite{johnson_itk_2013}and its Simple ITK API in python. The approach is available as a 3D Slicer python plugin \cite{fedorov_3d_2012} (\figurename~\ref{fig:slicer}) or as a stand-alone application to be run in batch mode. RAPSODI can be downloaded from \url{https://github.com/pimed/Slicer-RadPathFusion}. We measured the performance of the approach on an Intel i7-8700 CPU, 3.70GHz, 64GB Memory Computer. 

\subsection*{Digital Phantom for Radiology-Pathology Registration}

\begin{figure*}[t]
\centering
\begin{subfigure}[b]{0.33\textwidth}
\centering
\includegraphics[trim={0 1.5cm 0 0},clip, width=\linewidth]{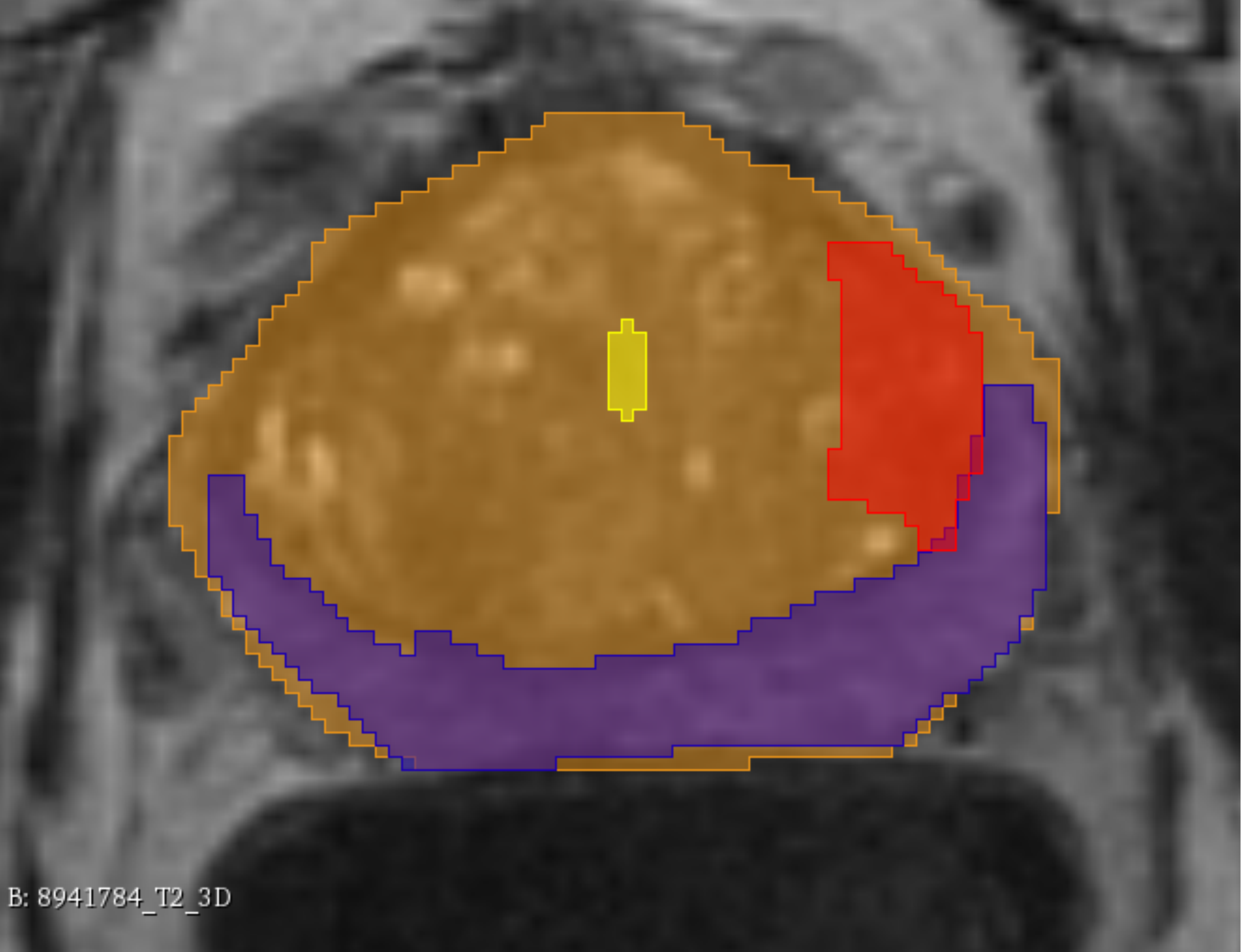}
\caption{}
\end{subfigure}
\begin{subfigure}[b]{0.33\textwidth}
\centering
\includegraphics[trim={0 1.5cm 5cm 0},clip, width=0.8\linewidth]{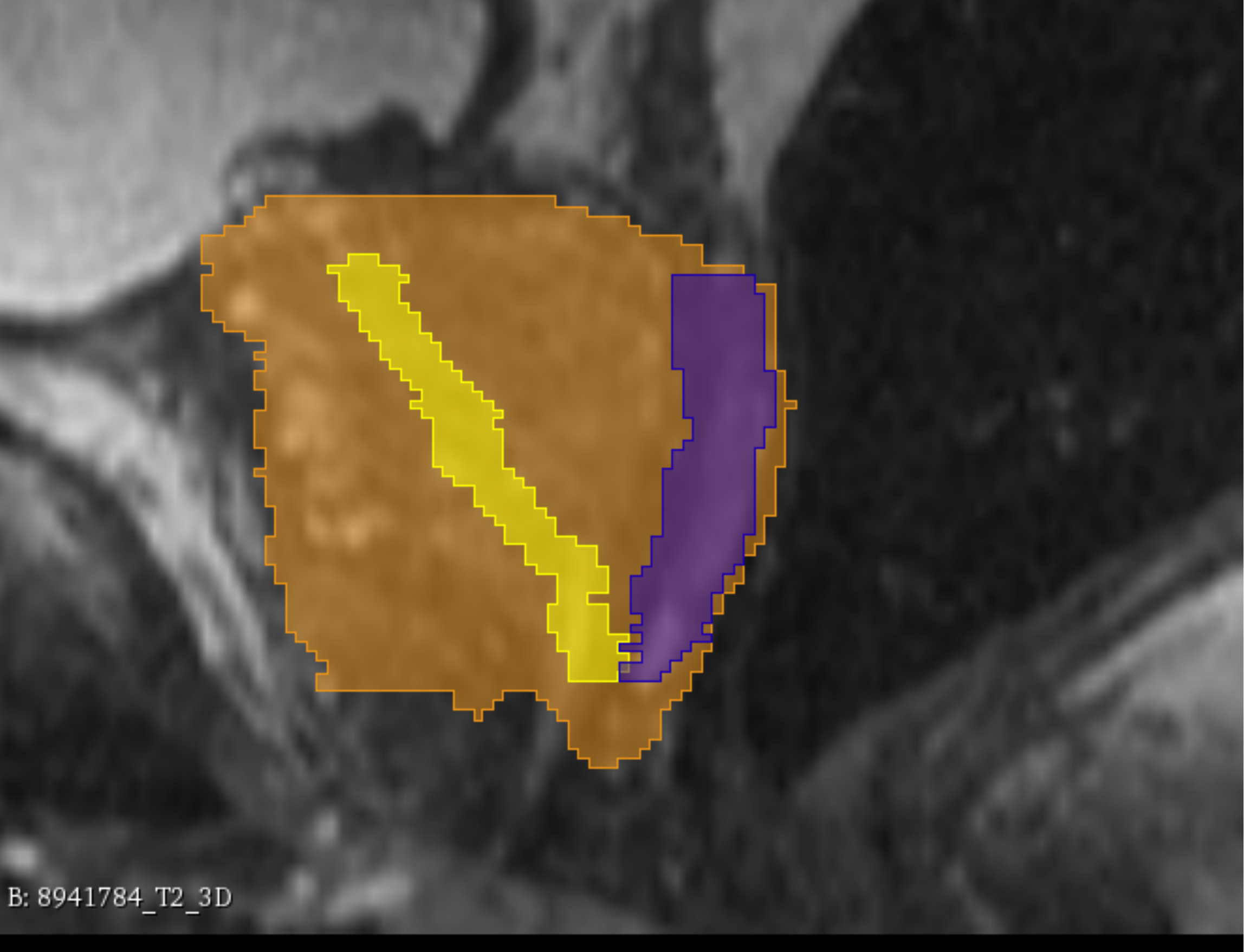}
\caption{}
\end{subfigure}
\begin{subfigure}[b]{0.30\textwidth}
\centering
\includegraphics[trim={5cm 2cm 3cm 2cm},clip,width=\linewidth]{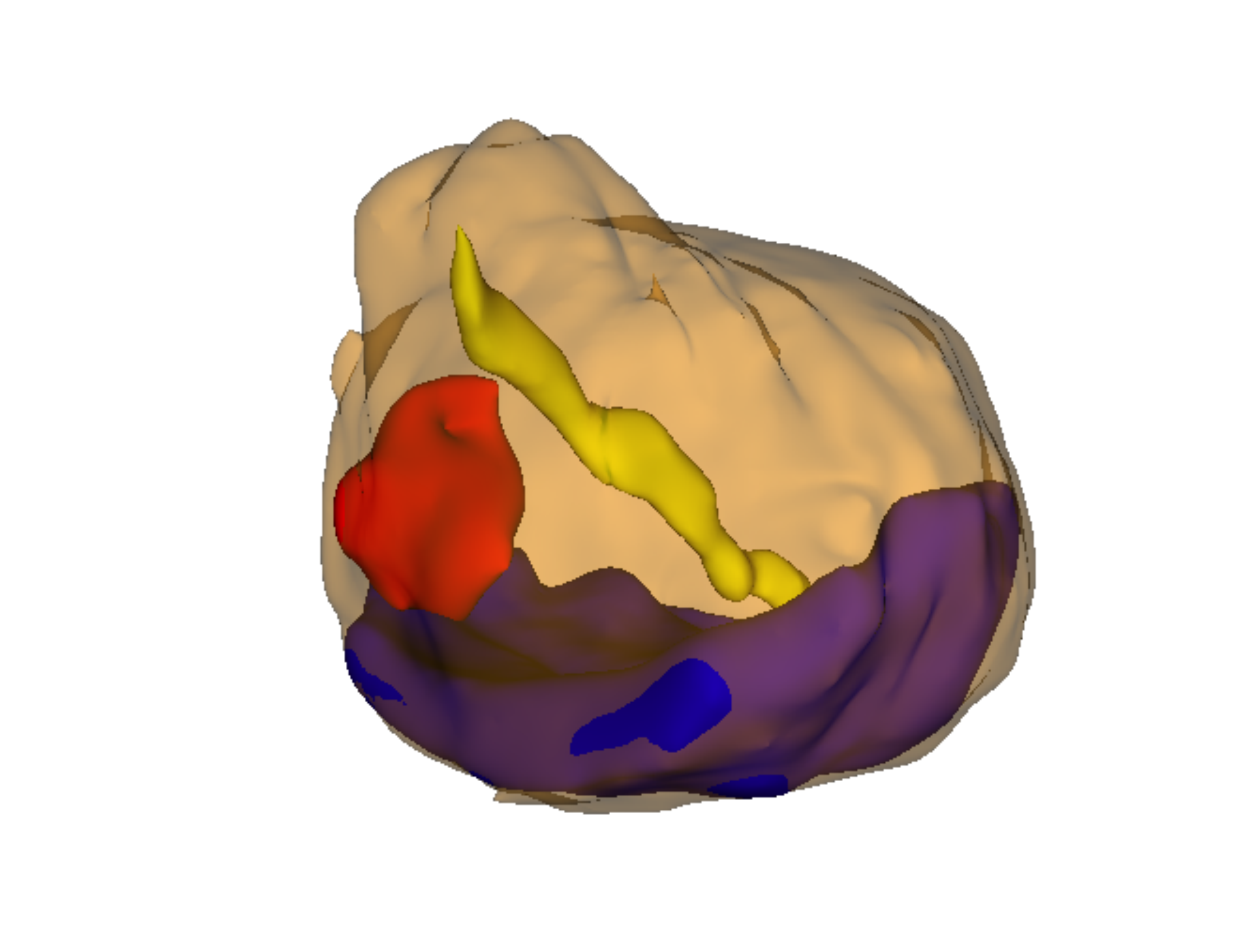}
\caption{}
\end{subfigure}

\hspace{0.05cm}
\begin{subfigure}[b]{0.31\textwidth}
\centering
\includegraphics[trim={0 1.8cm 0 0},clip, width=\linewidth]{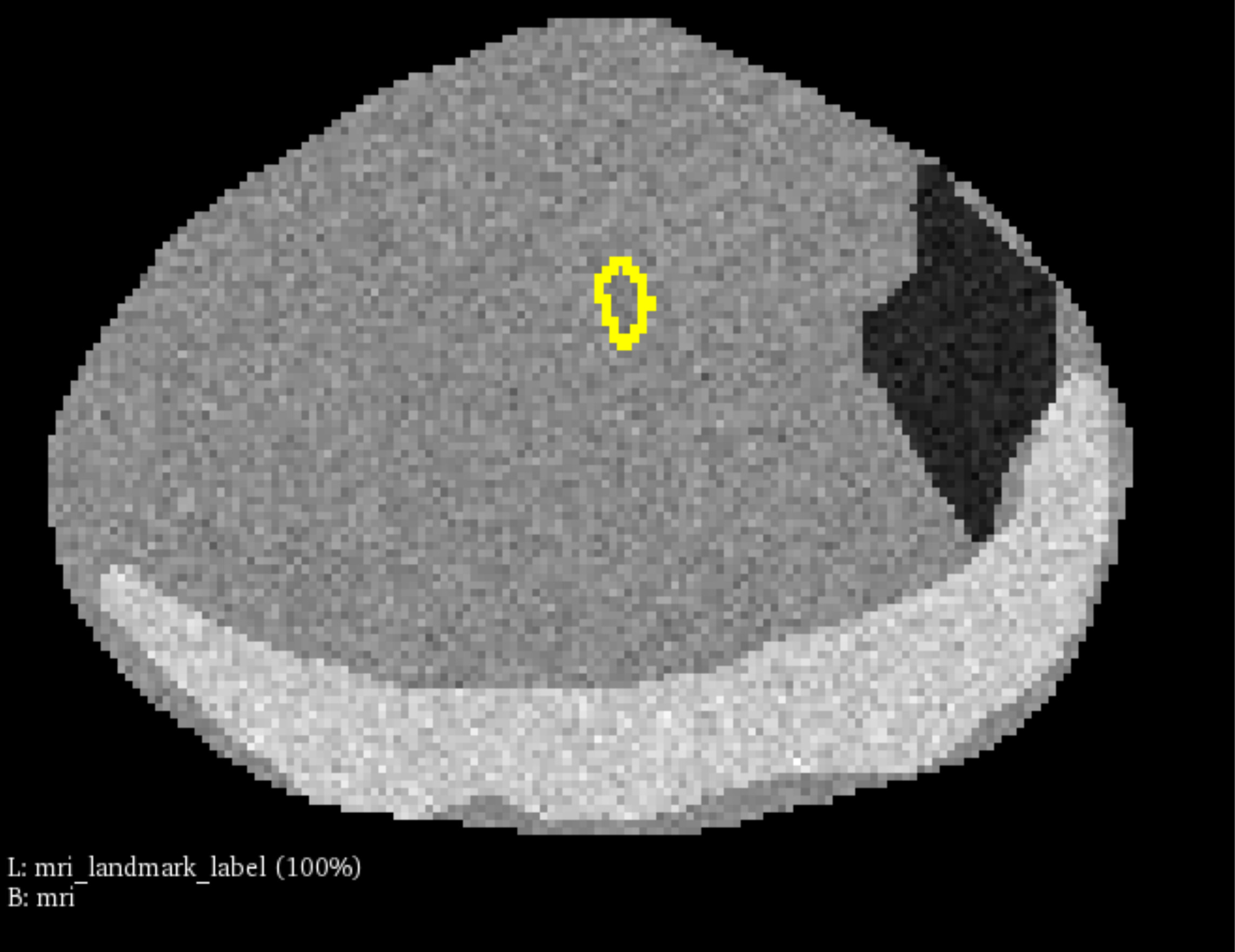}
\caption{}
\end{subfigure}
\hspace{0.10cm}
\begin{subfigure}[b]{0.31\textwidth}
\centering
\includegraphics[trim={0 1.8cm 0 0},clip, width=\linewidth]{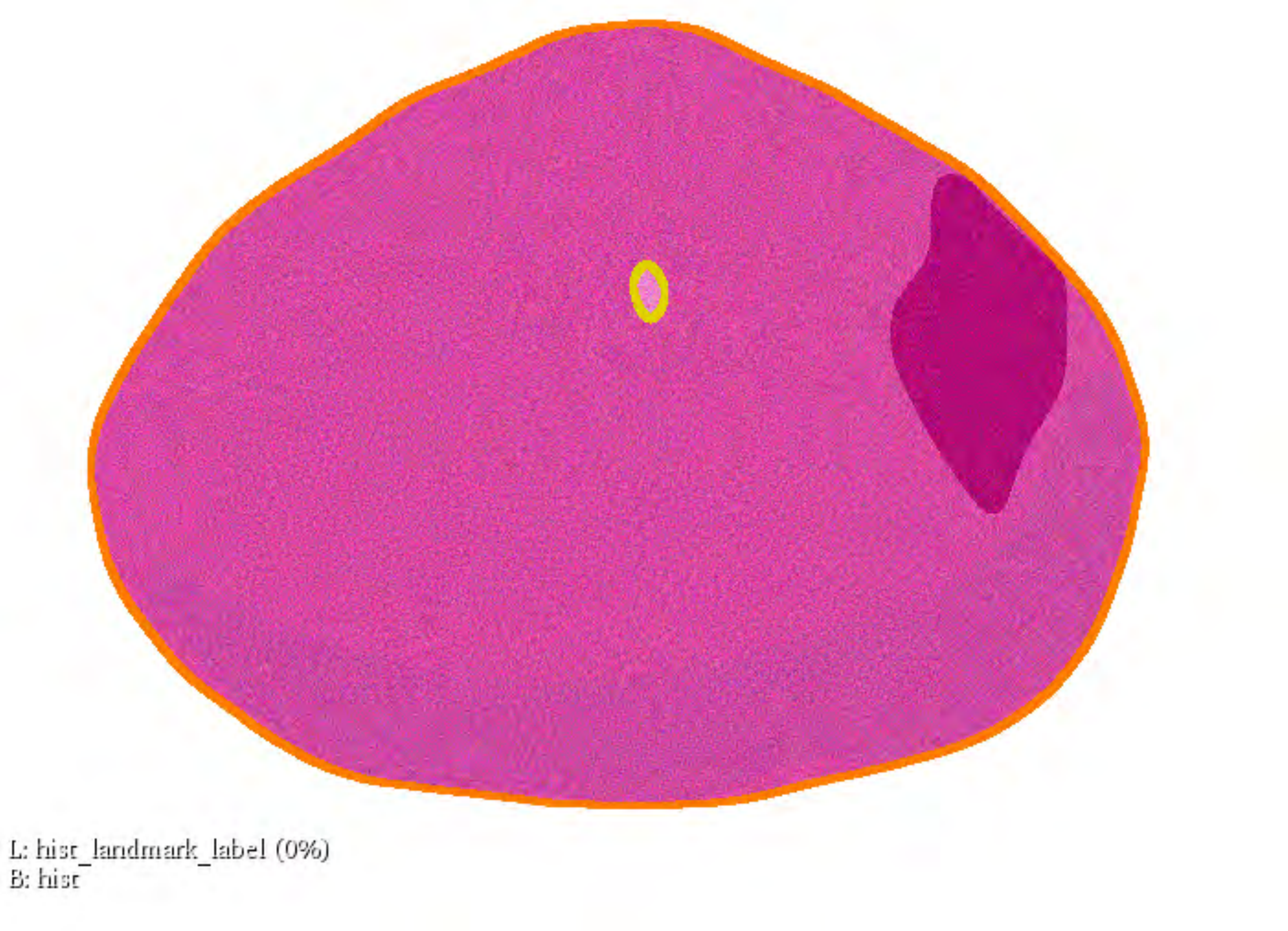}
\caption{}
\end{subfigure}
\hspace{0.15cm}
\begin{subfigure}[b]{0.31\textwidth}
\centering
\includegraphics[trim={0 1.75cm 0 0},clip, width=\linewidth]{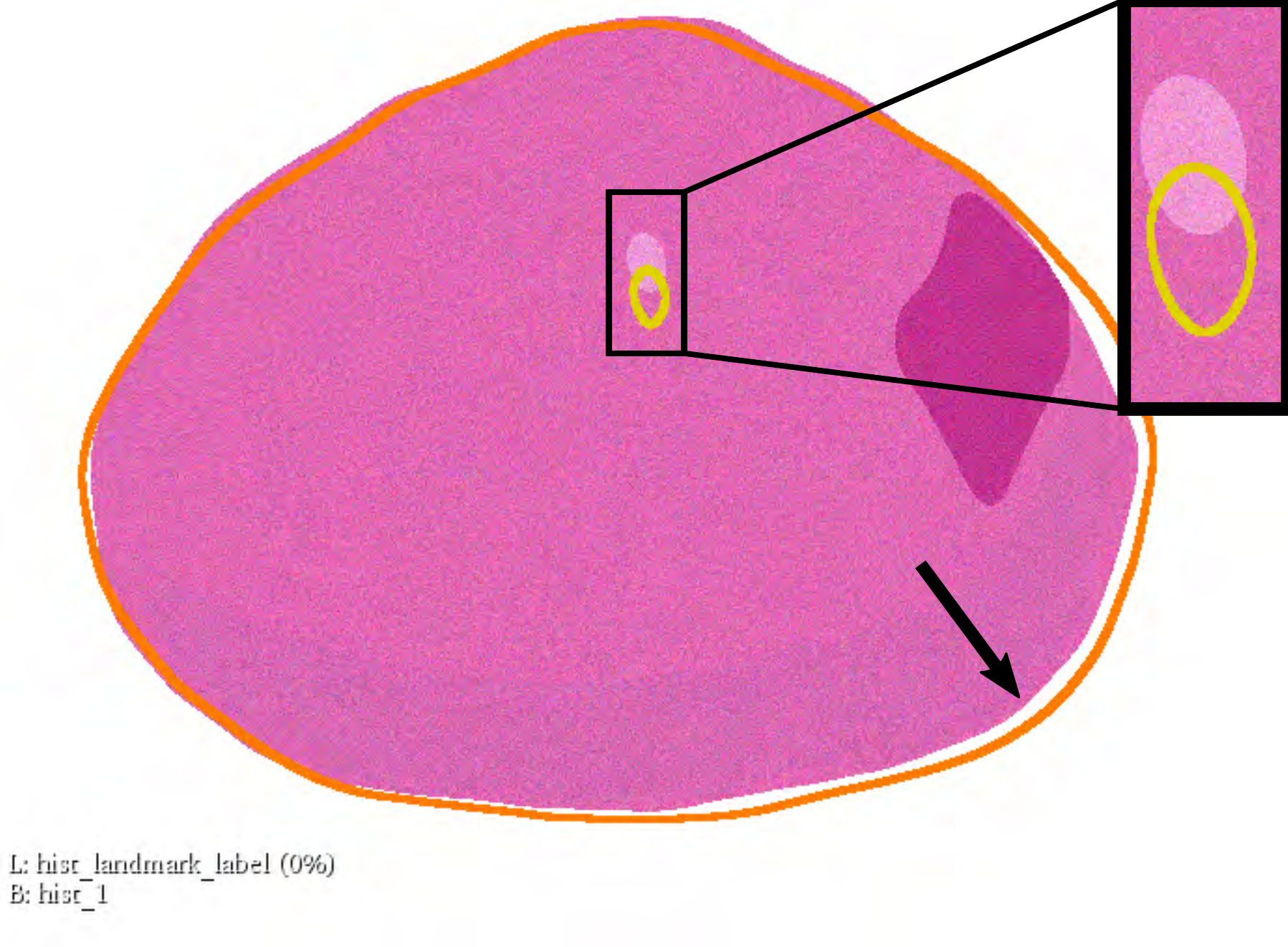}
\caption{}
\end{subfigure}

\caption{Radiology-Pathology Digital Phantom. The expert annotation of the  prostate (orange), peripheral zone (blue), urethra (yellow) and cancer(red) on 3D T2 MRI, shown in the (a) Axial; (b) Sagittal; (c) 3D Views, were used to create a digital phantom of the prostate: (d) Slice in the MRI phantom, (e) Corresponding slice in the Pathology phantom, and (f) imperfect corresponding slice that is 2mm apart from (d) in the sagittal plane (yellow and orange are the outlines of the urethra and prostate(d)). Note the urethra misalignment (inset) and the border differences (arrow). Color legend: yellow - urethra, orange - prostate, blue - peripheral zone, cancer - red}
\label{fig:Phantom}
\end{figure*}

We created a digital phantom to assess the quality of the alignment during the development of RAPSODI and to evaluate its performance when ground truth exists. The phantom is used to simulate artifacts known to affect the histopathology sample. We constructed the phantom by first outlining different prostatic regions, peripheral zone, cancer, and urethra in a 3D T2w MRI (\figurename{s}~\ref{fig:Phantom}a-c). Then, we synthesized the phantom T2 MRI by filling the segmented regions with the average intensities from the input T2 image (\figurename~\ref{fig:Phantom}d). Moreover, we created the pathology phantom based on the histopathology images already registered to the T2w MRI (data not shown), by averaging their color intensities within the segmented regions (\figurename{s}~\ref{fig:Phantom}e-f). Our simulations included Gaussian noise on both the MRI and histopathology phantom slices.

Using the T2w and pathology phantom, we tested three conditions: 1) the influence of the rotation angle when mounting the tissue slice on the glass slide, 2) the influence of shrinkage caused by fixation of the tissue during histology processing, and 3) the influence of imperfect slice correspondences between the MRI and histopathology slices, e.g., \figurename{s}~\ref{fig:Phantom}d-e have a perfect correspondence, while \figurename{s}~\ref{fig:Phantom}d-f are 2mm apart from each other.

To evaluate RAPSODI, we used one or multiple conditions and evaluated different quantitative metrics. Ten experiments were run for each condition to assess the mean and variance in performance of RAPSODI. 
When a random rotation of $r$ was assigned to the histopathology phantom, it resulted in applying a random angle ranging between $-r$ and $r$ to each slide and running 10 experiments with different noise and random angle conditions. When rotations were applied alone, no translation or scaling were applied. When a shrinkage factor $s$ is applied, all histopathology slices are shrunk by $s$ relative to their original appearance. Moreover, along with shrinking the images, we also apply a random translation of as much as 5\% relative to the entire image in either x or y directions. Thereby, the experiments that include rotation and shrinkage also include random translation, and when combined with the imperfect slice correspondences represent the closest condition to the real data.

\subsection*{Quantitative Evaluation}

The accuracy of the radiology-histopathology registration was evaluated using the Dice similarity coefficient, which assesses the overlap of the prostate outlined on T2w MRI and the outline of the prostate from the histopathology reconstruction: 

\begin{equation}
 Dice (\mathcal{H}, \mathcal{M}) = \frac{1}{D}\sum^D_{i=1} \frac{2 \times |\mathcal{H}^{Pr}_i \cap  \mathcal{M}^{Pr}_i|}{|\mathcal{H}^{Pr}_i| +  |\mathcal{M}^{Pr}_i|}   
\label{eq:Dice}
\end{equation}

where $D$ is the number of slices in the histopathology specimen, $\mathcal{H}^{Pr}_i$ represents the slice $i$ in the prostate segmentation on histopathology, while $\mathcal{M}^{Pr}_i$ represents the slice $i$ in the prostate segmentation on MRI. 

Additionally, we evaluated the Hausdorff distance between the prostate boundary, to asses how far the boundary is after performing the registration:

\begin{equation}
\begin{split}
    Hausdorff^{Pr} (\mathcal{H}, \mathcal{M}) = 
    max\{sup_{h \in \mathcal{H}^{Pr}} ~ inf_{m \in \mathcal{M}^{Pr}}d(\mathcal{H}, 
    \mathcal{M}),\\sup_{m \in \mathcal{M}^{Pr}} inf_{h \in \mathcal{H}^{Pr}} d(\mathcal{H}, \mathcal{M}) \}
\end{split}
\label{eq:HDice}
\end{equation}

where $sup$ represents the supremum operator and $inf$ represents the infimum operators.

Moreover, we evaluated the landmark distance: 

\begin{equation}
    Dist (\mathcal{L}^H,\mathcal{L}^M) = \frac{1}{X}\sum^X_{j=1}|\mathcal{L}^H_j,\mathcal{L}^M_j|_2
\label{eq:Dist}
\end{equation}

where $|.|_2$ represents the Euclidean distance of the center of mass of the j landmark $\mathcal{L}^H_j$ on histopathology and center of mass of the j landmark $\mathcal{L}^M_j$ on MRI, while X represents the number of landmarks. Similarly, we computed the urethra distances, using per-slice correspondences, and slices where the urethra was visible on both MRI and histopathology slices.

\begin{figure*}[t]
\begin{subfigure}[b]{0.24\textwidth}
\centering
\includegraphics[trim={0 0 0 0},clip,width=\linewidth]{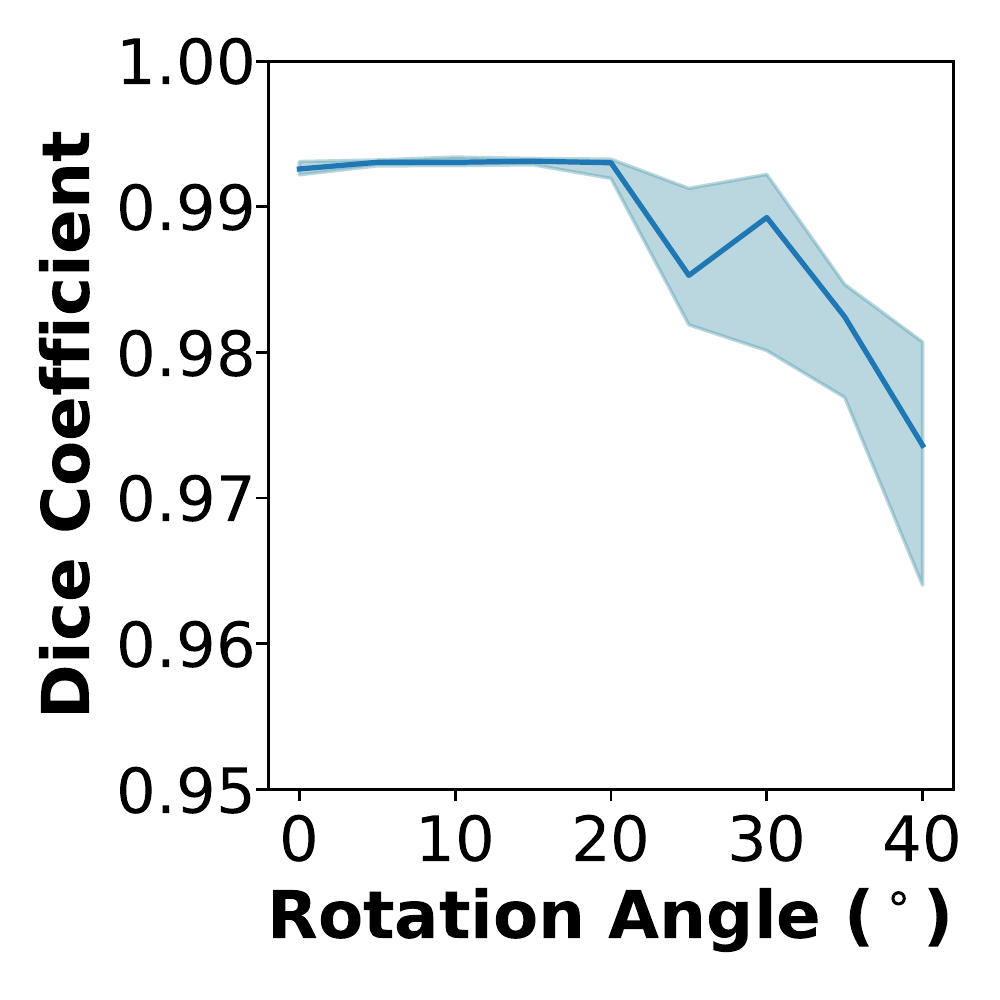}
\caption{}
\end{subfigure}
\begin{subfigure}[b]{0.24\textwidth}
\centering
\includegraphics[trim={0 0 0 0},clip,width=\linewidth]{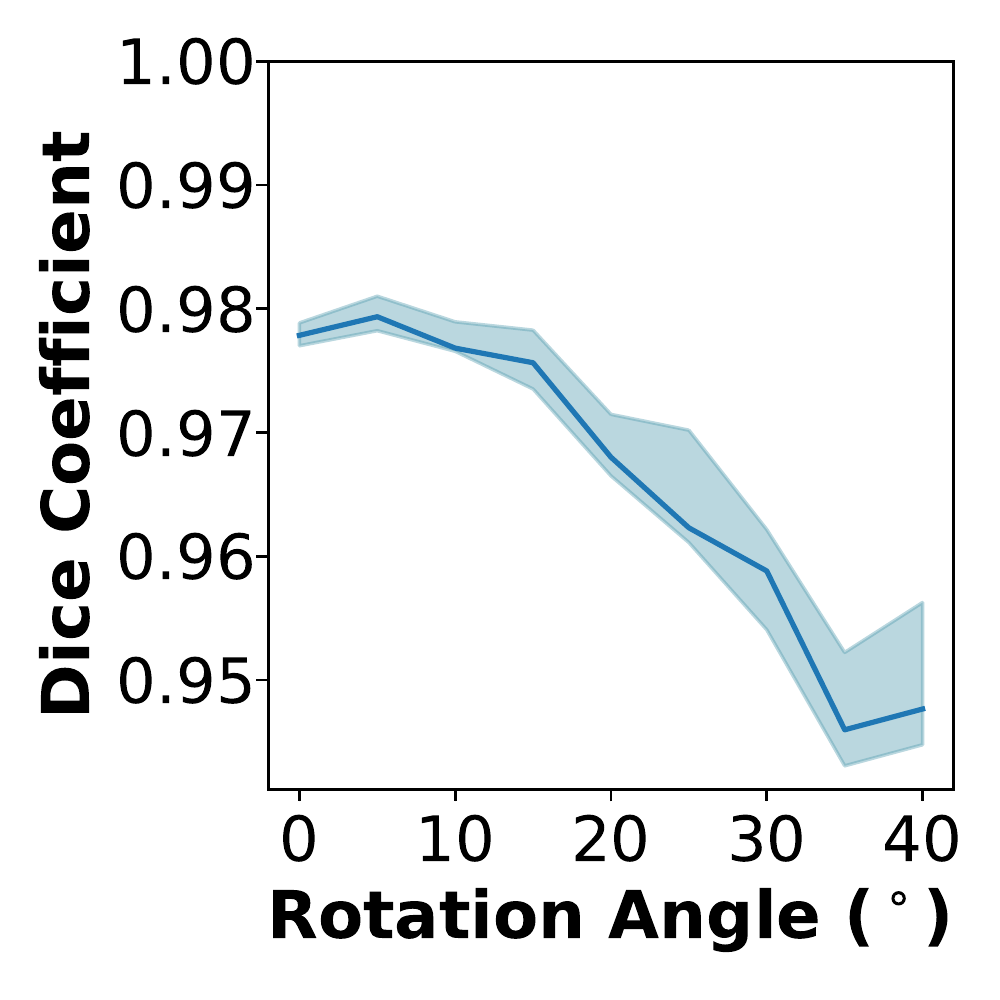}
\caption{}
\end{subfigure}
\begin{subfigure}[b]{0.24\textwidth}
\centering
\includegraphics[trim={0 0 0 0},clip,width=\linewidth]{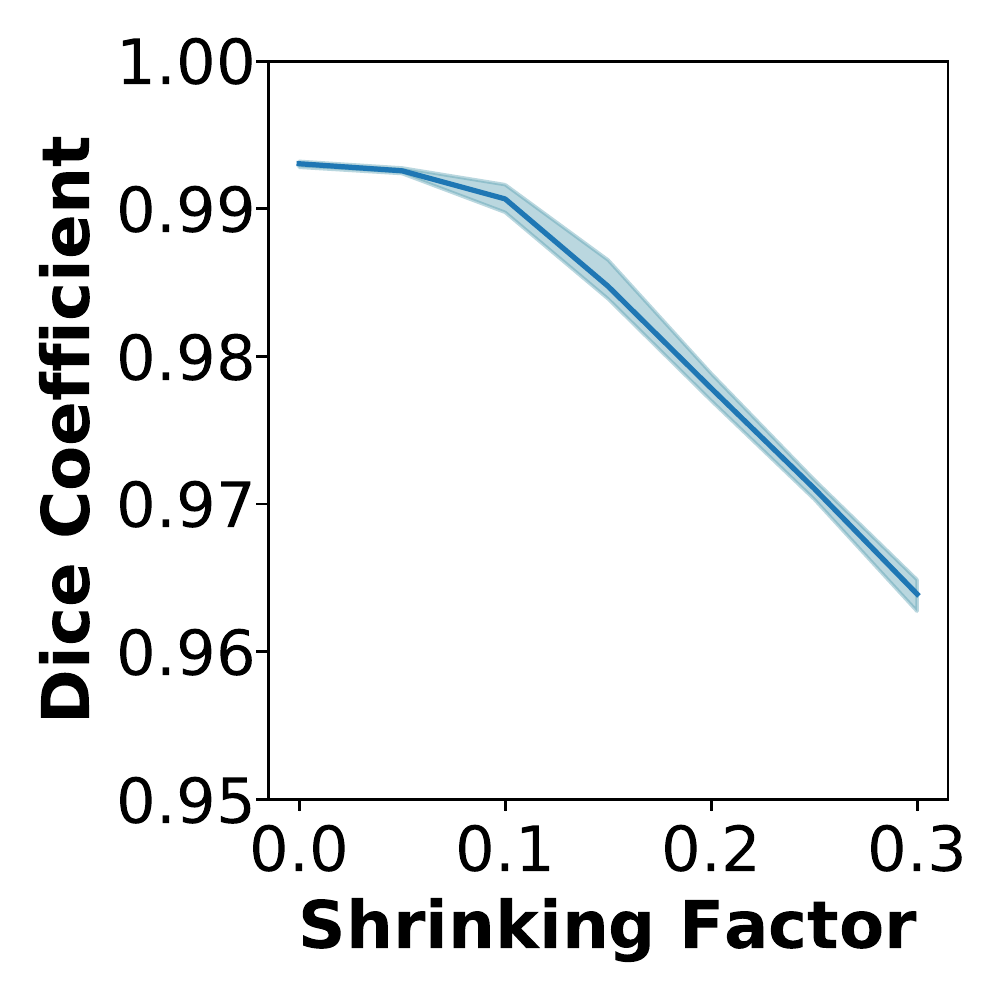}
\caption{}
\end{subfigure}
\begin{subfigure}[b]{0.24\textwidth}
\centering
\includegraphics[trim={0 0 0 0},clip,width=\linewidth]{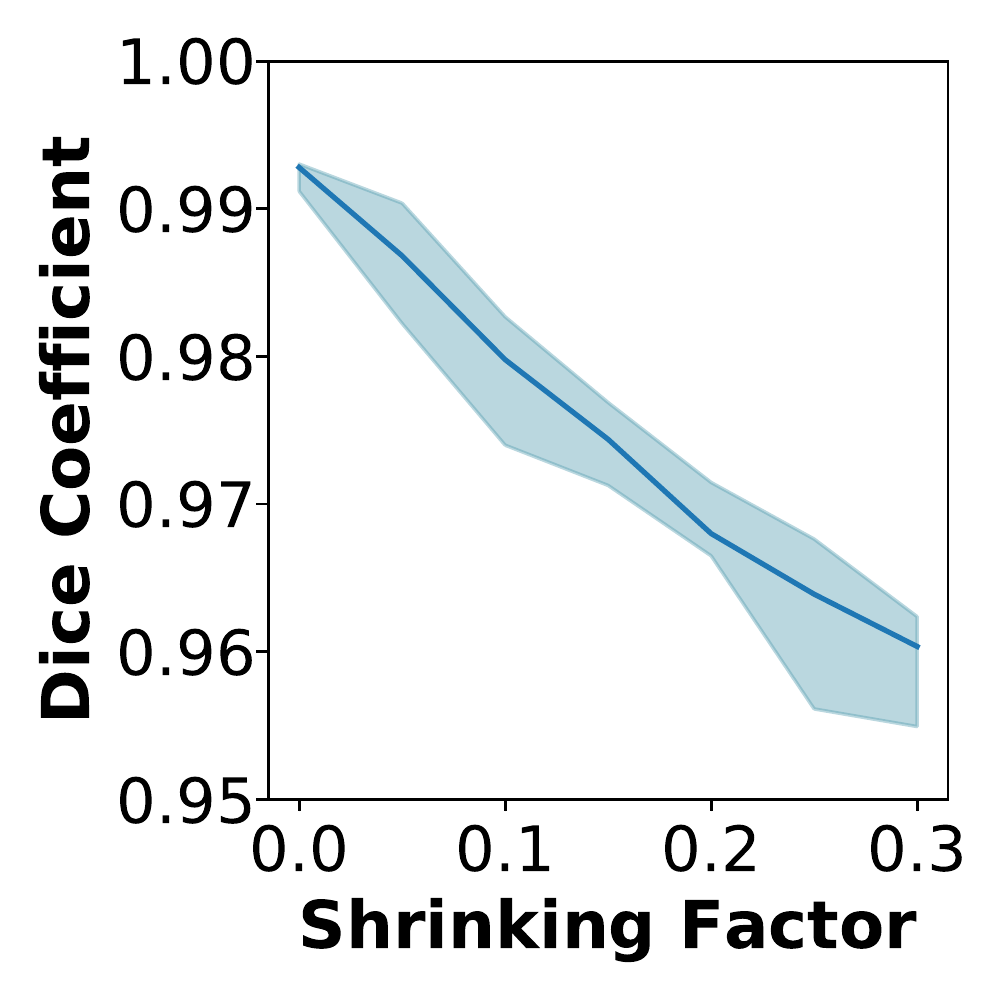}
\caption{}
\end{subfigure}

\begin{subfigure}[b]{0.24\textwidth}
\centering
\includegraphics[trim={0 0 0 0},clip,width=\linewidth]{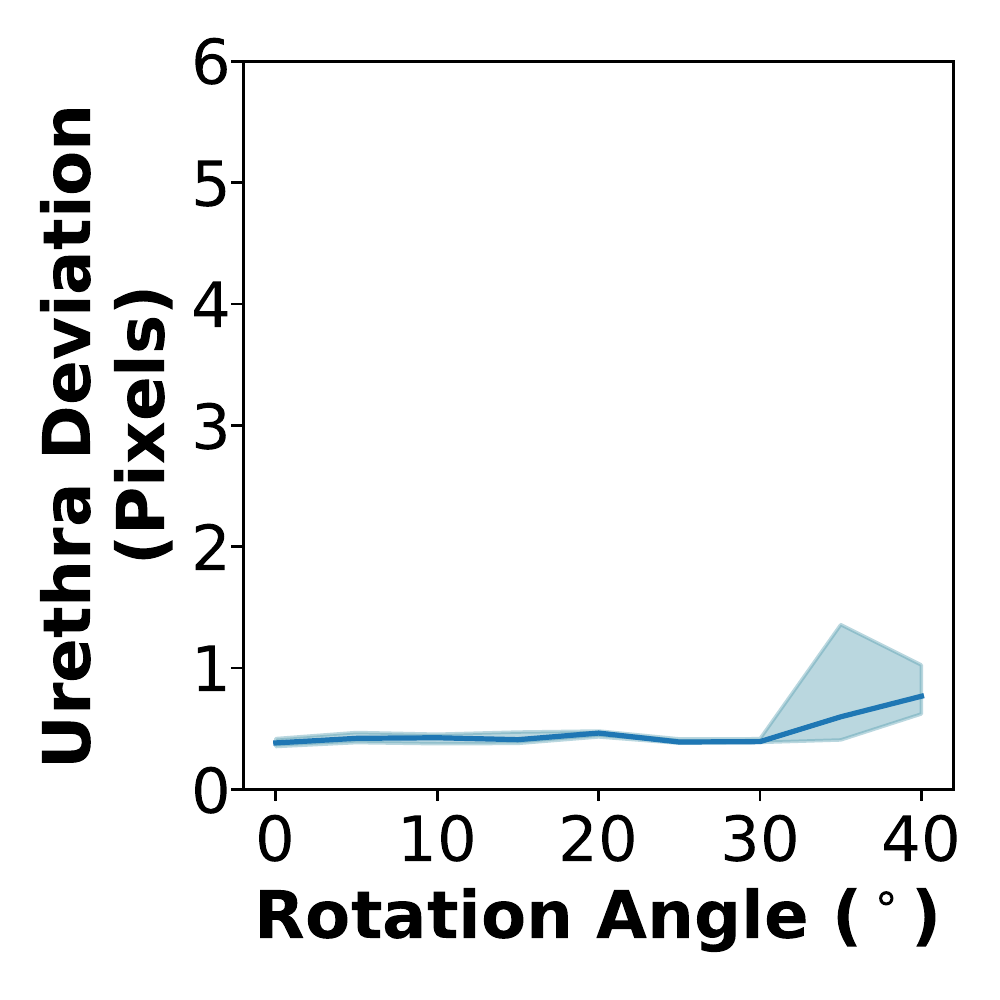}
\caption{}
\end{subfigure}
\begin{subfigure}[b]{0.24\textwidth}
\centering
\includegraphics[trim={0 0 0 0},clip,width=\linewidth]{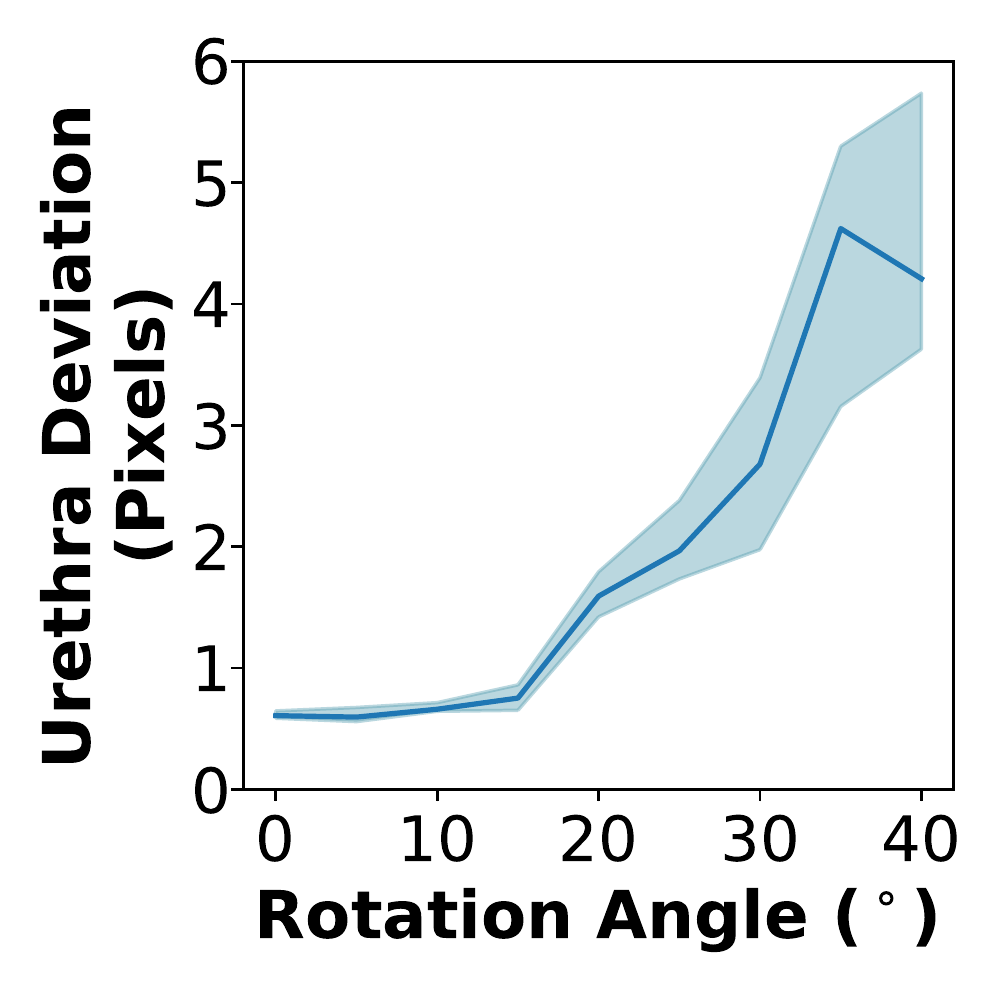}
\caption{}
\end{subfigure}
\begin{subfigure}[b]{0.24\textwidth}
\centering
\includegraphics[trim={0 0 0 0},clip,width=\linewidth]{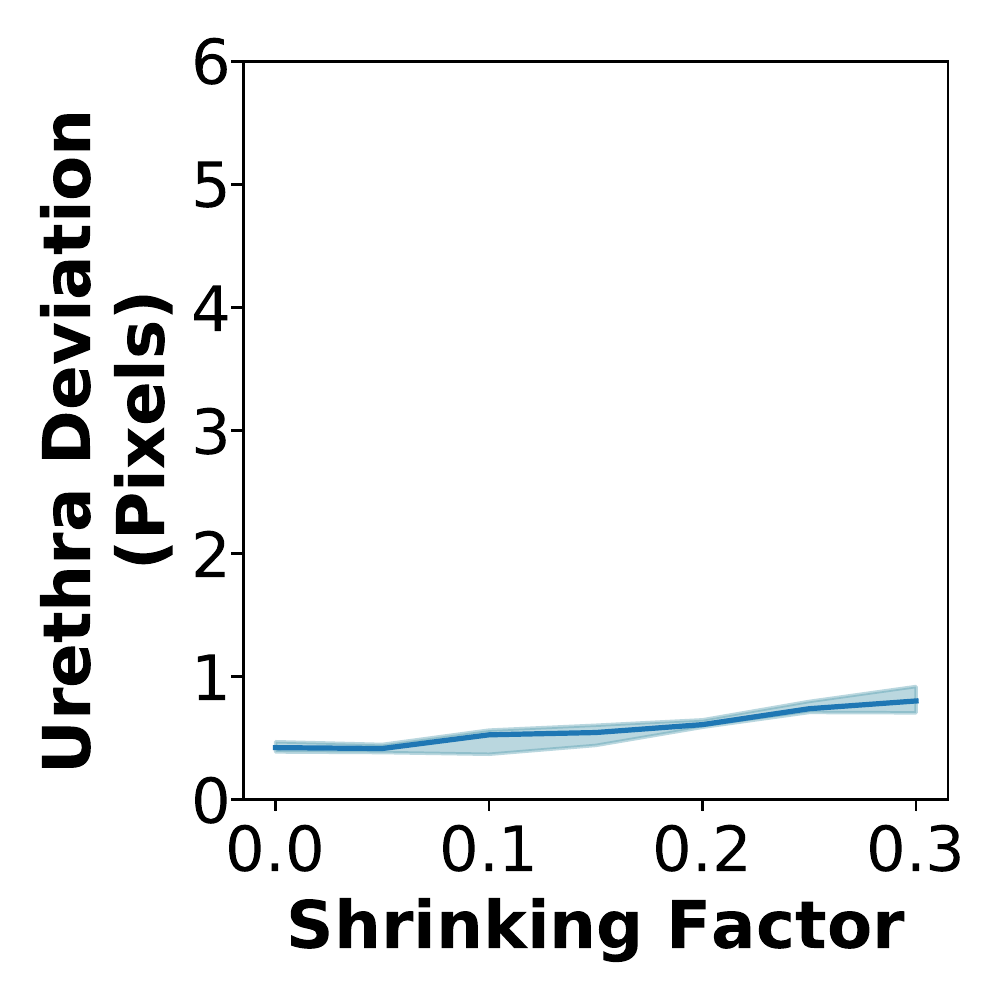}
\caption{}
\end{subfigure}
\begin{subfigure}[b]{0.24\textwidth}
\centering
\includegraphics[trim={0 0 0 0},clip,width=\linewidth]{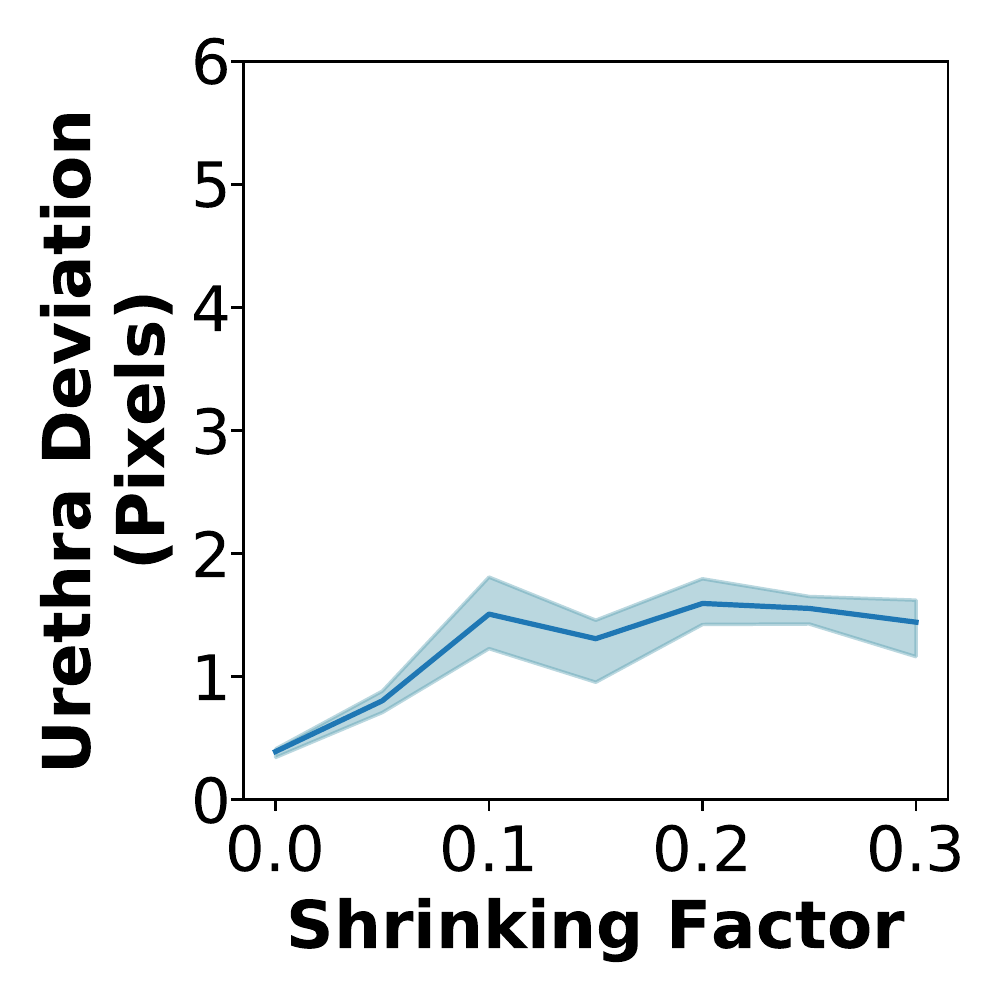}
\caption{}
\end{subfigure}

\begin{subfigure}[b]{0.24\textwidth}
\centering
\includegraphics[trim={0 0 0 0},clip,width=\linewidth]{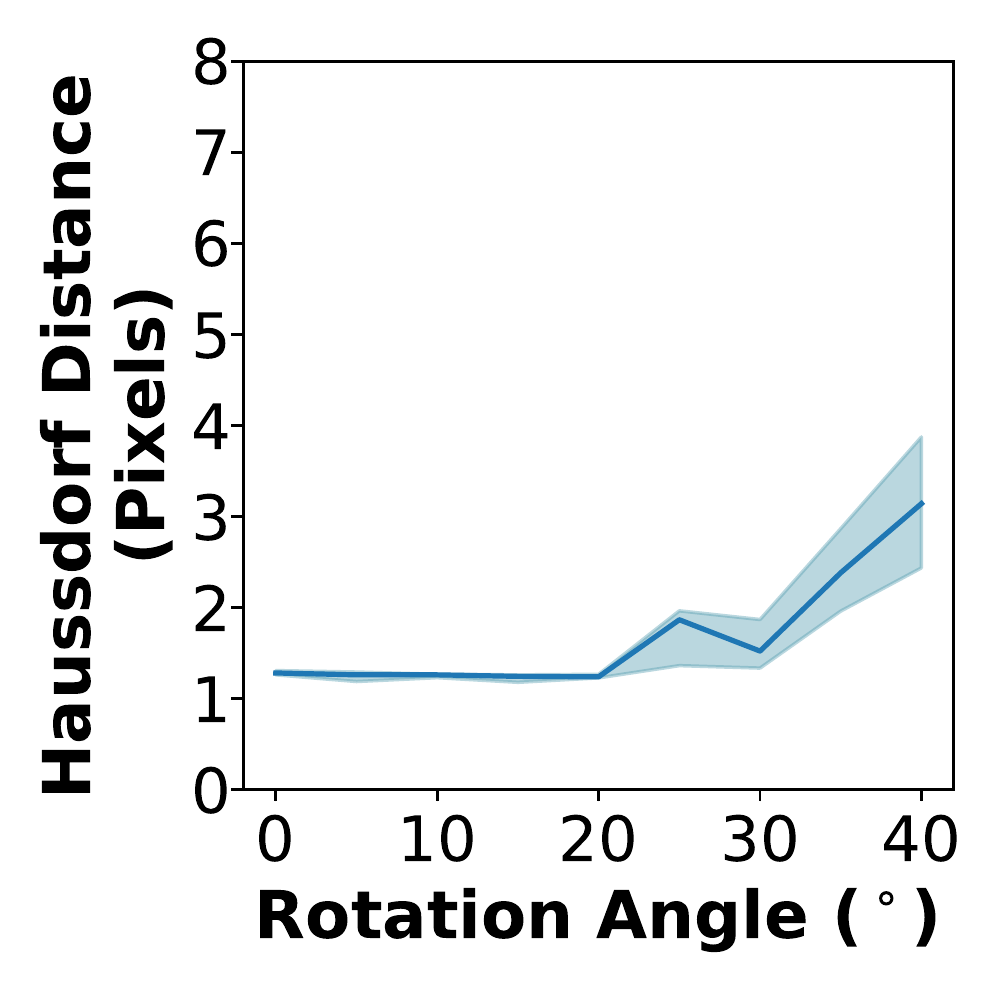}
\caption{}
\end{subfigure}
\begin{subfigure}[b]{0.24\textwidth}
\centering
\includegraphics[trim={0 0 0 0},clip,width=\linewidth]{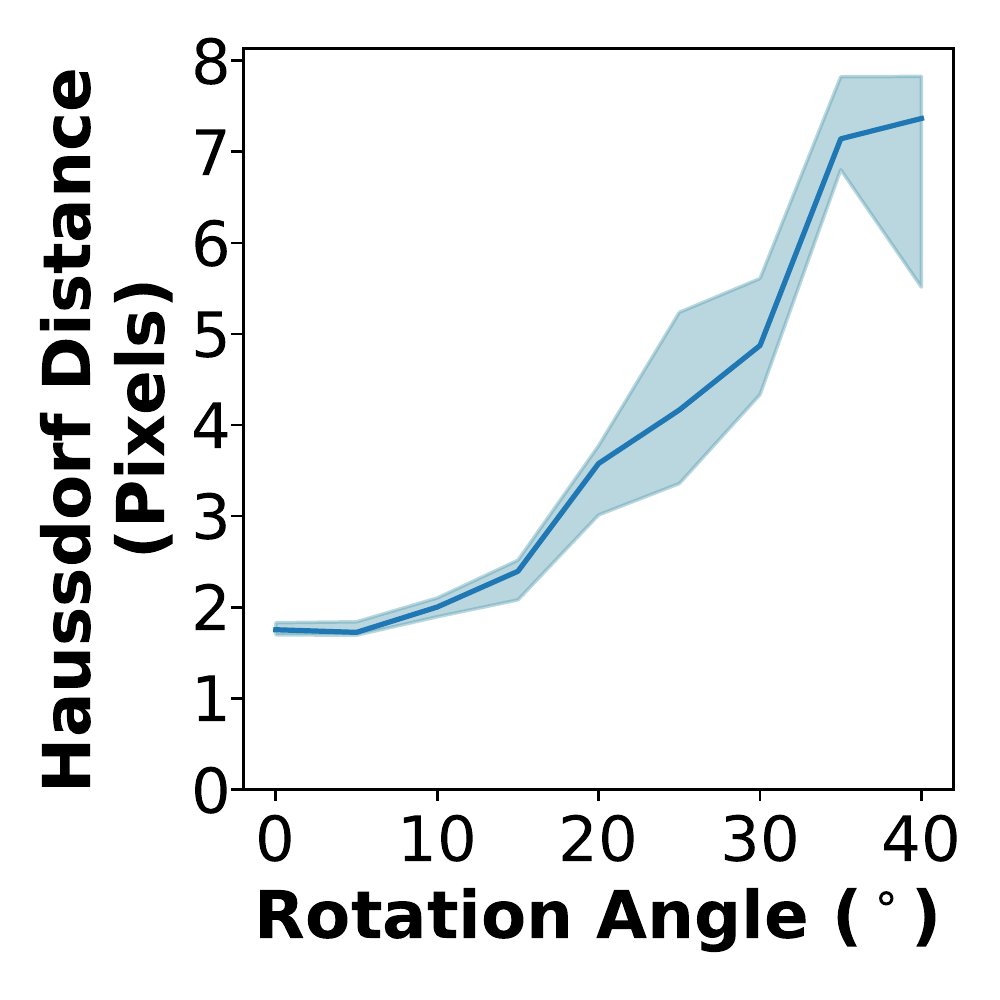}
\caption{}
\end{subfigure}
\begin{subfigure}[b]{0.24\textwidth}
\centering
\includegraphics[trim={0 0 0 0},clip,width=\linewidth]{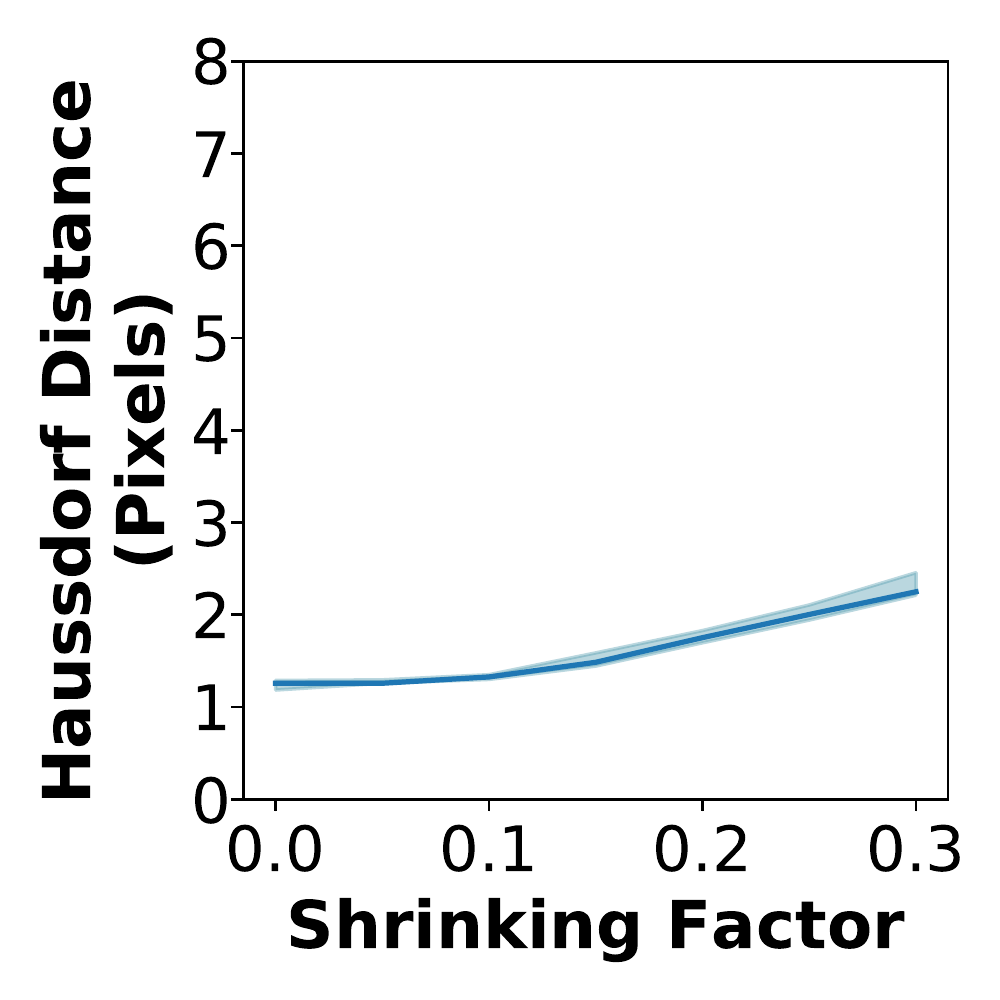}
\caption{}
\end{subfigure}
\begin{subfigure}[b]{0.24\textwidth}
\centering
\includegraphics[trim={0 0 0 0},clip,width=\linewidth]{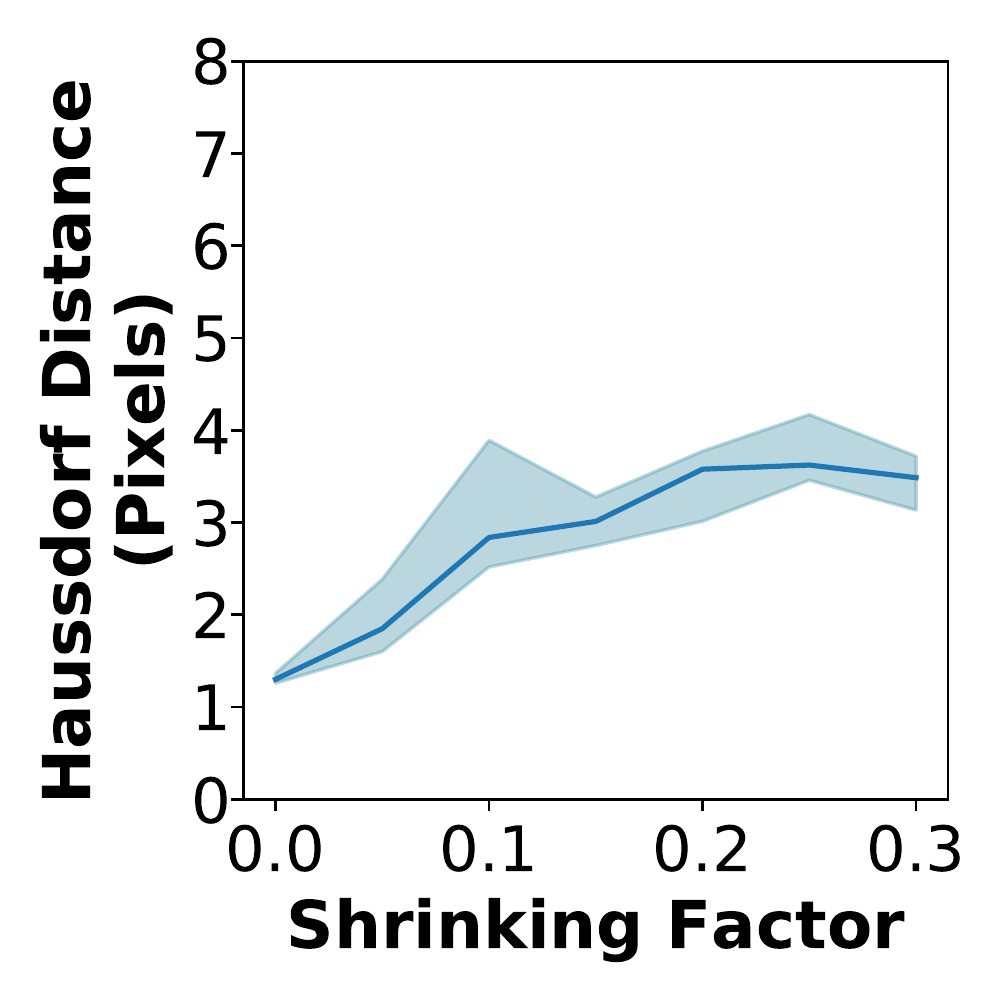}
\caption{}
\end{subfigure}

\caption{RAPSODI results for the registration of histopathology and T2w MRI slices in the digital phantom in terms of (a-d) Dice Coefficient, (e-h) Urethra Deviation, (i-l) Hausdorff Distance. (a,e,i) Effect of the rotation (X-axis) on the quantitative metrics when no shrinkage is applied; (b,f,j) Constant 20\% shrinkage is applied along with a randomly assigned rotations; (c,g,k)  Effect of tissue shrinkage, while no rotation was applied; (d,h,l) A rotation angle of 20$^\circ$ is applied along with the shrinkage.}
\label{fig:PhantomResults}
\end{figure*}

\section*{Results}

\subsection*{Phantom Study}

The phantom study is used to assess the average performance and variability of RAPSODI under conditions known to affect the tissue during the histopathology preparation. We ran our registration approach for 480 different conditions, to estimate the trends of the evaluation metrics as well as to their variations. \figurename~\ref{fig:PhantomResults} summarizes our results in which we tested the effect of the rotation of histopathology slices while mounting on glass slide (range: 0-40$^\circ$), and the effect of shrinkage (range: 0-30\%) when perfect slice correspondences exist between the histopathology and the MRI images in the phantom. Our approach is able to perfectly recover rotation angles ranging between 0-20$^\circ$ or shrinkage of 0-10\% when applied alone (\figurename~\ref{fig:PhantomResults}), indicated by the perfect $\tilde{}$ 1 dice coefficient and the sub-pixel error. When combined, either using 20\% shrinkage and random rotation (\figurename{s}~\ref{fig:PhantomResults}b,f,j) or using 20$^\circ$ rotation and shrinkage (\figurename{s}~\ref{fig:PhantomResults}d,h,l), sub-pixel accuracy was observed for angles ranging between 0-15$^\circ$ or shrinkage of 0-5\%. Beyond these conditions, RAPSODI is still able to recover induced rotation and shrinkage, yet with some misalignment as the initial starting conditions are far from the correct solution.

\begin{figure}[h]
\centering
\begin{subfigure}[b]{0.24\textwidth}
\centering
\includegraphics[trim={0 0 0 0},clip,width=\linewidth]{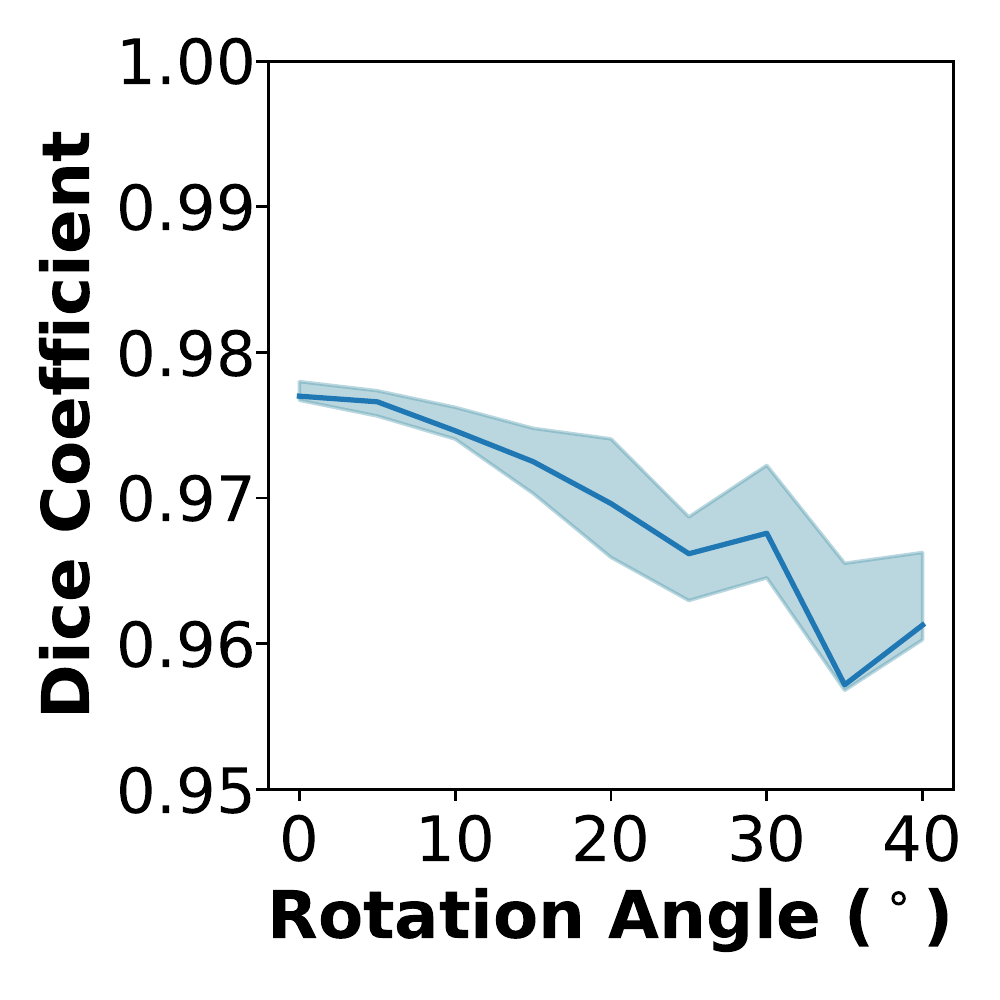}
\caption{}
\end{subfigure} 
\begin{subfigure}[b]{0.24\textwidth}
\centering
\includegraphics[trim={0 0 0 0},clip,width=\linewidth]{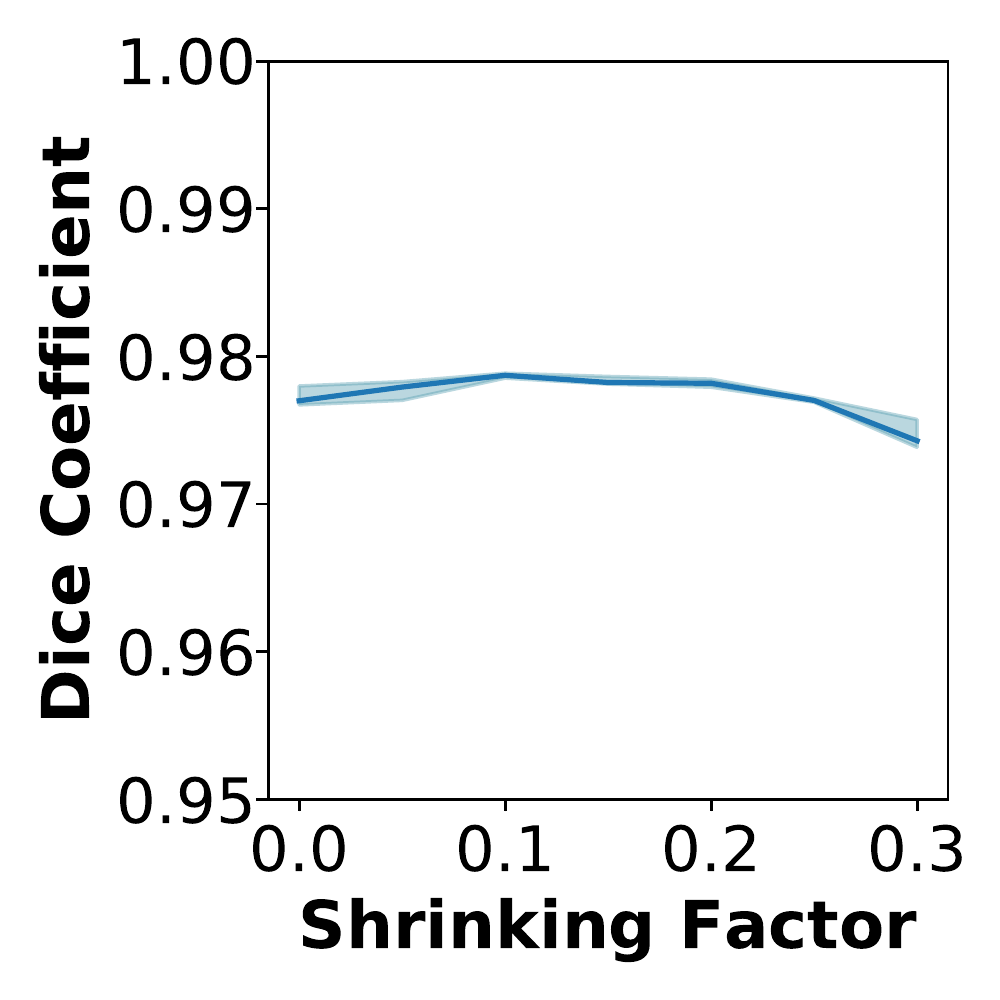}
\caption{}
\end{subfigure}

\begin{subfigure}[b]{0.24\textwidth}
\centering
\includegraphics[trim={0 0 0 0},clip,width=\linewidth]{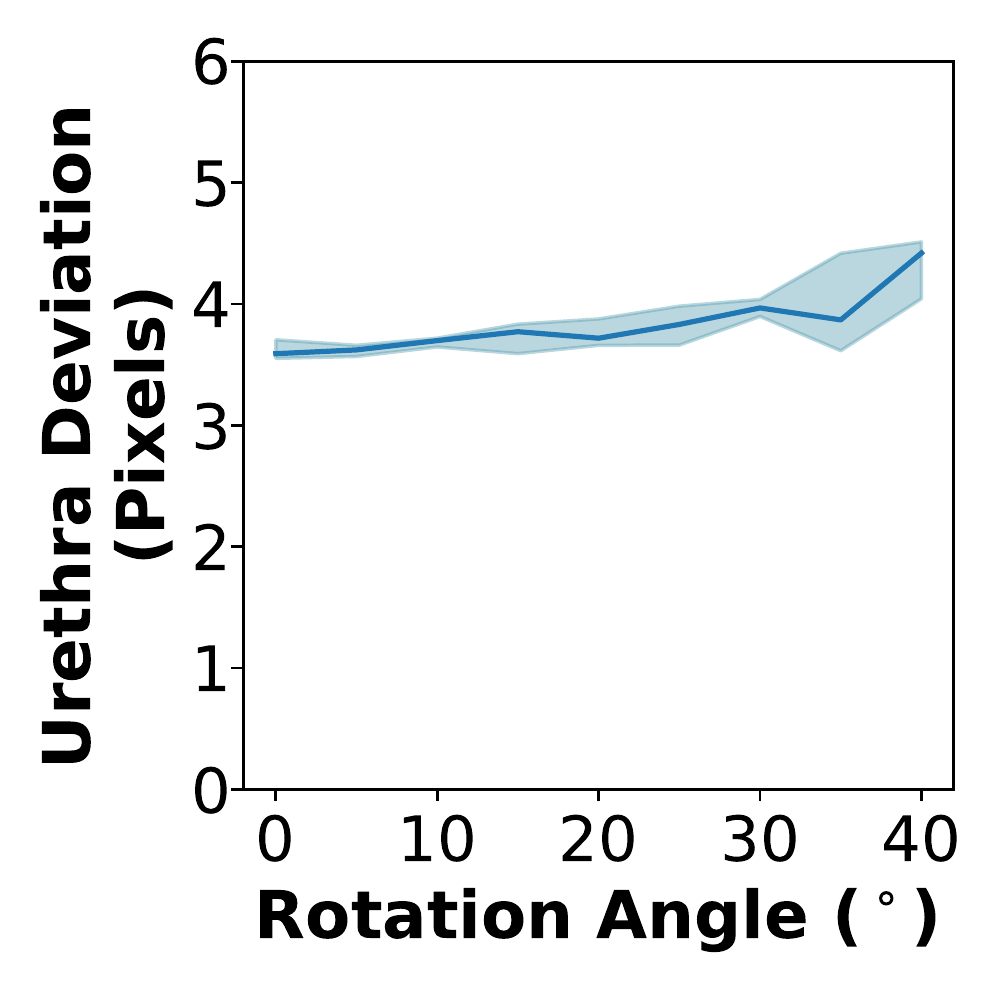}
\caption{}
\end{subfigure}
\begin{subfigure}[b]{0.24\textwidth}
\centering
\includegraphics[trim={0 0 0 0},clip,width=\linewidth]{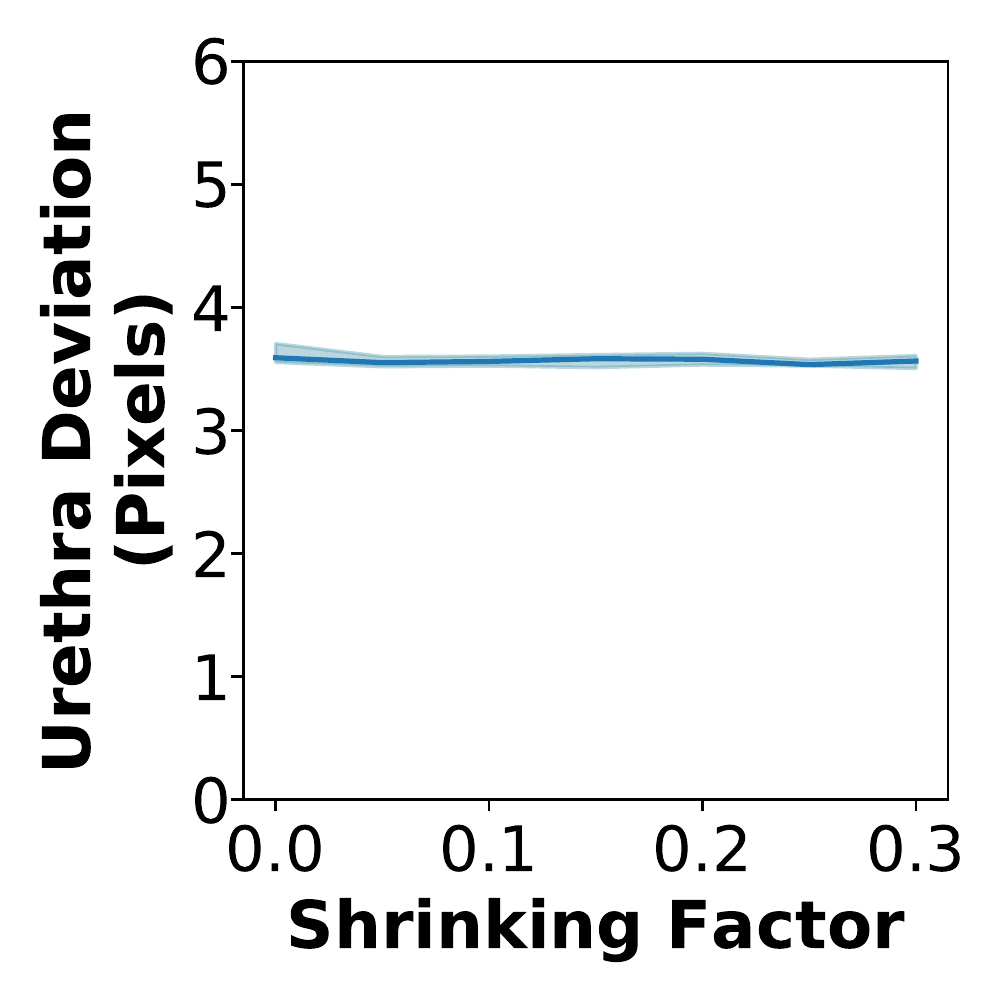}
\caption{}
\end{subfigure}

\begin{subfigure}[b]{0.24\textwidth}
\centering
\includegraphics[trim={0 0 0 0},clip,width=\linewidth]{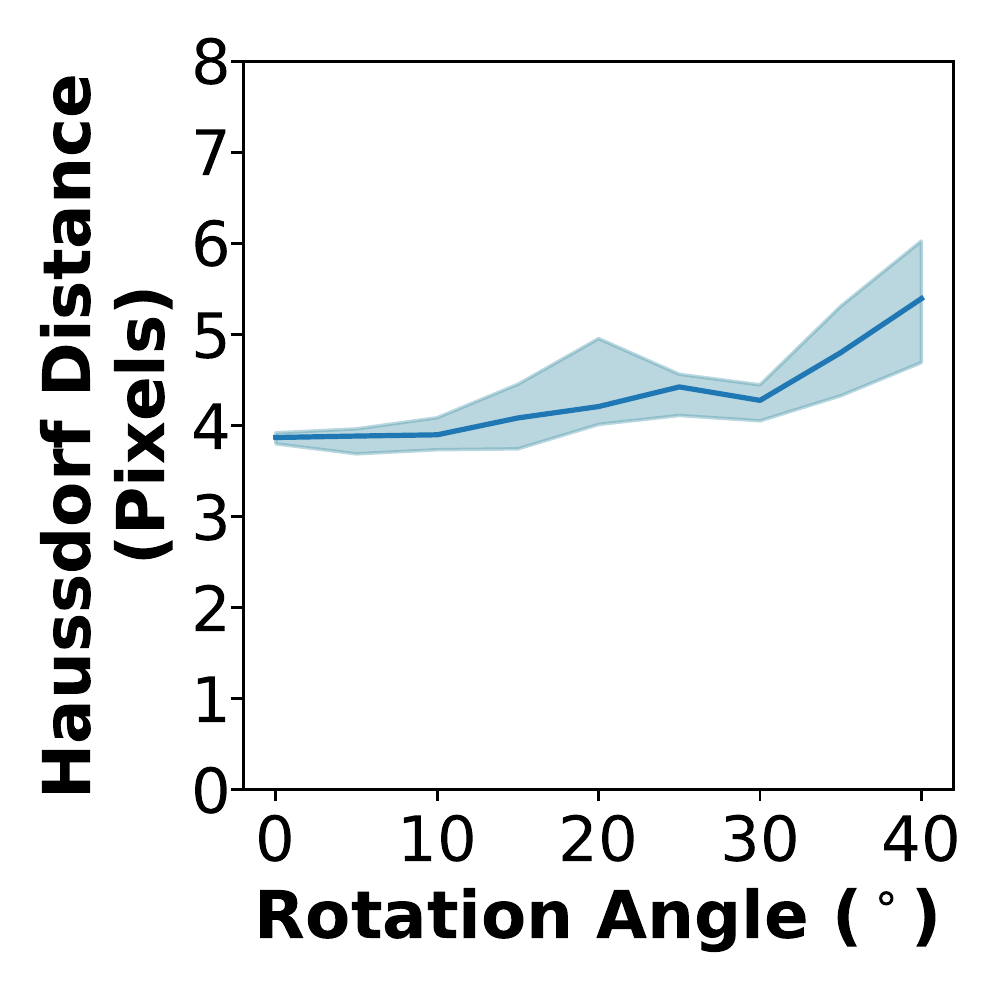}
\caption{}
\end{subfigure}
\begin{subfigure}[b]{0.24\textwidth}
\centering
\includegraphics[trim={0 0 0 0},clip,width=\linewidth]{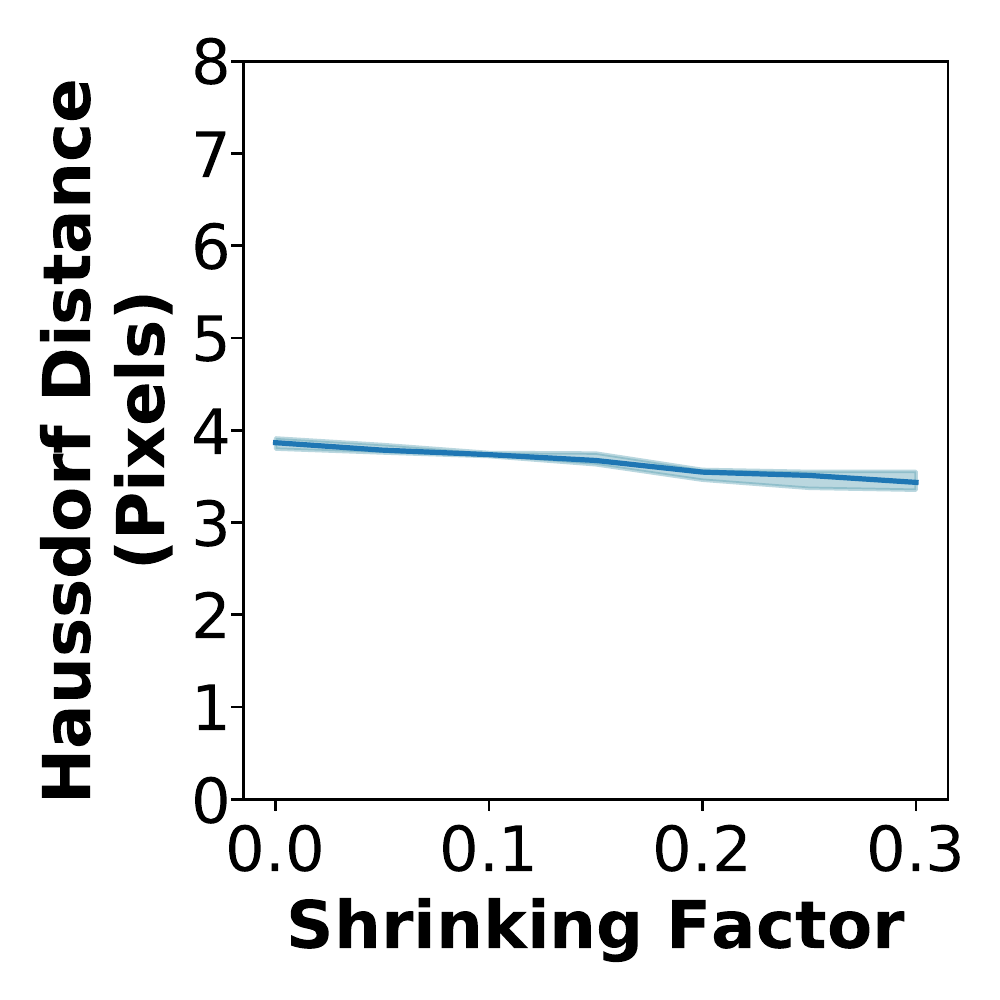}
\caption{}
\end{subfigure}
\caption{RAPSODI results for the registration of histopathology and T2w MRI slices in the digital phantom where an imperfect correspondence between the histopathology and T2w MRI slices exist (they are 2 mm apart from each other in the Sagittal and coronal planes, e.g., \figurename~\ref{fig:Phantom}d,f): (a-b) Dice Coefficient; (c-d) Urethra Deviation; (e-f) Hausdorff Distance. (a,c,e) Experiment where only the rotation angle was varied between 0-40$^\circ$; (b,d,f) The histopathology images were shrunk by 0-30\% of the original size.}
\label{fig:PhantomResults3}
\end{figure}

Moreover, the limitations of the registration may be observed when perfect correspondences are lacking between the histopathology and MRI slices (\figurename~\ref{fig:PhantomResults3}). Not surprising, the landmark and prostate border deviation are as large as $\sim$ 4 pixels (1.6 mm), as these features are not perfectly matching. Yet we can observe the relative stability of the approach for rotations ranging 0-30$^\circ$ and all tested shrinkage factors ranging between 0-30\%, as the induced rotation and shrinkage are properly recovered.

\subsection*{Qualitative Results}

\begin{figure*}
\begin{subfigure}[b]{0.31\textwidth}
\centering
\includegraphics[trim={0.5cm 2cm 1.5cm 0},clip,width=\linewidth]{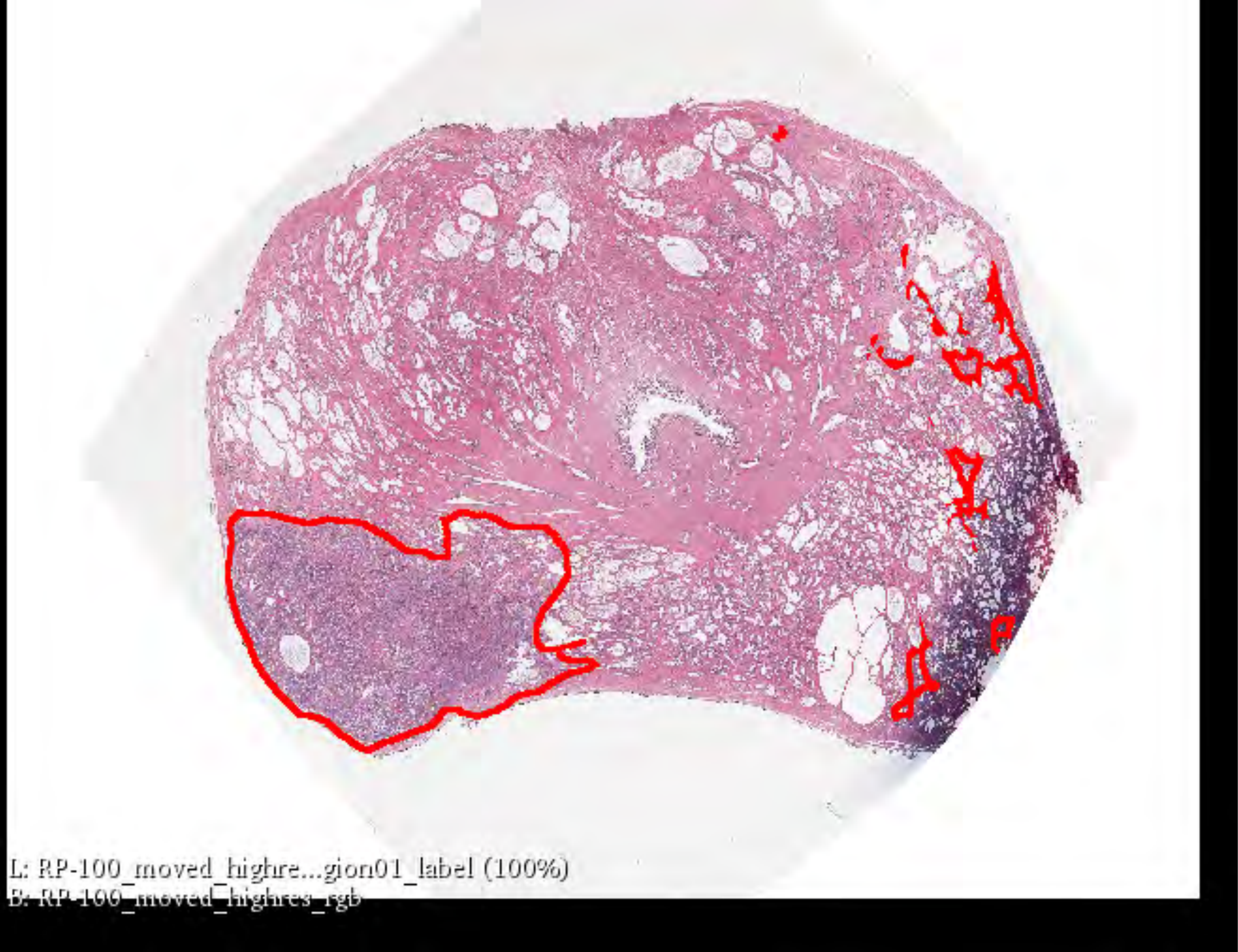}
\end{subfigure}
\begin{subfigure}[b]{0.31\textwidth}
\centering
\includegraphics[trim={0.5cm 2cm 0.5cm 0},clip,width=\linewidth]{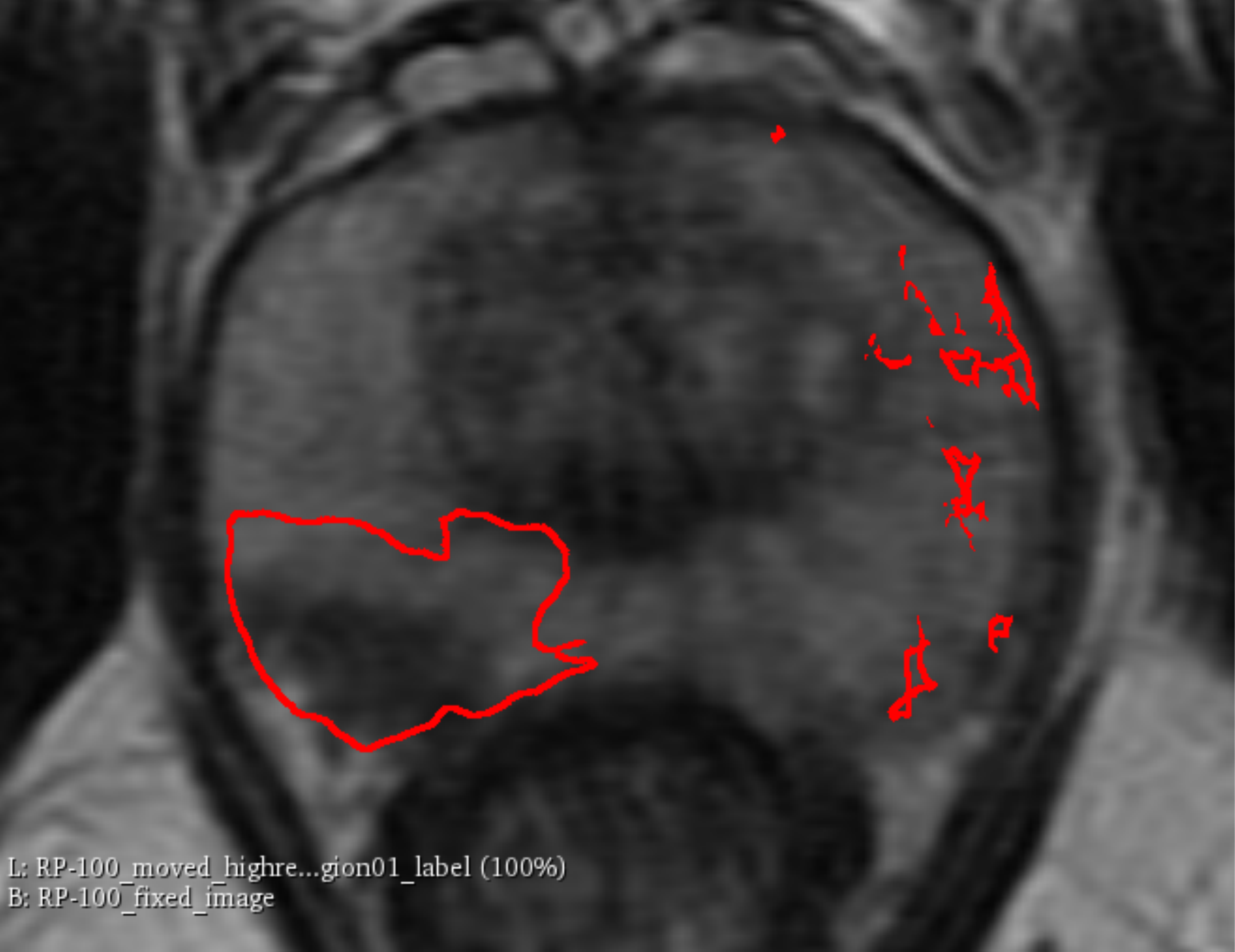}
\end{subfigure}
\begin{subfigure}[b]{0.31\textwidth}
\centering
\includegraphics[trim={0.5cm 2cm 0.5cm 0},clip,width=\linewidth]{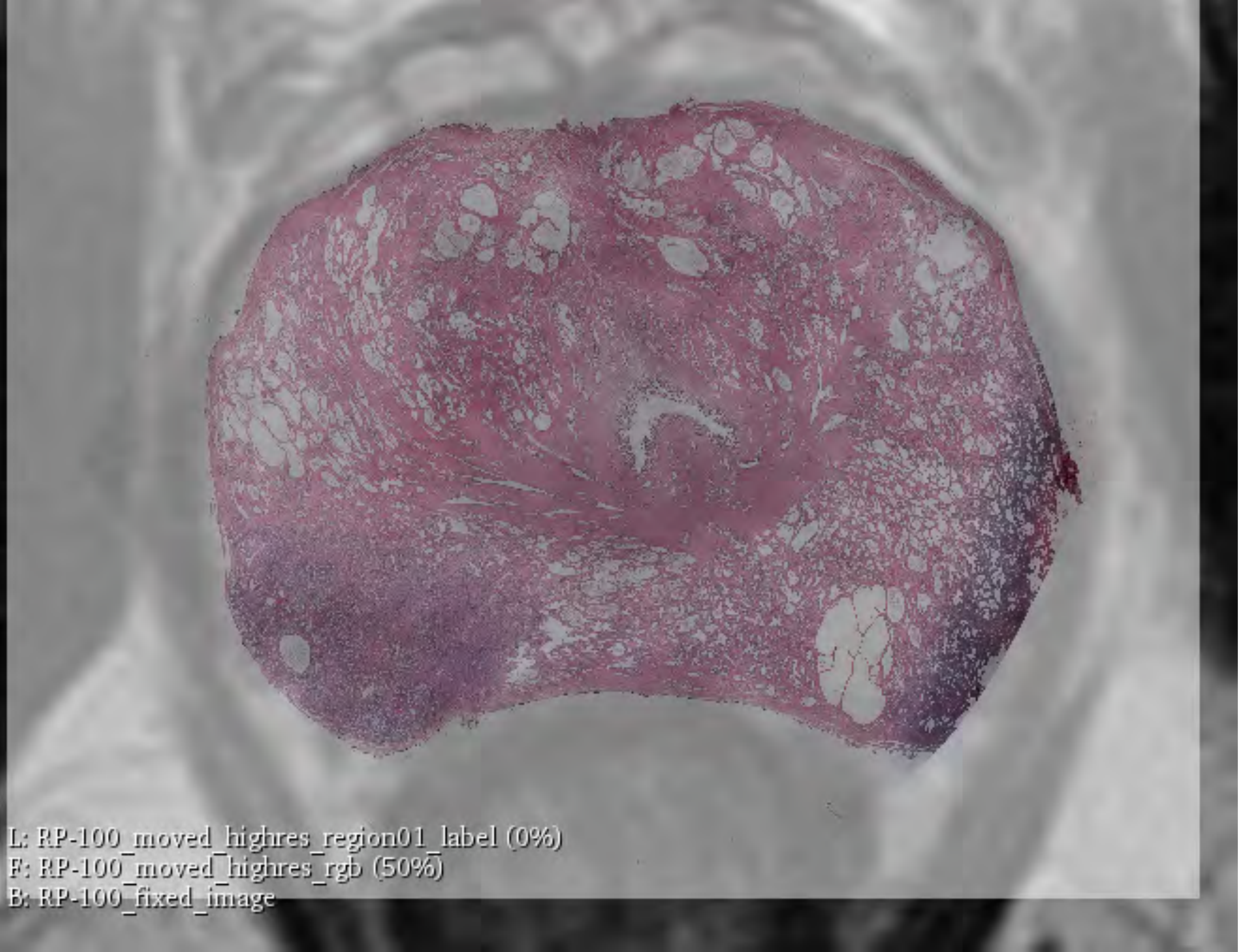}
\end{subfigure}

\begin{subfigure}[b]{0.31\textwidth}
\centering
\includegraphics[trim={0.5cm 2cm 1.5cm 0},clip,width=\linewidth]{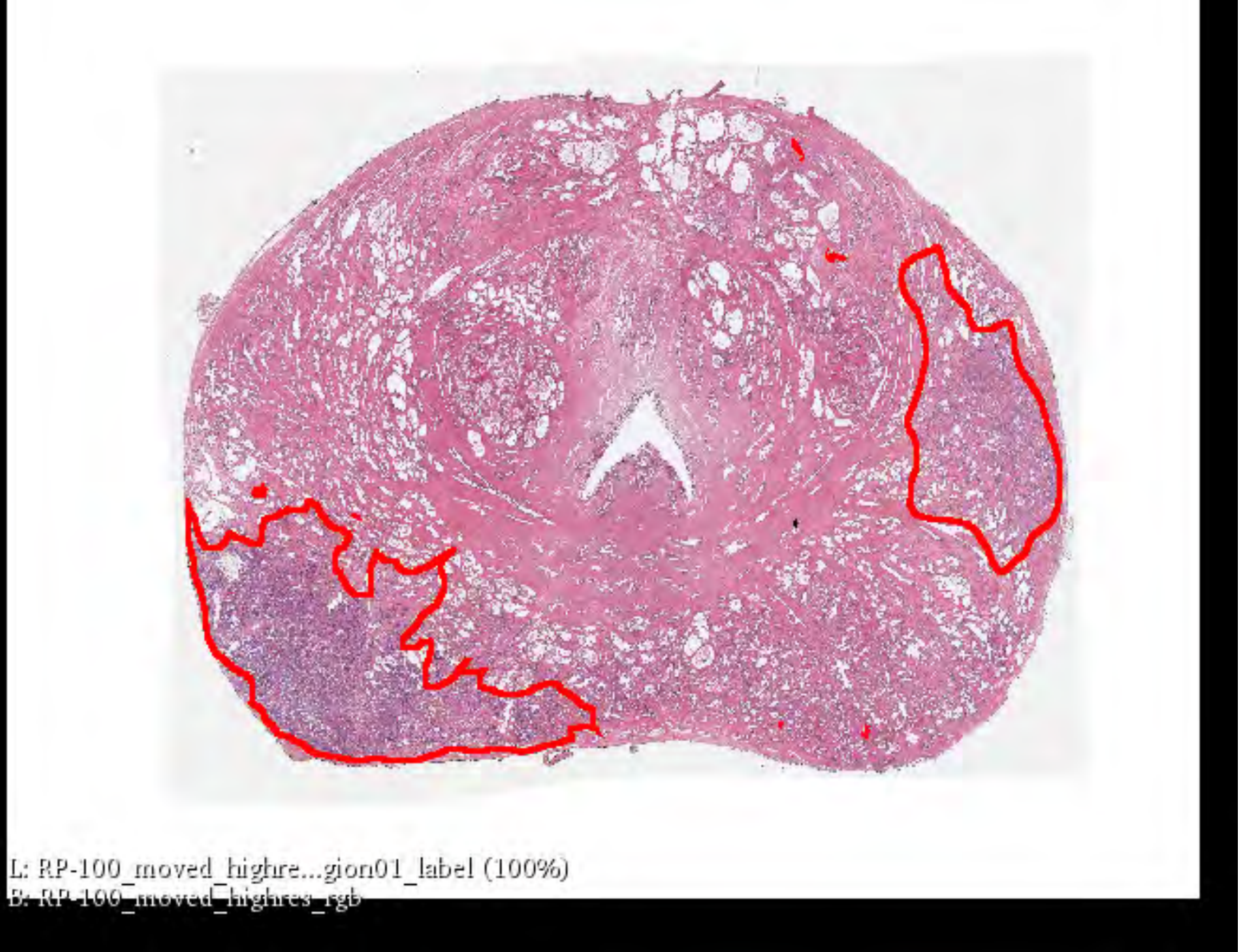}
\end{subfigure}
\begin{subfigure}[b]{0.31\textwidth}
\centering
\includegraphics[trim={0.5cm 2cm 0.5cm 0},clip,width=\linewidth]{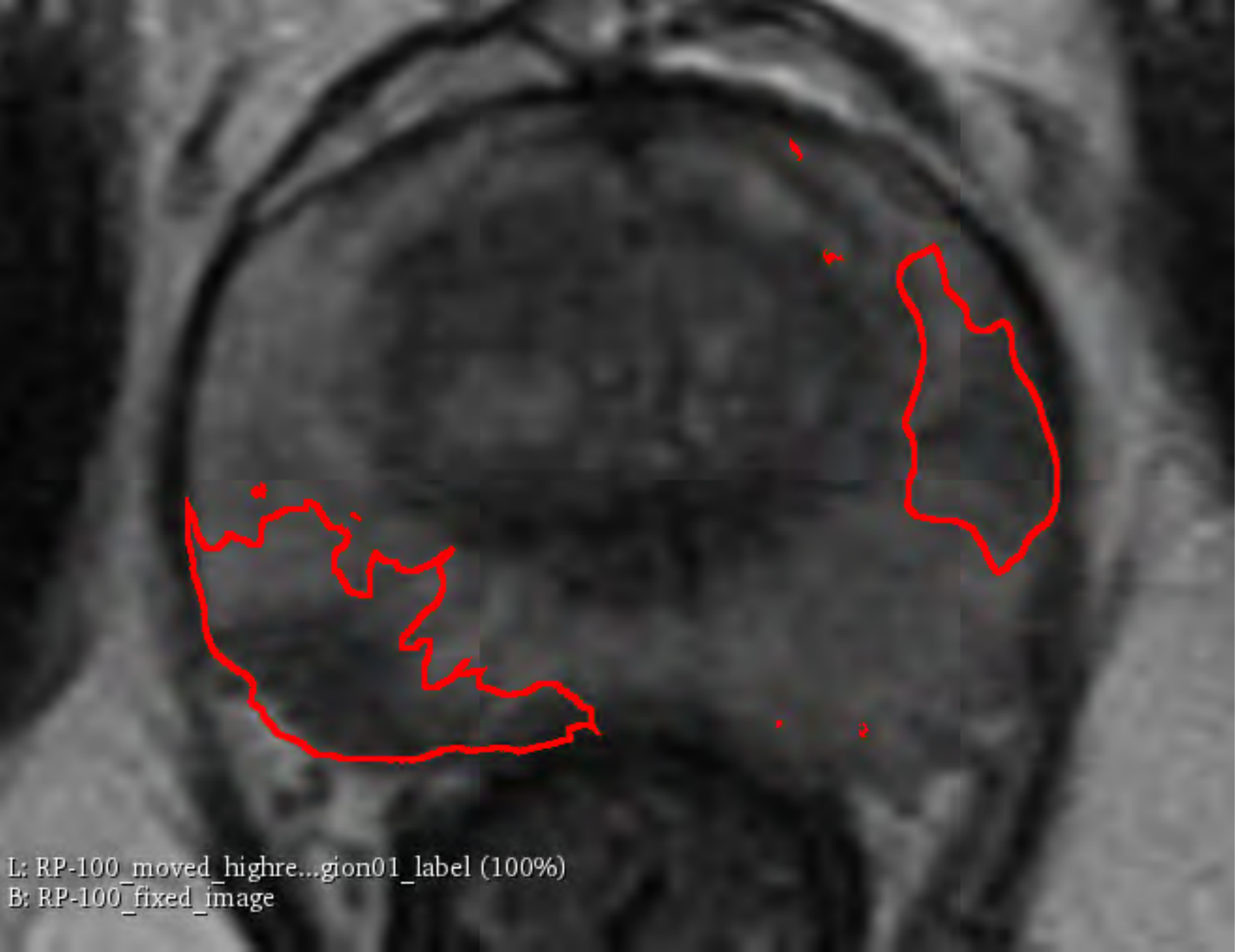}
\end{subfigure}
\begin{subfigure}[b]{0.31\textwidth}
\centering
\includegraphics[trim={0.5cm 2cm 0.5cm 0},clip,width=\linewidth]{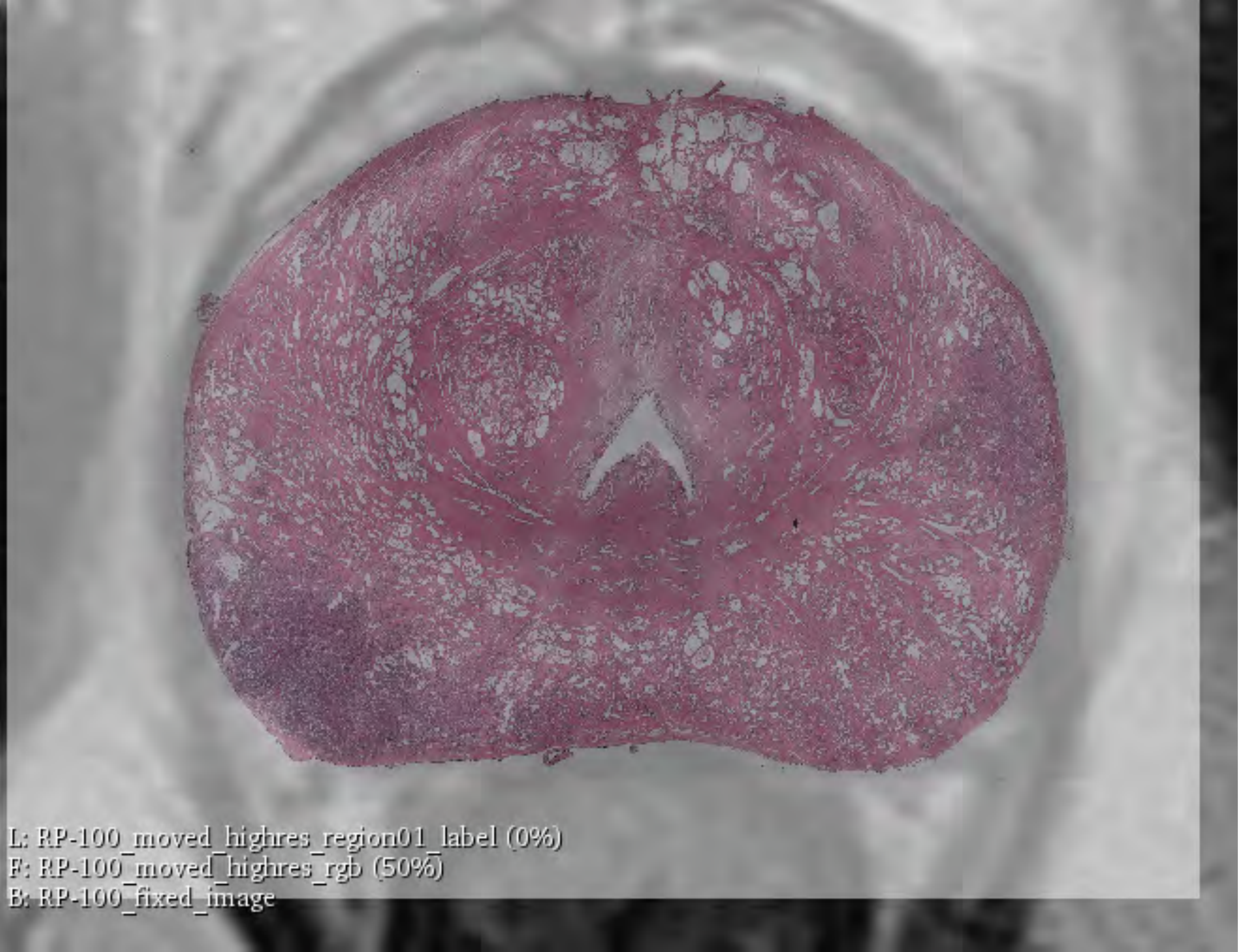}
\end{subfigure}

\begin{subfigure}[b]{0.31\textwidth}
\centering
\includegraphics[trim={0.5cm 2cm 1.5cm 0},clip,width=\linewidth]{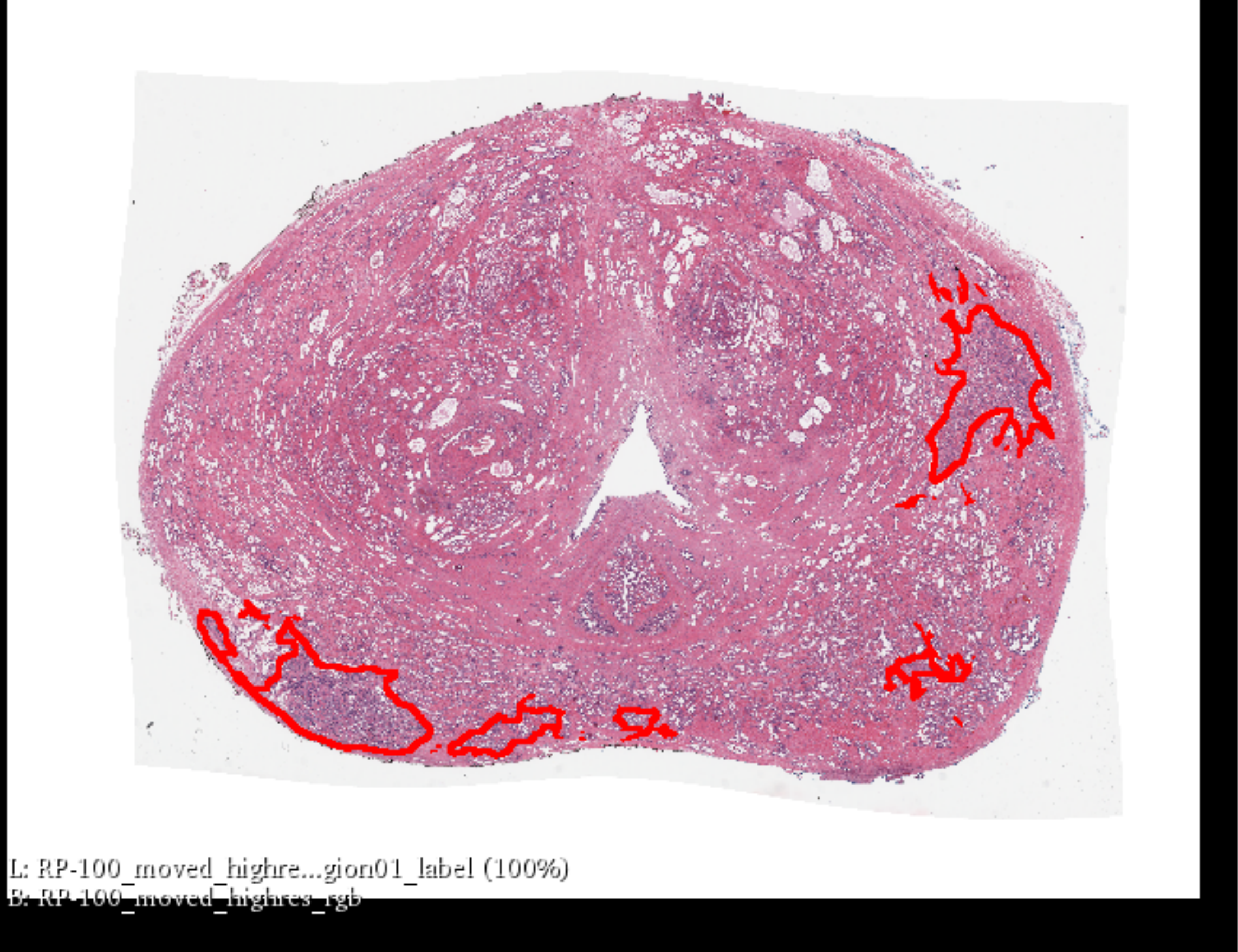}
\end{subfigure}
\begin{subfigure}[b]{0.31\textwidth}
\centering
\includegraphics[trim={0.5cm 2cm 0.5cm 0},clip,width=\linewidth]{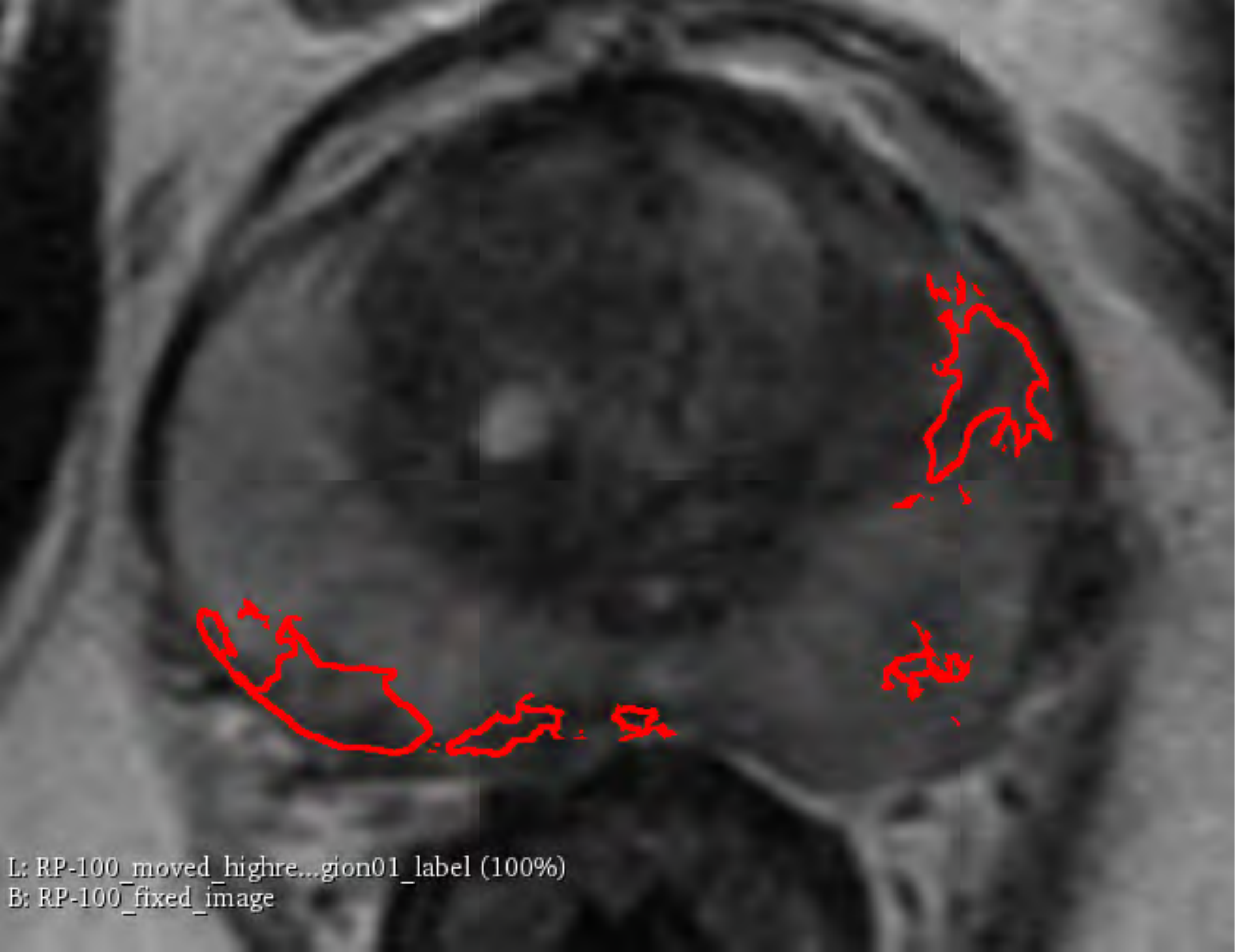}
\end{subfigure}
\begin{subfigure}[b]{0.31\textwidth}
\centering
\includegraphics[trim={0.5cm 2cm 0.5cm 0},clip,width=\linewidth]{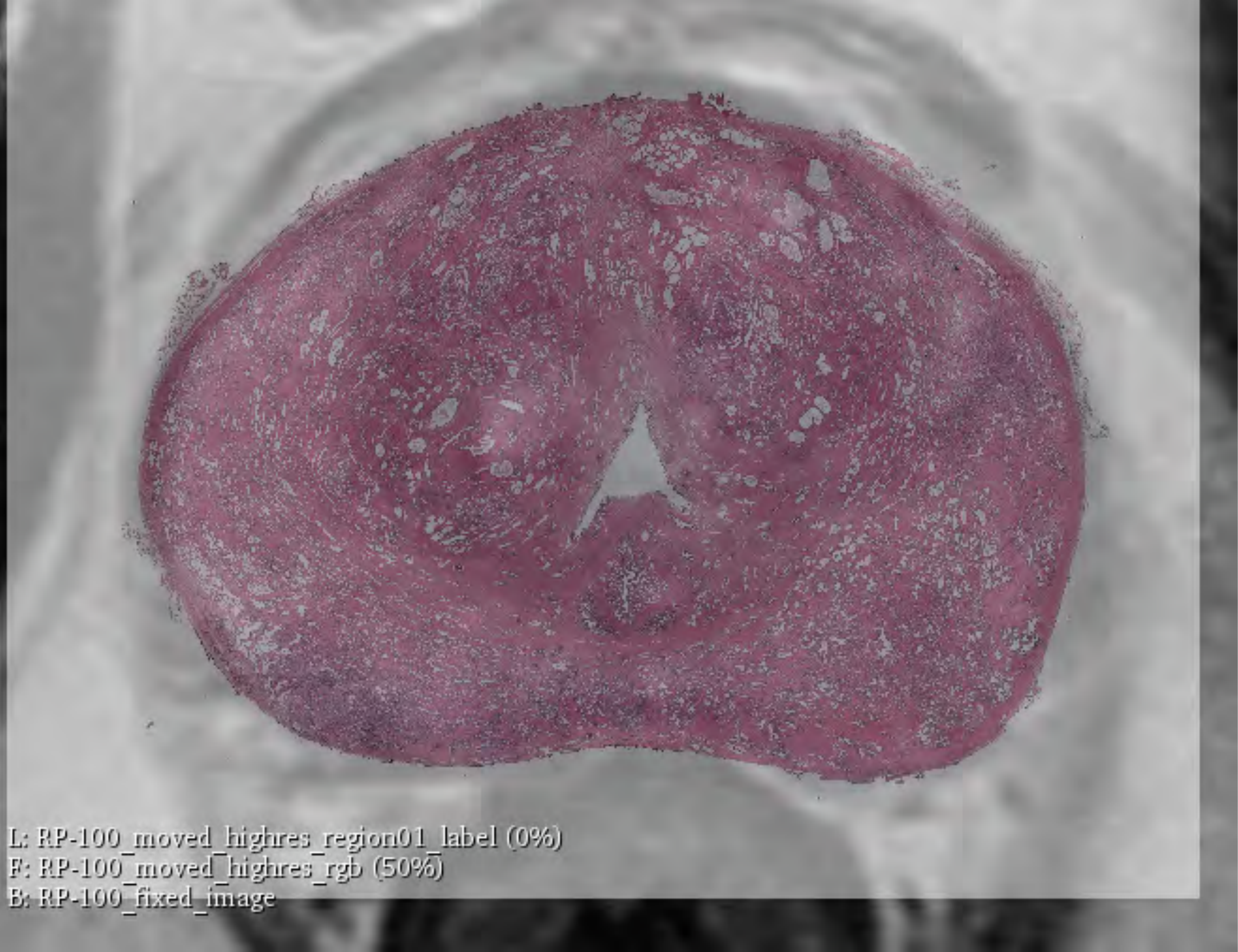}
\end{subfigure}

\begin{subfigure}[b]{0.31\textwidth}
\centering
\includegraphics[trim={0.5cm 2cm 1.5cm 0},clip,width=\linewidth]{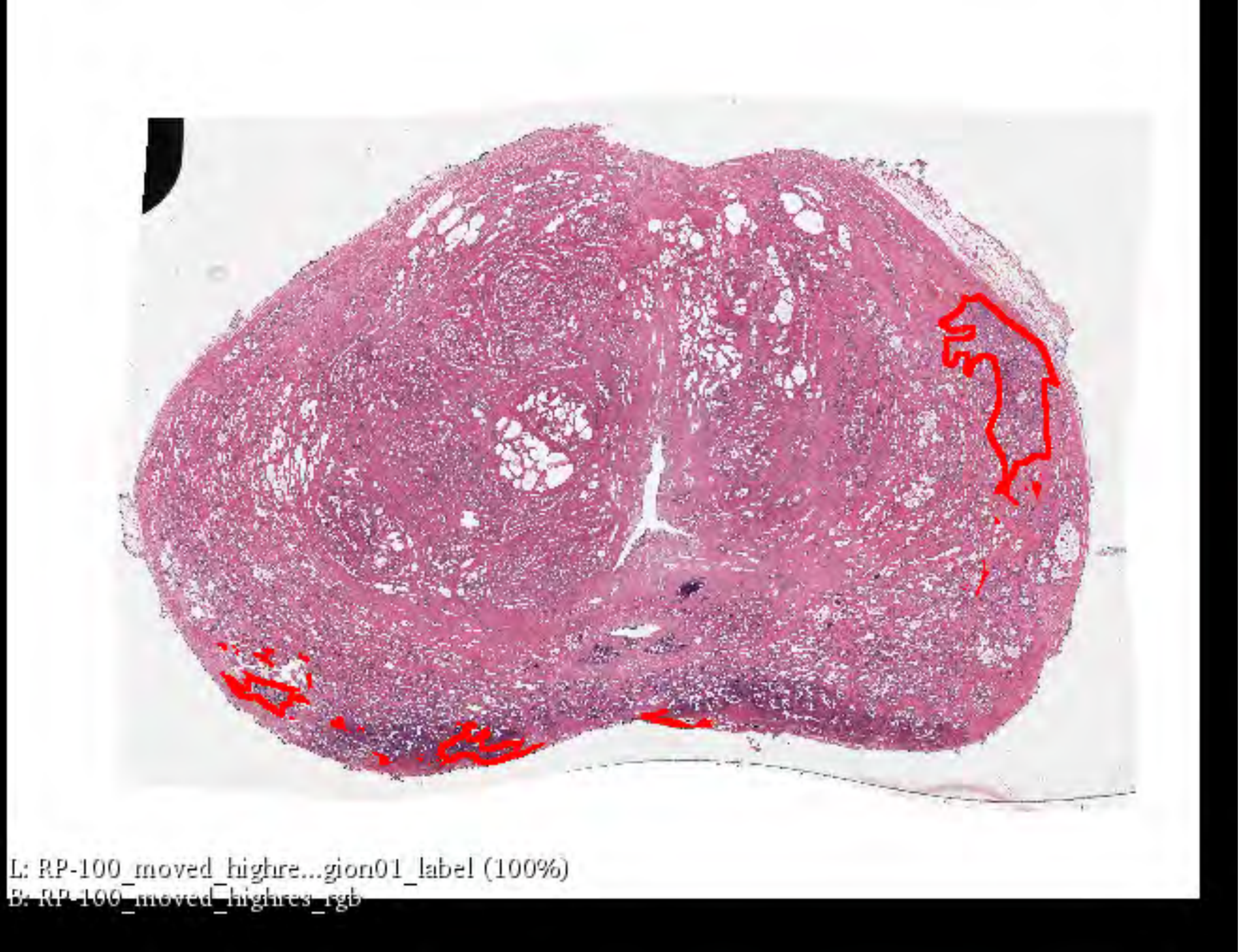}
\end{subfigure}
\begin{subfigure}[b]{0.31\textwidth}
\centering
\includegraphics[trim={0.5cm 2cm 0.5cm 0},clip,width=\linewidth]{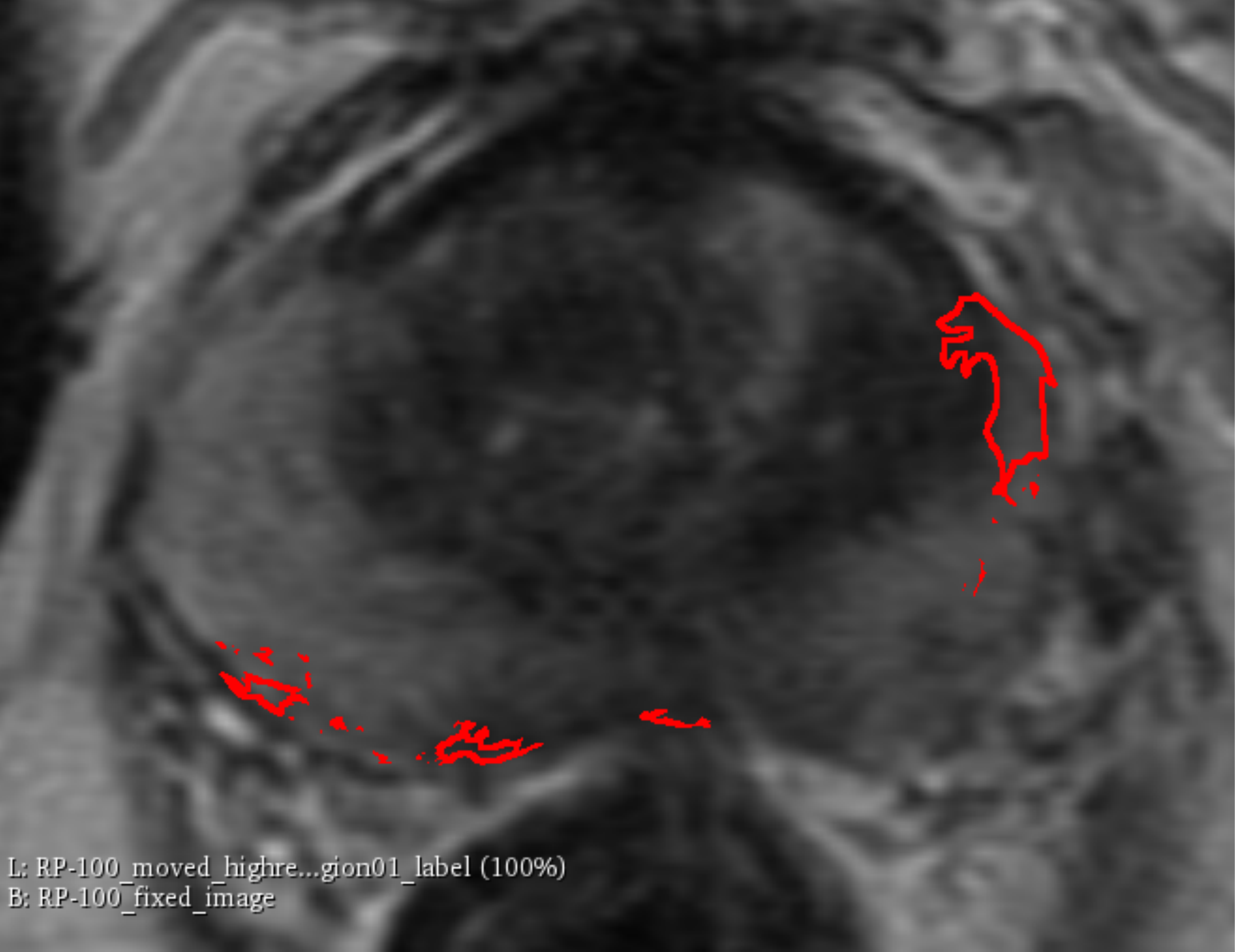}
\end{subfigure}
\begin{subfigure}[b]{0.31\textwidth}
\centering
\includegraphics[trim={0.5cm 2cm 0.5cm 0},clip,width=\linewidth]{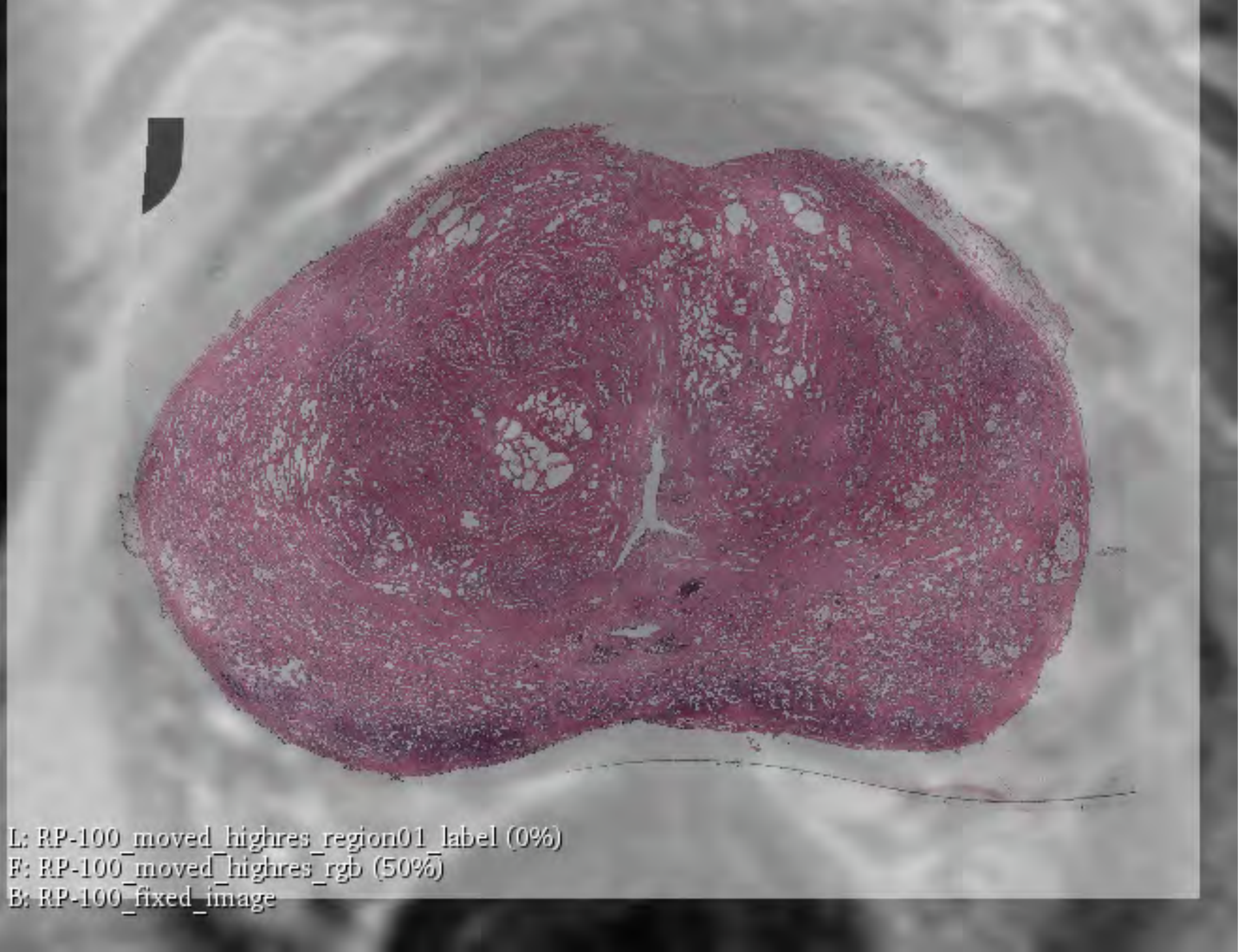}
\end{subfigure}

\begin{subfigure}[b]{0.31\textwidth}
\centering
\includegraphics[trim={0.5cm 2cm 1.5cm 0},clip,width=\linewidth]{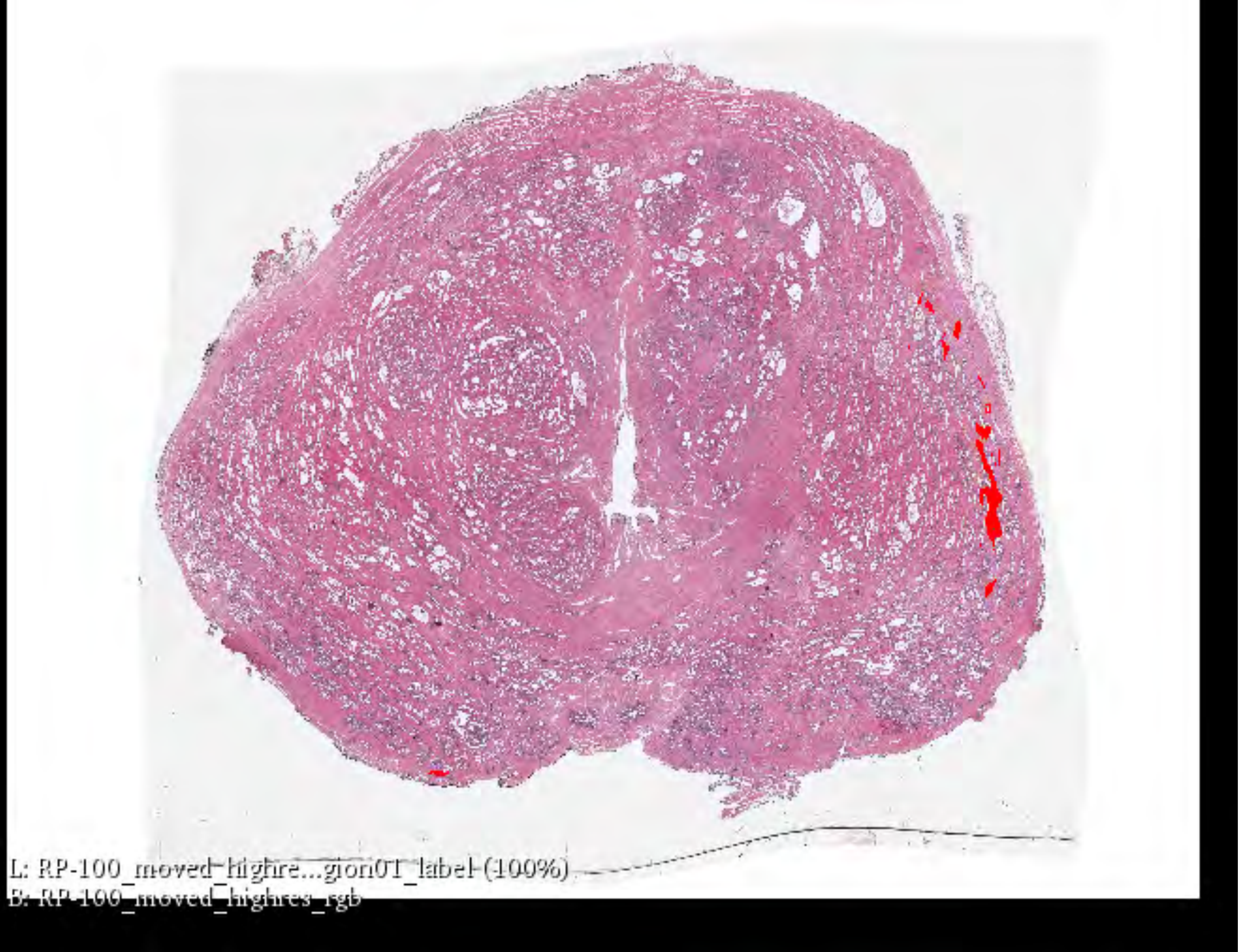}
\end{subfigure}
\begin{subfigure}[b]{0.31\textwidth}
\centering
\includegraphics[trim={0.5cm 2cm 0.5cm 0},clip,width=\linewidth]{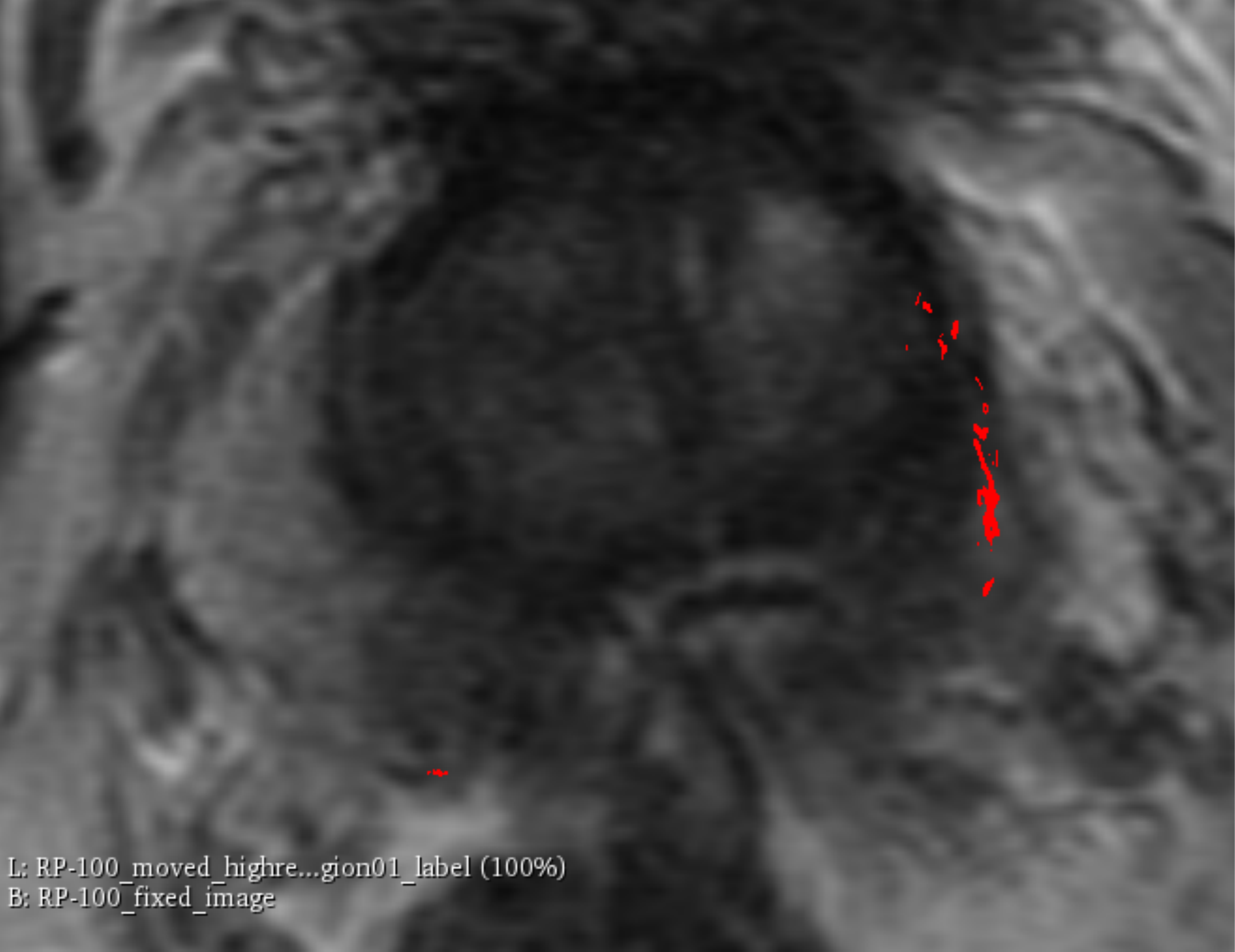}
\end{subfigure}
\begin{subfigure}[b]{0.31\textwidth}
\centering
\includegraphics[trim={0.5cm 2cm 0.5cm 0},clip,width=\linewidth]{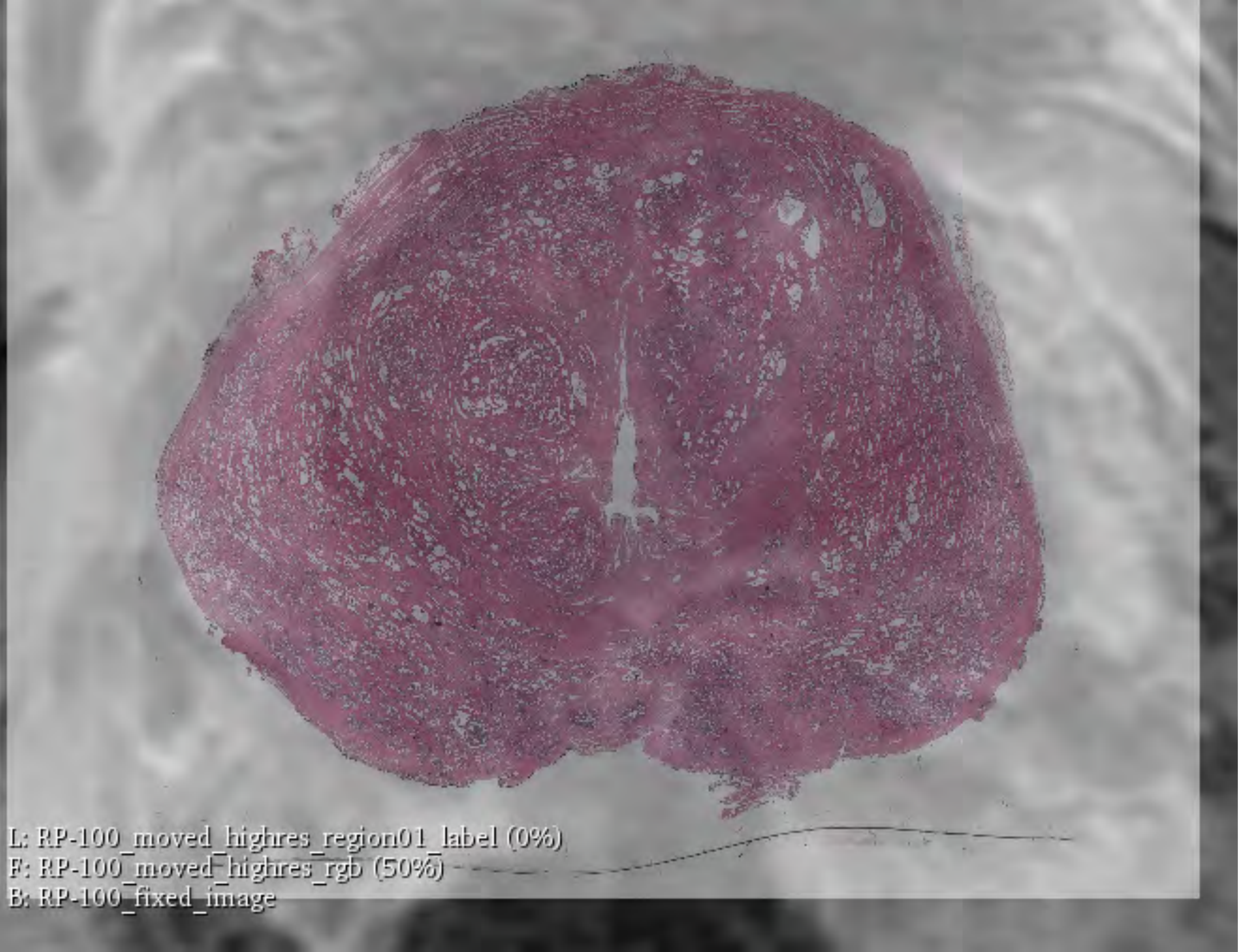}
\end{subfigure}

\caption{Qualitative results showing the registration for all the histopathology slices from apex to base in one subject in Cohort C1. (Left) Histopathology slices with cancer outlines (red); (Middle) Corresponding T2w MRI with cancer outlines obtained via RAPSODI; (Right) overlay of the registered histopathology images and corresponding T2w MRI with histopathology images shown transparent.}
\label{fig:ResultsQualitativeC1}
\end{figure*}

\begin{figure*}[t]
\begin{subfigure}[b]{0.48\textwidth}
\centering
\includegraphics[trim={0 0.5cm 0 0},clip,width=\linewidth]{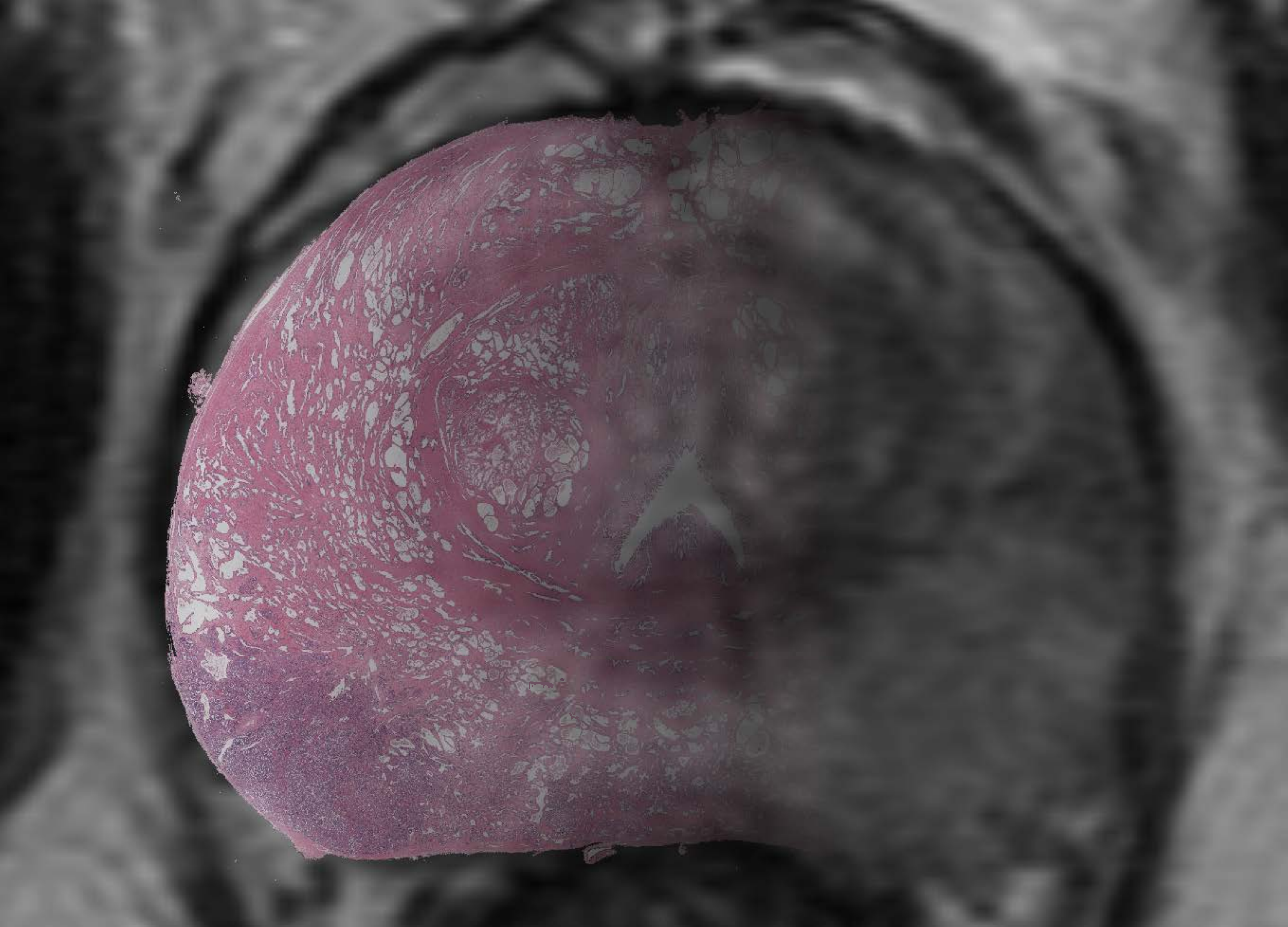}
\caption{}
\end{subfigure}
\begin{subfigure}[b]{0.48\textwidth}
\centering
\includegraphics[trim={0 0.5cm 0 0},clip,width=\linewidth]{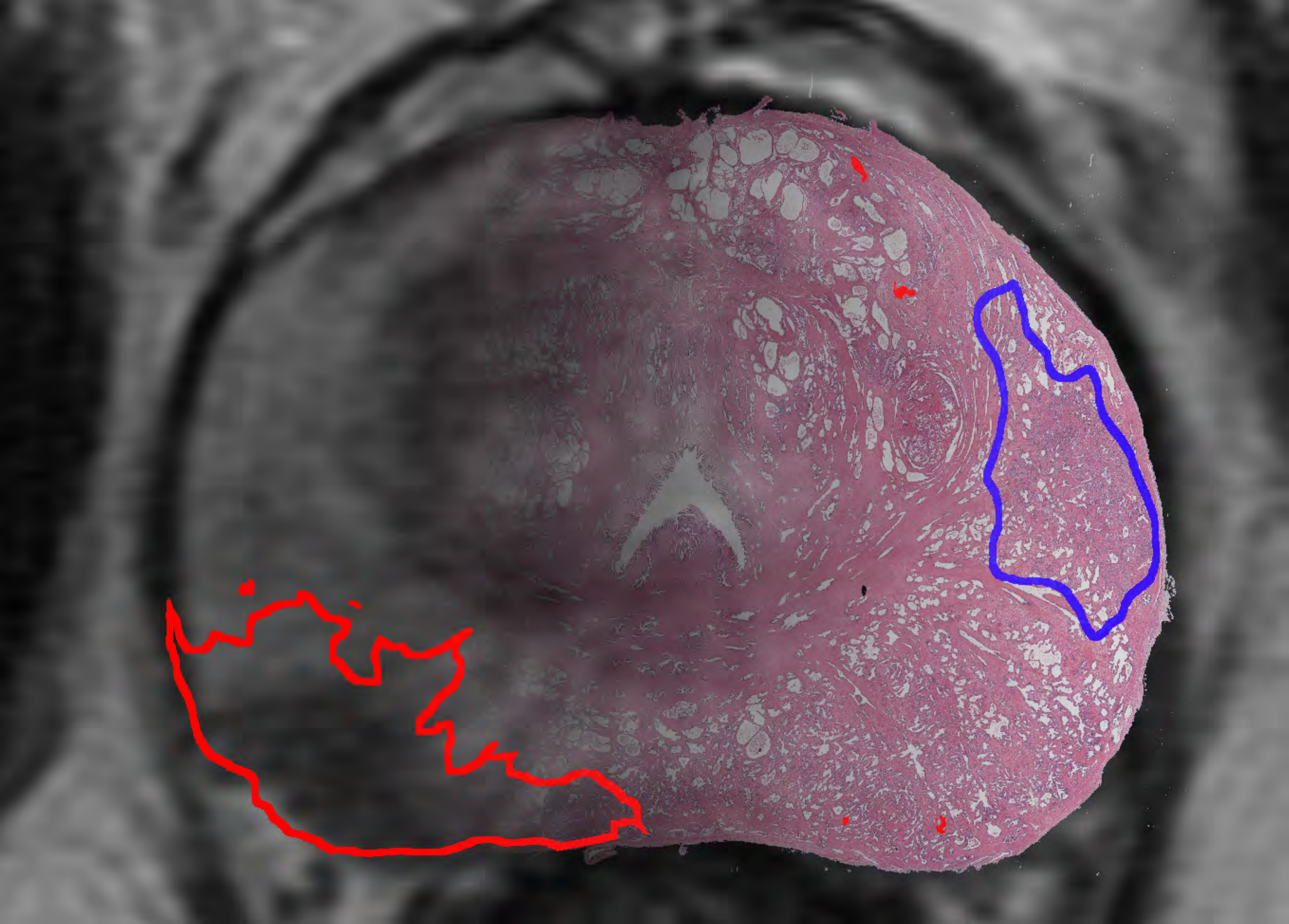}
\caption{}
\end{subfigure}
\caption{
Overlay of registered histopathology and T2w images (same as slice as shown in \figurename~\ref{fig:ResultsQualitativeC1} Raw 2). Histopathology shown with a progressive transparency from (a) right-left, and (d) left-right with cancer outlines (red – Gleason Group 3, blue-Gleason group 2).}
\label{fig:ResultsQualitative2}
\end{figure*}

We applied RAPSODI to register the histopathology slices and T2w MRI in our radical prostatectomy cohorts of 89 patients. \figurename~\ref{fig:ResultsQualitativeC1} shows the qualitative results for a subject in cohort C1 that had a Dice Coefficient of 0.98 and a Hausdorff distance of 1.75 mm of the prostate border ($\sim$ 4 pixels). 
\figurename~\ref{fig:ResultsQualitative2} shows the same slice as \figurename{s}~\ref{fig:ResultsQualitativeC1} Raw 2, with the histopathology slice shown with progressive transparency from right-left (\figurename~\ref{fig:ResultsQualitative2}a) and left to right (\figurename~\ref{fig:ResultsQualitative2}b) to emphasize the alignment of the two modalities. The qualitative and quantitative evaluation suggest that a good alignment was obtained for this subject. 

The accurate registration allowed us to map the extent of two cancer foci with different Gleason groups (\figurename~\ref{fig:ResultsQualitative2}, blue - Gleason group 2; red - Gleason Group 3). Although the higher grade cancer is visible on MRI, its MRI visible borders are smaller than the histopathology projected lesion, confirming previous work showing that MRI underestimates actual tumor size \cite{priester_magnetic_2017}. The fusion enabled the mapping of the Gleason Group 2 cancer, which is not clearly visible on MRI, and would have been otherwise difficult to outline on MRI.

\begin{figure*}[th]

\begin{subfigure}[b]{0.24\textwidth}
\centering
\includegraphics[trim={0.5cm 3cm 0.5cm 1cm},clip,width=\linewidth]{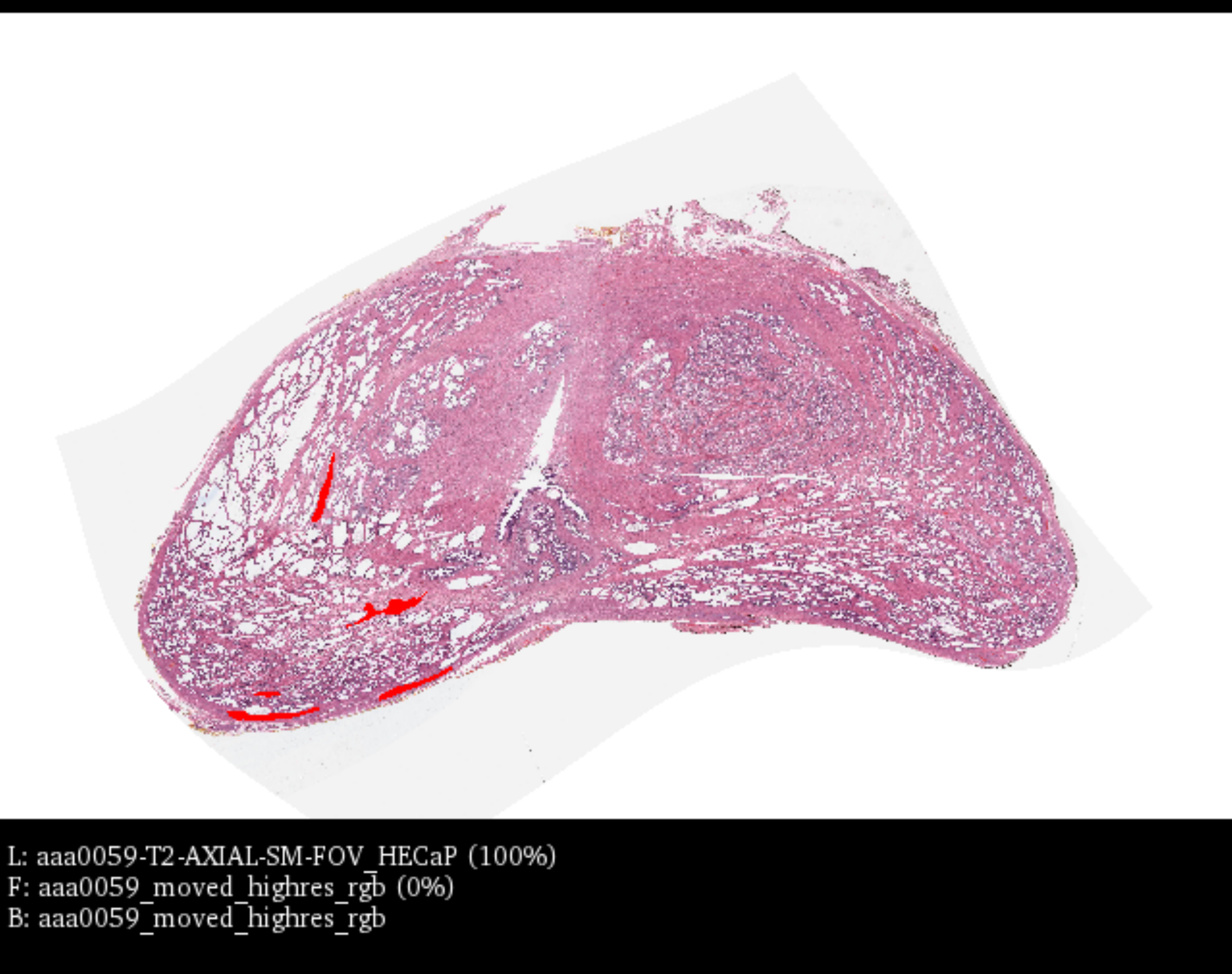}
\end{subfigure}
\begin{subfigure}[b]{0.24\textwidth}
\centering
\includegraphics[trim={0.5cm 3cm 0.5cm 1cm},clip,width=\linewidth]{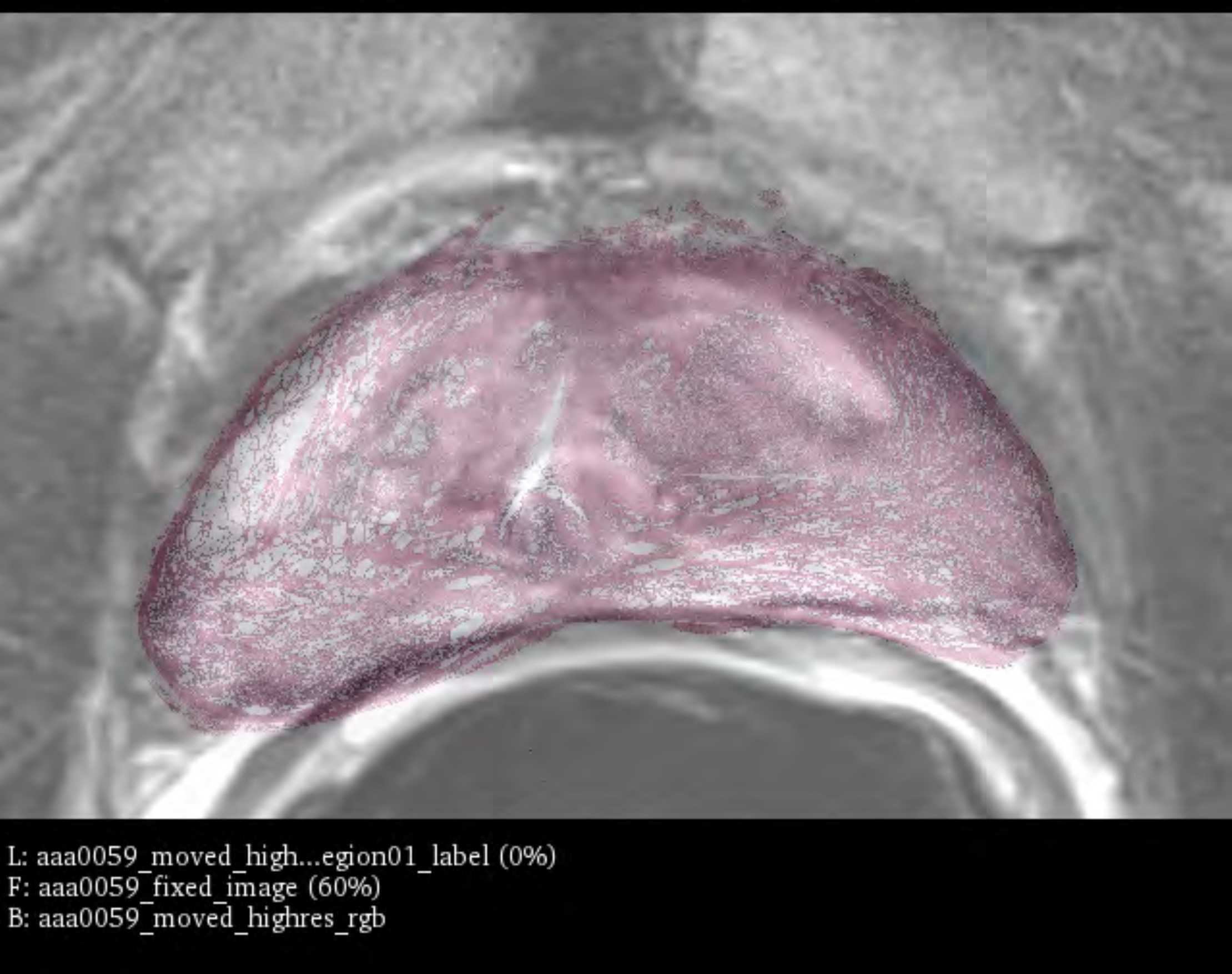}
\end{subfigure}
\begin{subfigure}[b]{0.24\textwidth}
\centering
\includegraphics[trim={0.5cm 3cm 0.5cm 1cm},clip,width=\linewidth]{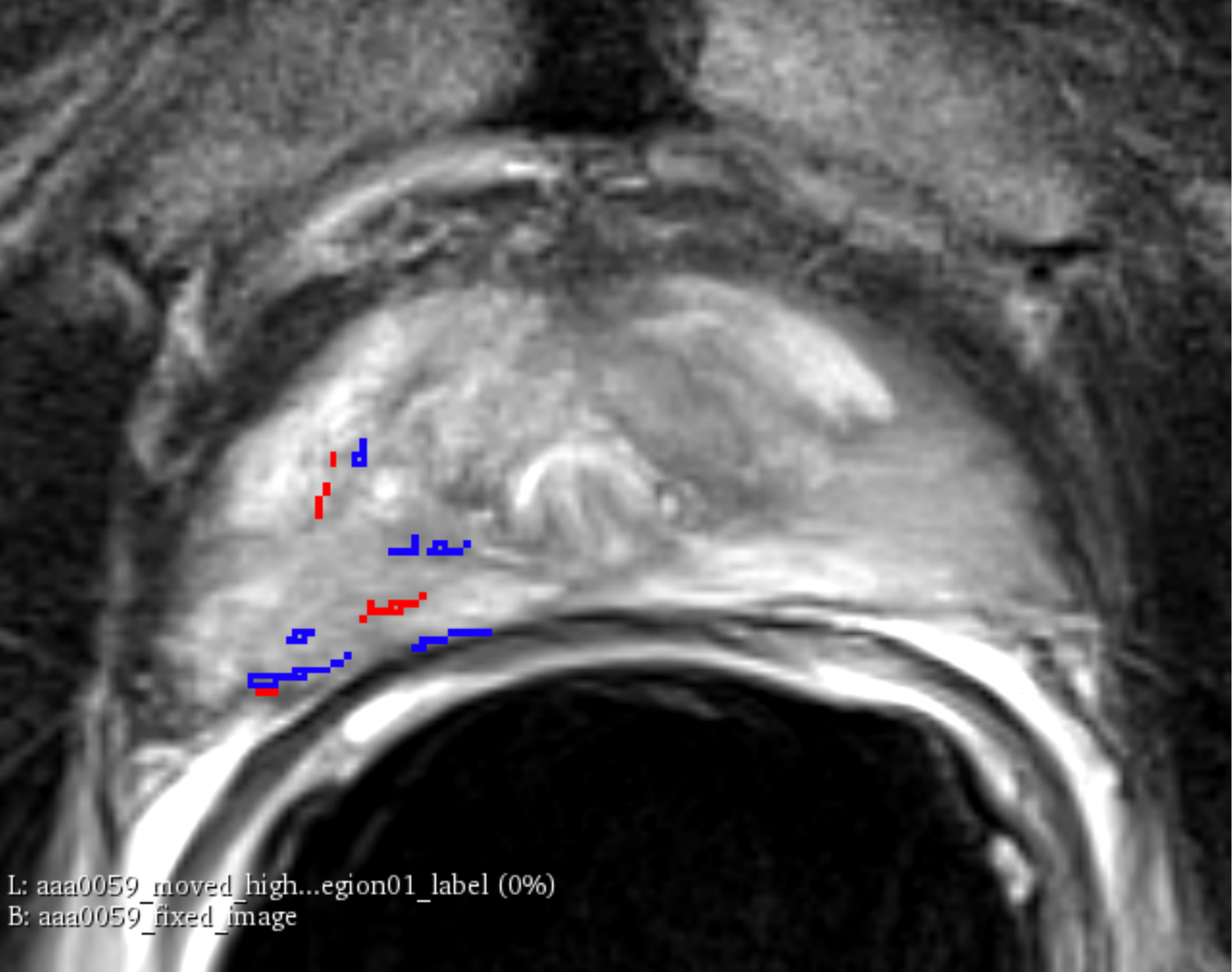}
\end{subfigure}
\begin{subfigure}[b]{0.24\textwidth}
\centering
\includegraphics[trim={2.5cm 4cm 12.5cm 9cm},clip,width=\linewidth]{figures/TCIA/MRI_w_outlines_0.pdf}
\end{subfigure}

\begin{subfigure}[b]{0.24\textwidth}
\centering
\includegraphics[trim={0.5cm 3cm 0.5cm 1cm},clip,width=\linewidth]{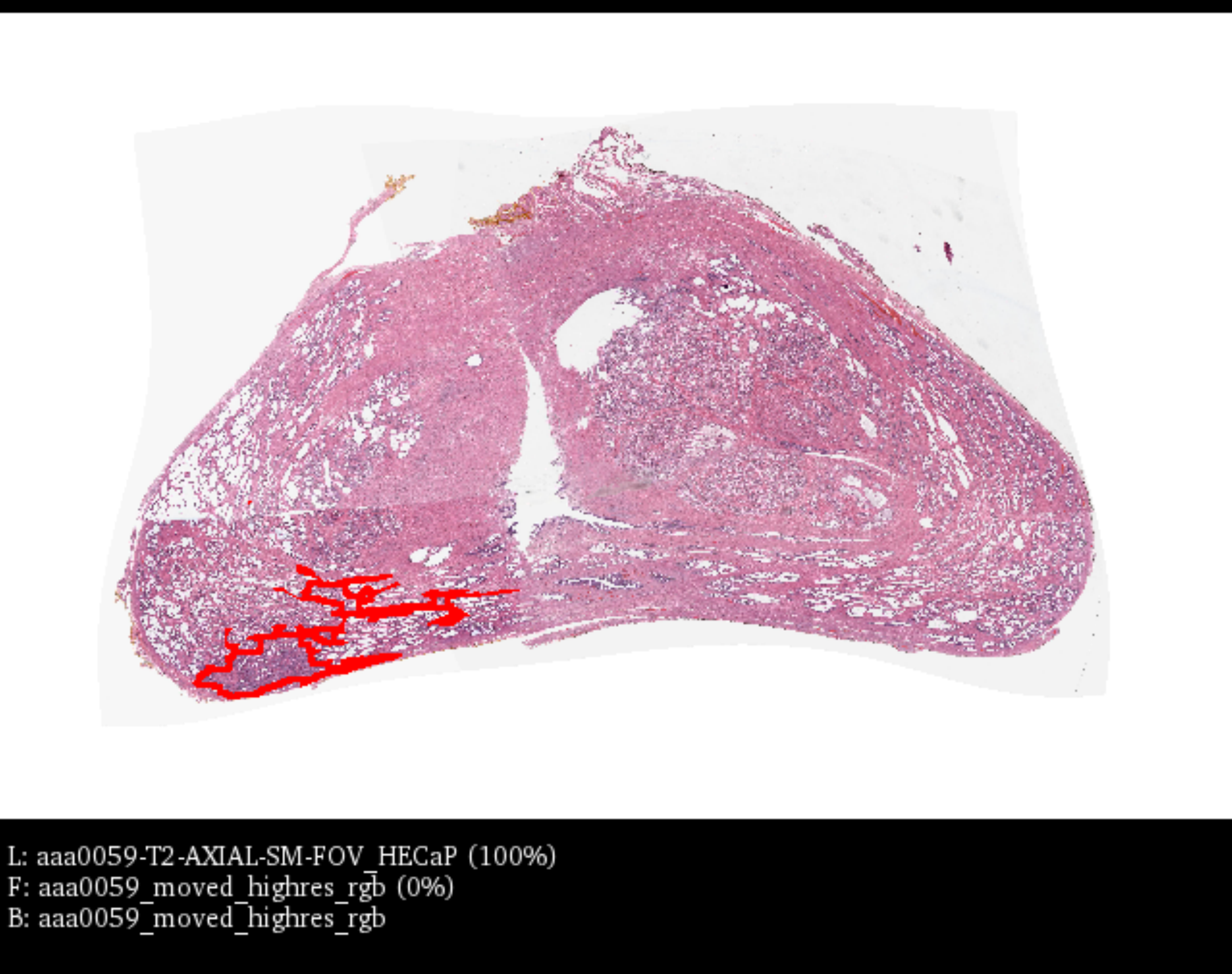}
\end{subfigure}
\begin{subfigure}[b]{0.24\textwidth}
\centering
\includegraphics[trim={0.5cm 3cm 0.5cm 1cm},clip,width=\linewidth]{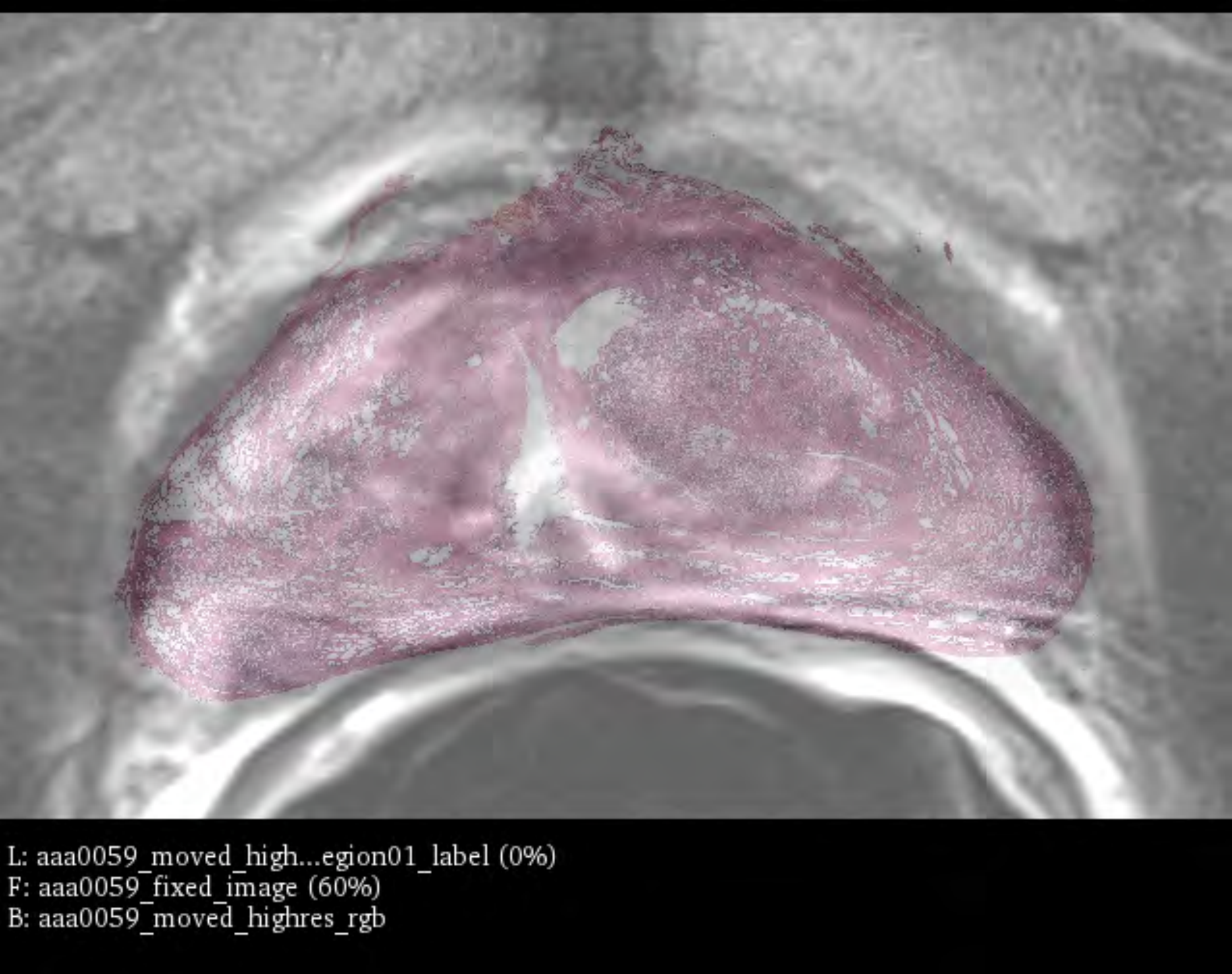}
\end{subfigure}
\begin{subfigure}[b]{0.24\textwidth}
\centering
\includegraphics[trim={0.5cm 3cm 0.5cm 1cm},clip,width=\linewidth]{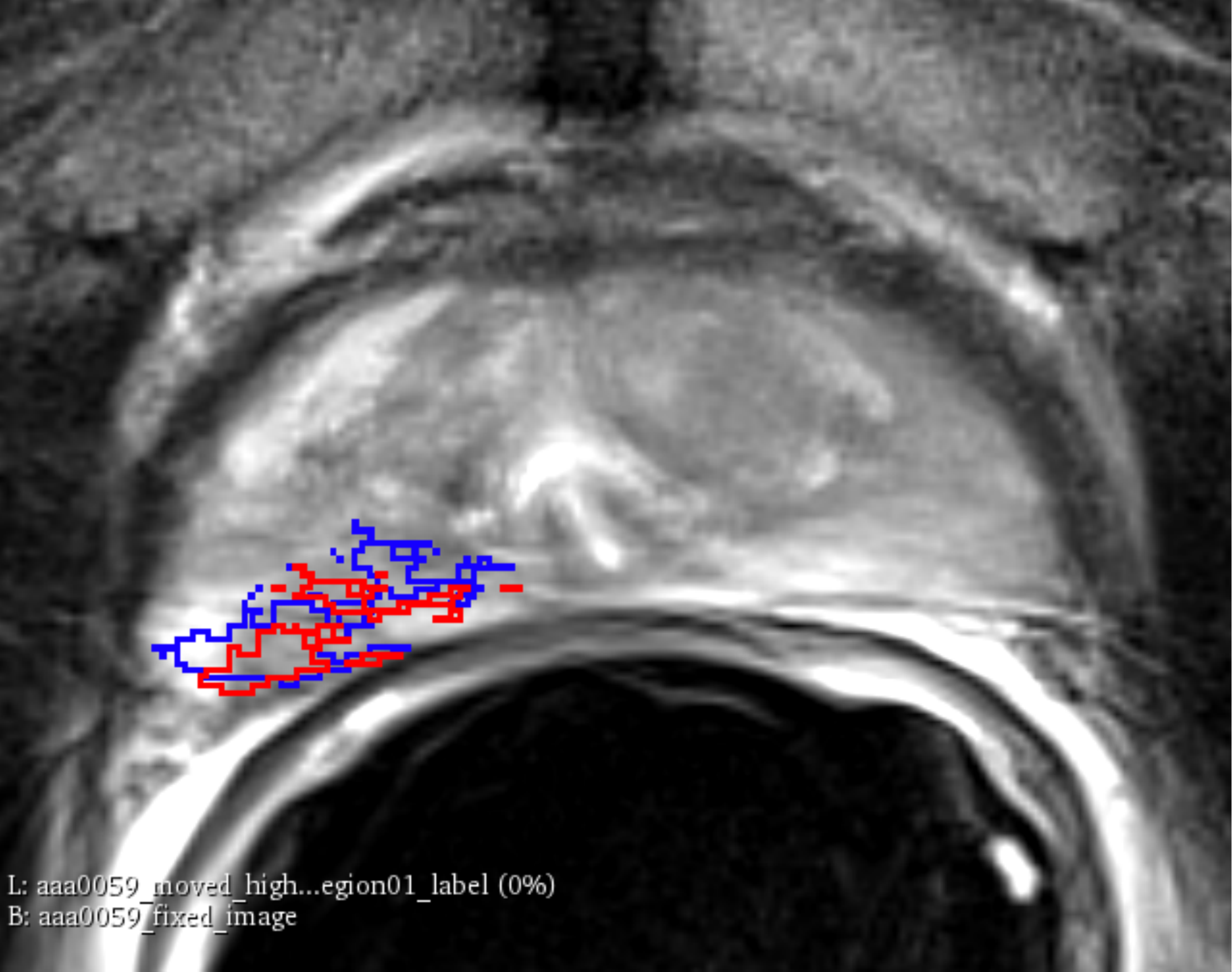}
\end{subfigure}
\begin{subfigure}[b]{0.24\textwidth}
\centering
\includegraphics[trim={2.5cm 4cm 12.5cm 9cm},clip,width=\linewidth]{figures/TCIA/MRI_w_outlines_1.pdf}
\end{subfigure}

\begin{subfigure}[b]{0.24\textwidth}
\centering
\includegraphics[trim={0.5cm 3cm 0.5cm 1cm},clip,width=\linewidth]{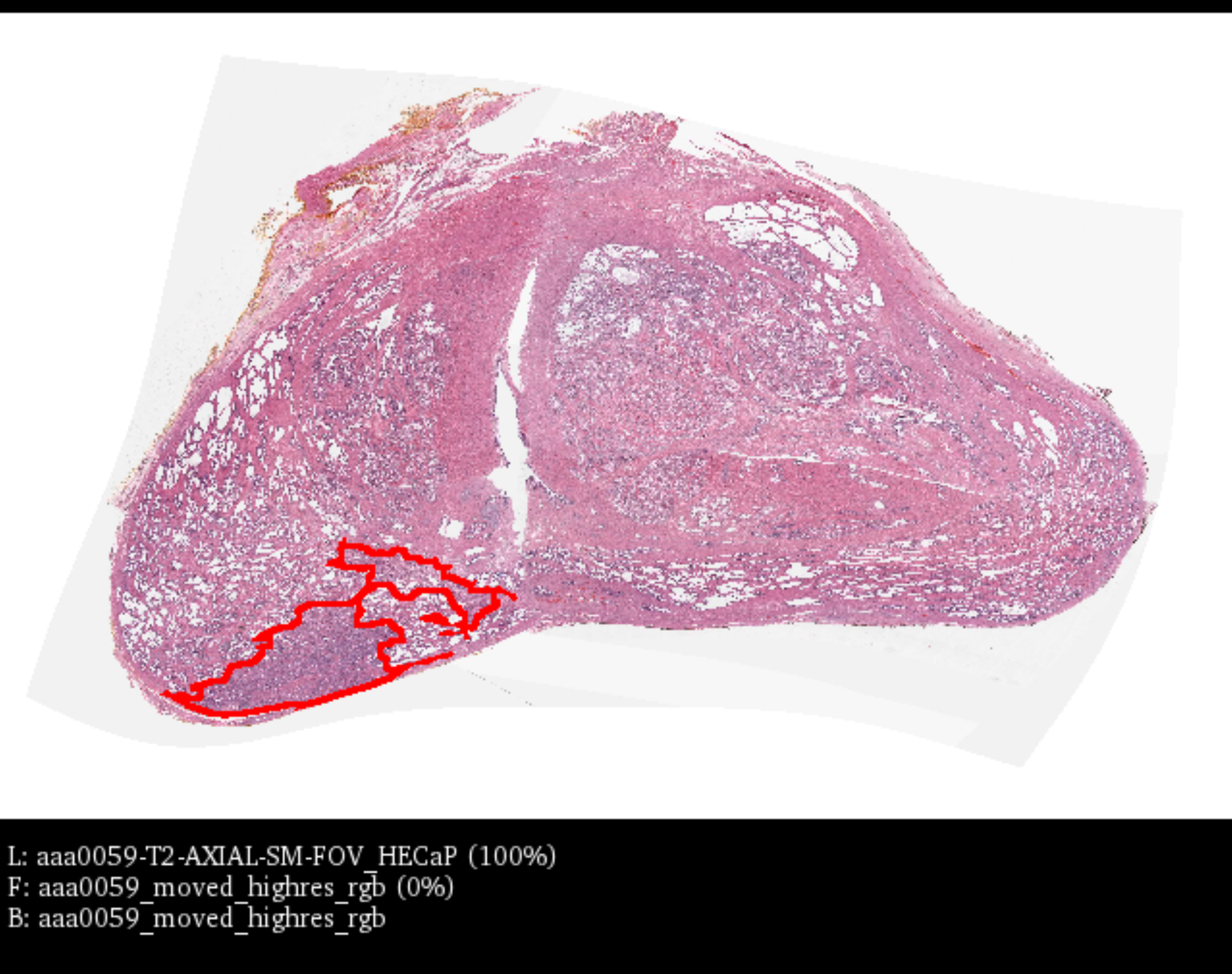}
\end{subfigure}
\begin{subfigure}[b]{0.24\textwidth}
\centering
\includegraphics[trim={0.5cm 3cm 0.5cm 1cm},clip,width=\linewidth]{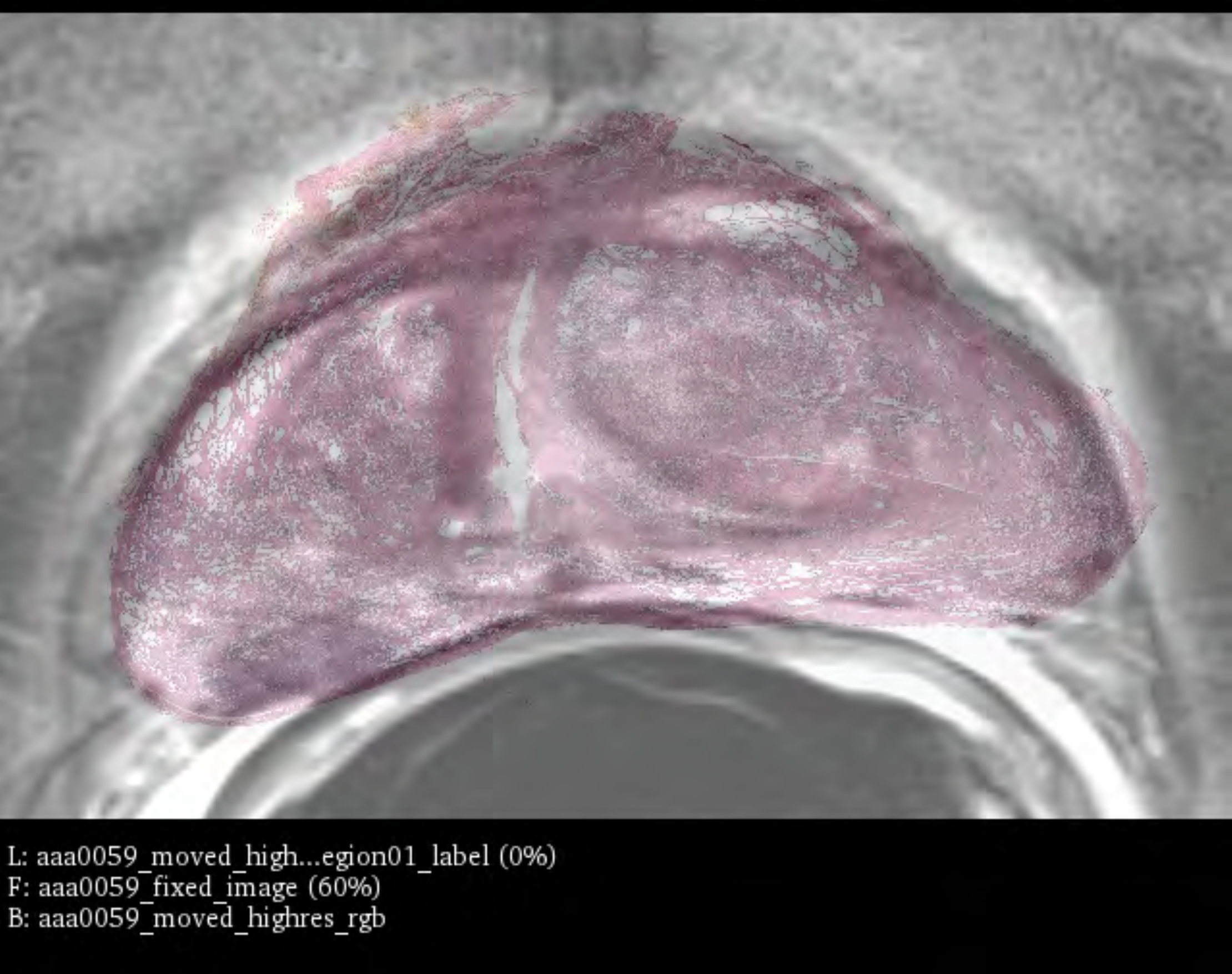}
\end{subfigure}
\begin{subfigure}[b]{0.24\textwidth}
\centering
\includegraphics[trim={0.5cm 3cm 0.5cm 1cm},clip,width=\linewidth]{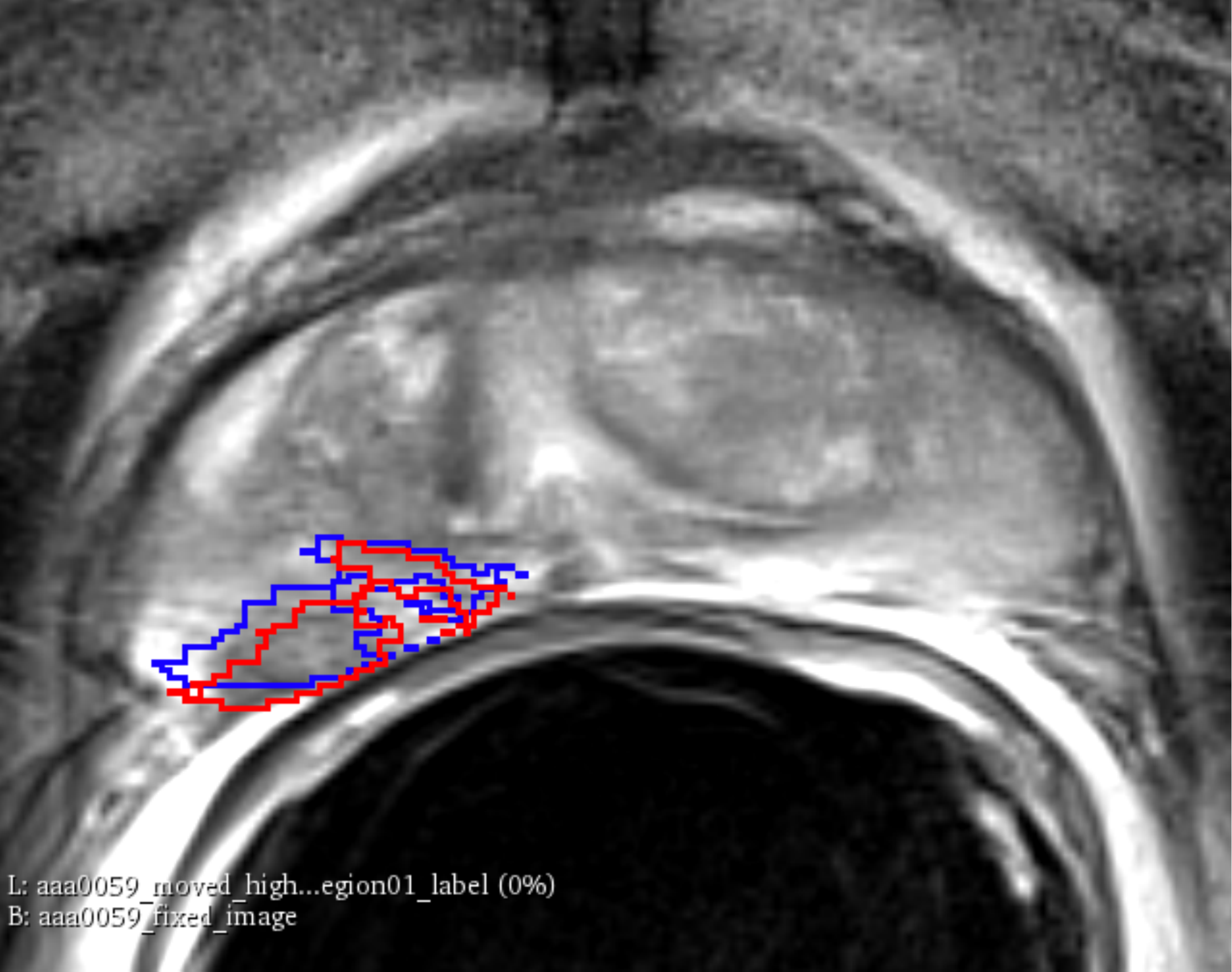}
\end{subfigure}
\begin{subfigure}[b]{0.24\textwidth}
\centering
\includegraphics[trim={2.5cm 4cm 12.5cm 9cm},clip,width=\linewidth]{figures/TCIA/MRI_w_outlines_2.pdf}
\end{subfigure}

\begin{subfigure}[b]{0.24\textwidth}
\centering
\includegraphics[trim={0.5cm 3cm 0.5cm 1cm},clip,width=\linewidth]{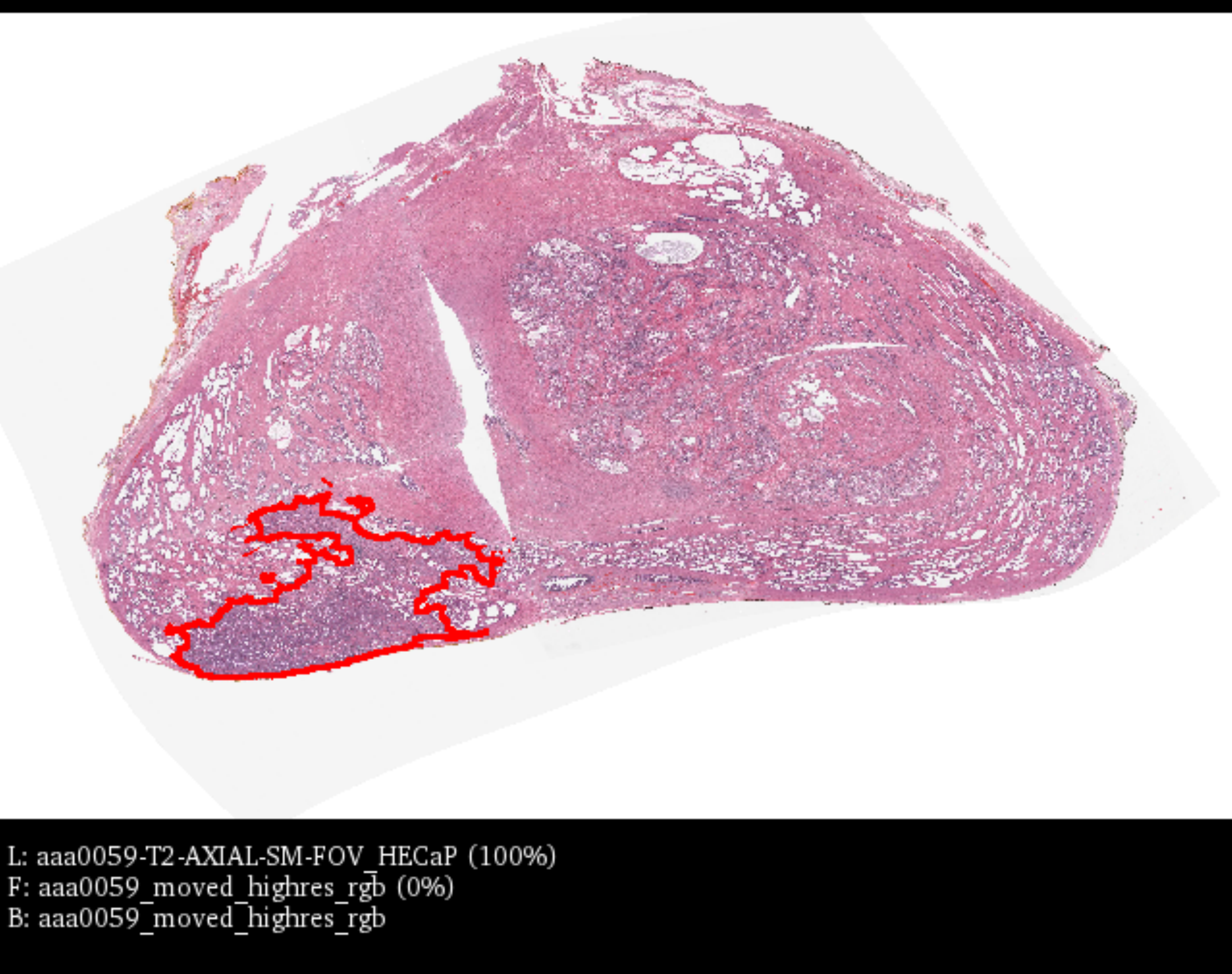}
\end{subfigure}
\begin{subfigure}[b]{0.24\textwidth}
\centering
\includegraphics[trim={0.5cm 3cm 0.5cm 1cm},clip,width=\linewidth]{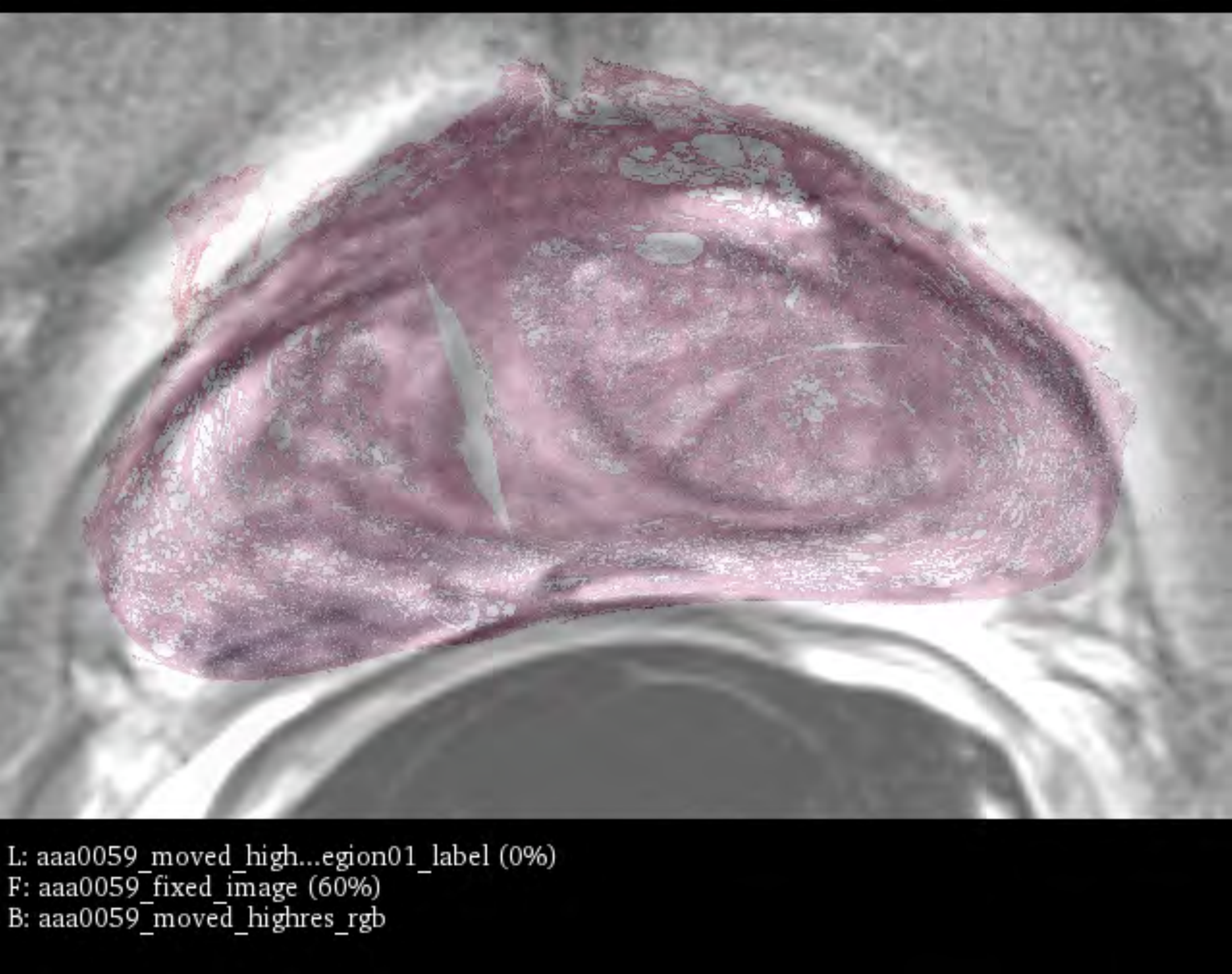}
\end{subfigure}
\begin{subfigure}[b]{0.24\textwidth}
\centering
\includegraphics[trim={0.5cm 3cm 0.5cm 1cm},clip,width=\linewidth]{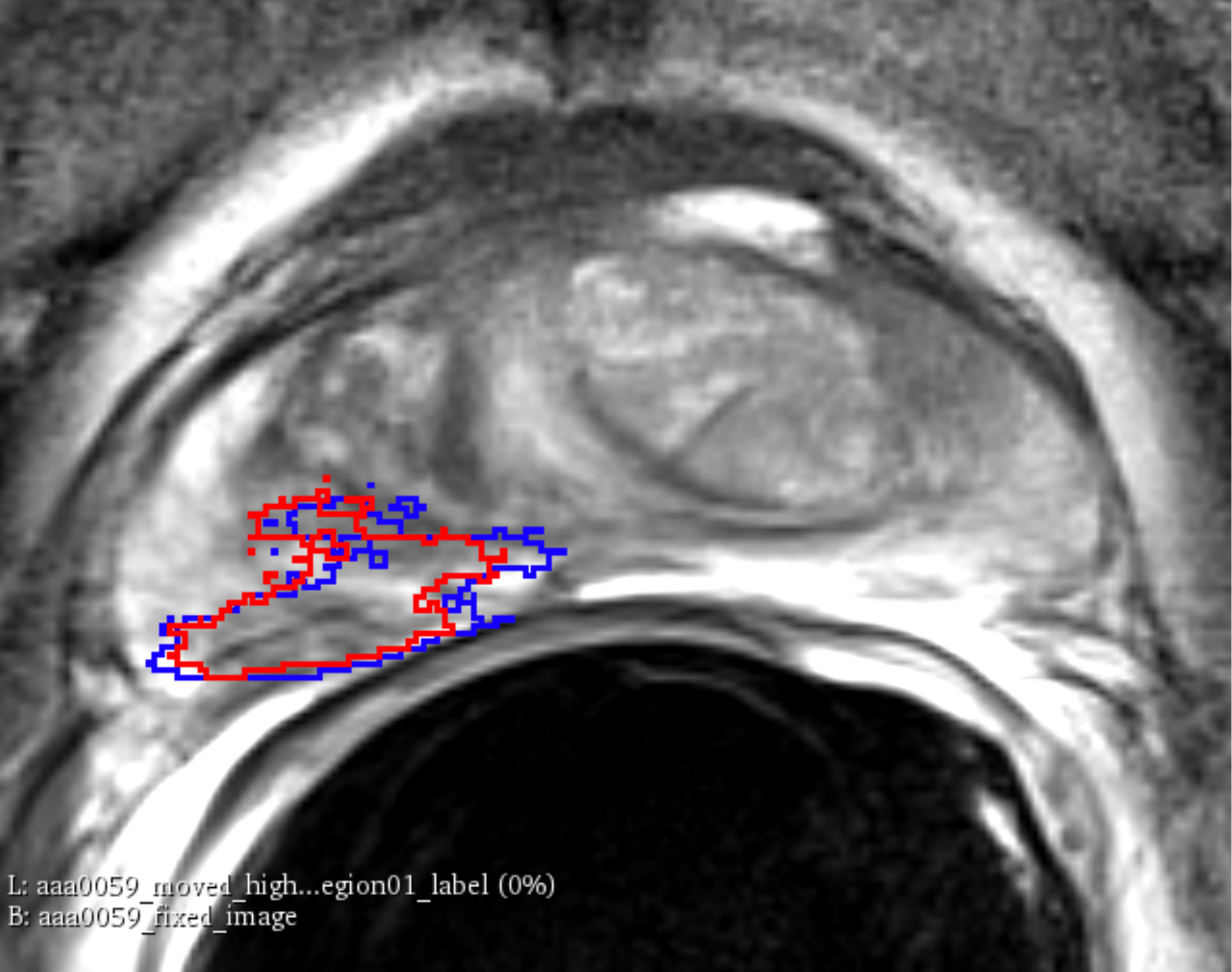}
\end{subfigure}
\begin{subfigure}[b]{0.24\textwidth}
\centering
\includegraphics[trim={2.5cm 4.5cm 12.5cm 8.5cm},clip,width=\linewidth]{figures/TCIA/MRI_w_outlines_3.pdf}
\end{subfigure}

\begin{subfigure}[b]{0.24\textwidth}
\centering
\includegraphics[trim={0.5cm 3cm 0.5cm 1cm},clip,width=\linewidth]{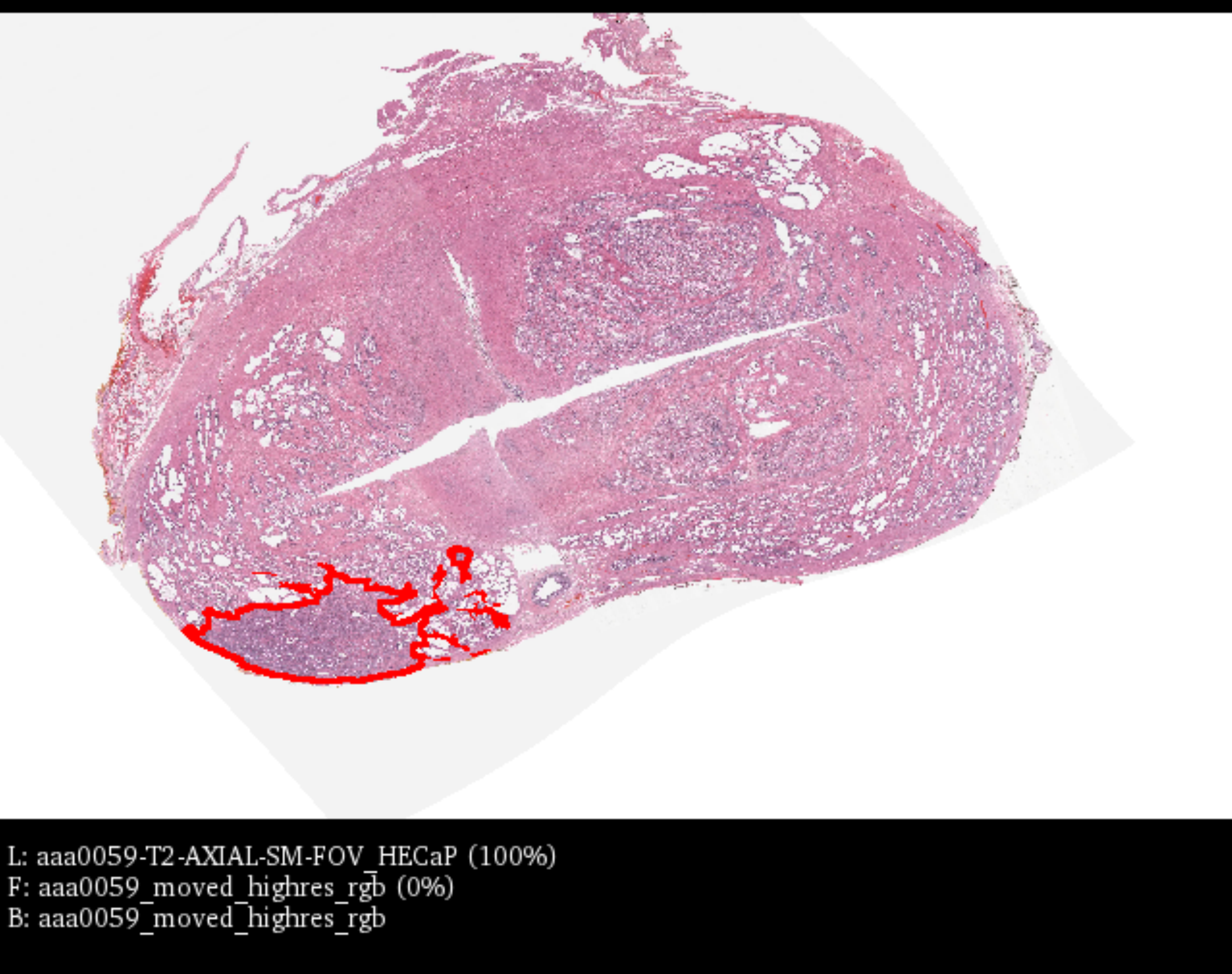}
\end{subfigure}
\begin{subfigure}[b]{0.24\textwidth}
\centering
\includegraphics[trim={0.5cm 3cm 0.5cm 1cm},clip,width=\linewidth]{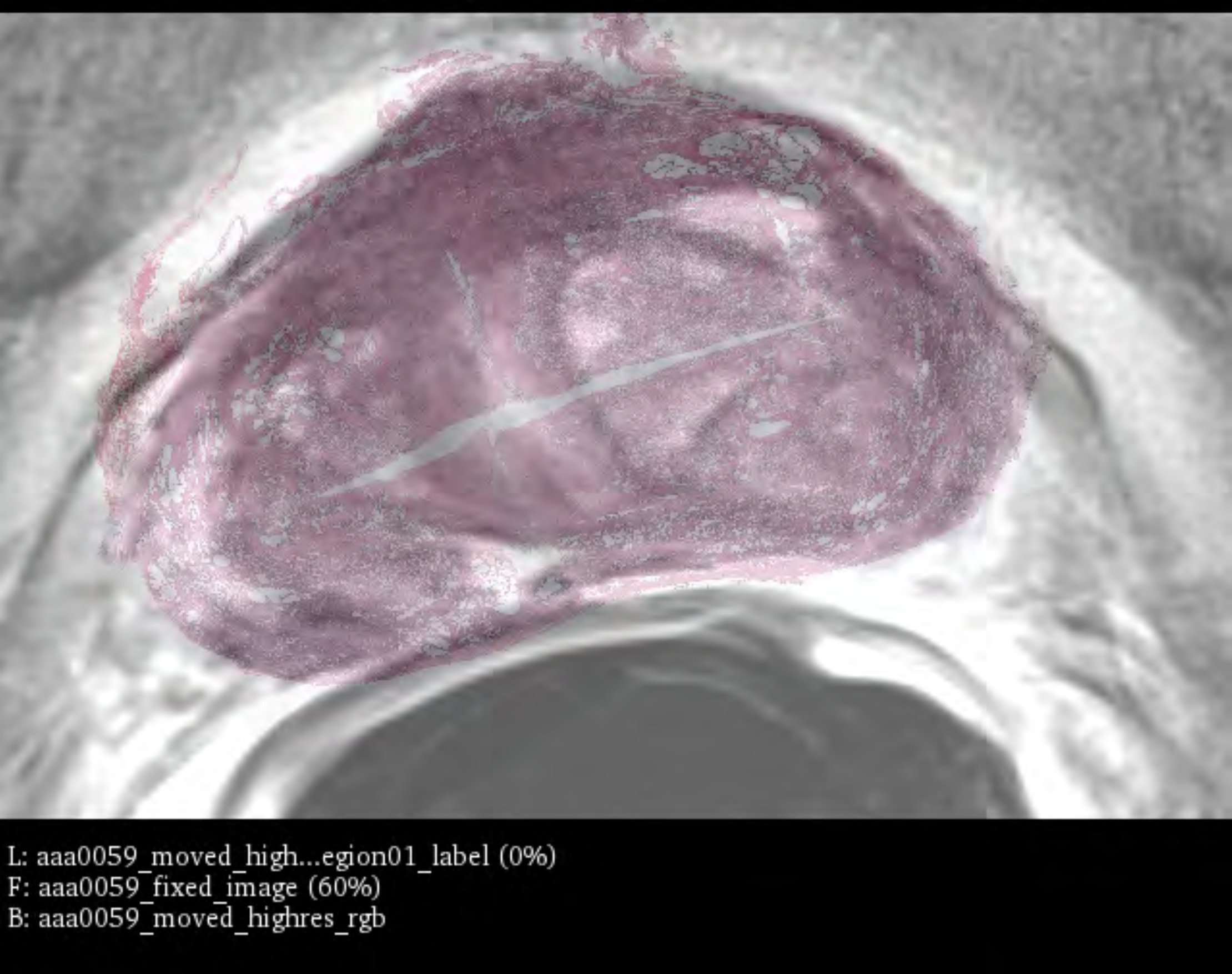}
\end{subfigure}
\begin{subfigure}[b]{0.24\textwidth}
\centering
\includegraphics[trim={0.5cm 3cm 0.5cm 1cm},clip,width=\linewidth]{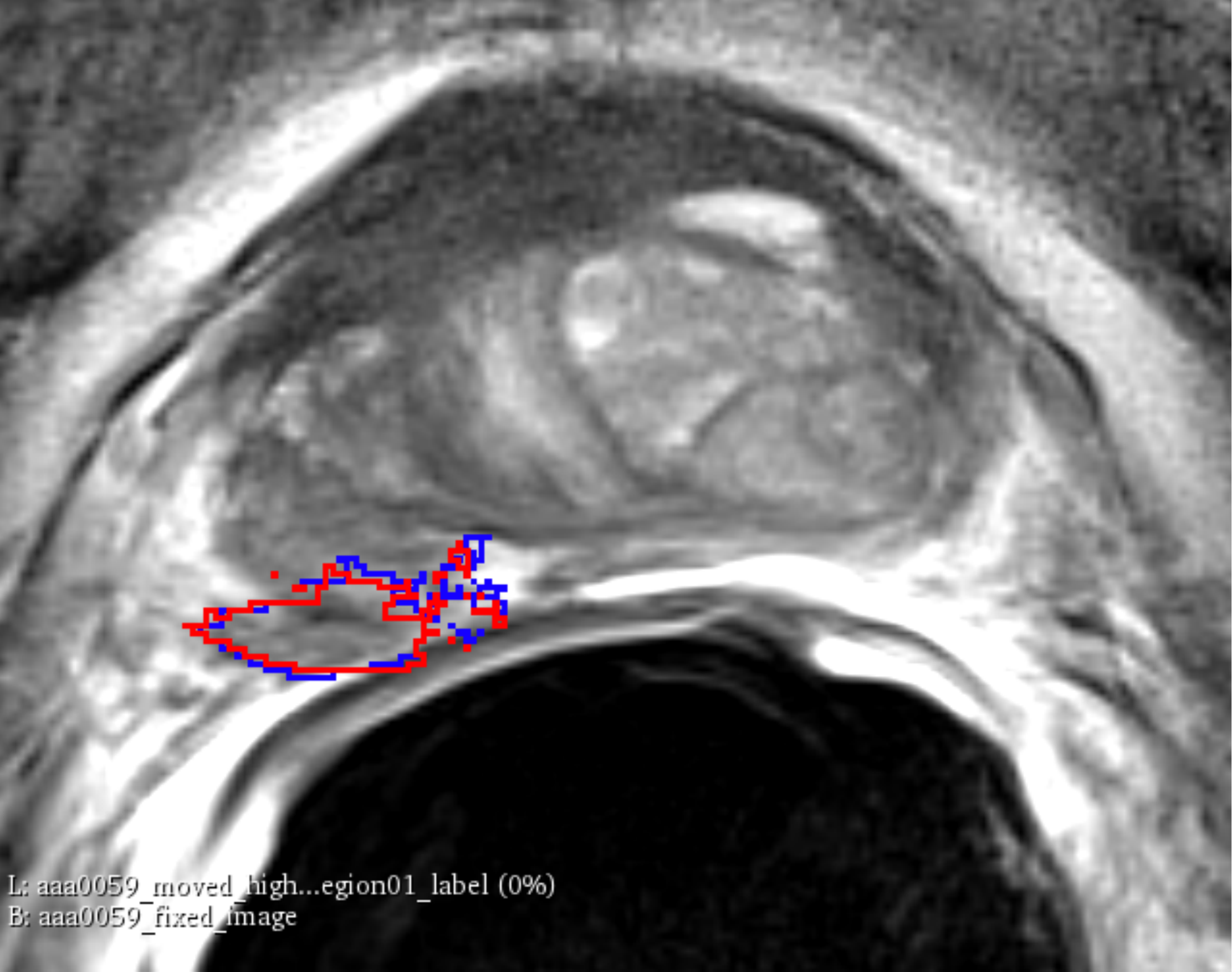}
\end{subfigure}
\begin{subfigure}[b]{0.24\textwidth}
\centering
\includegraphics[trim={2.5cm 4cm 12.5cm 9cm},clip,width=\linewidth]{figures/TCIA/MRI_w_outlines_4.pdf}
\end{subfigure}

\begin{subfigure}[b]{0.24\textwidth}
\centering
\includegraphics[trim={0.5cm 4.3cm 0.5cm 0.3cm},clip,width=\linewidth]{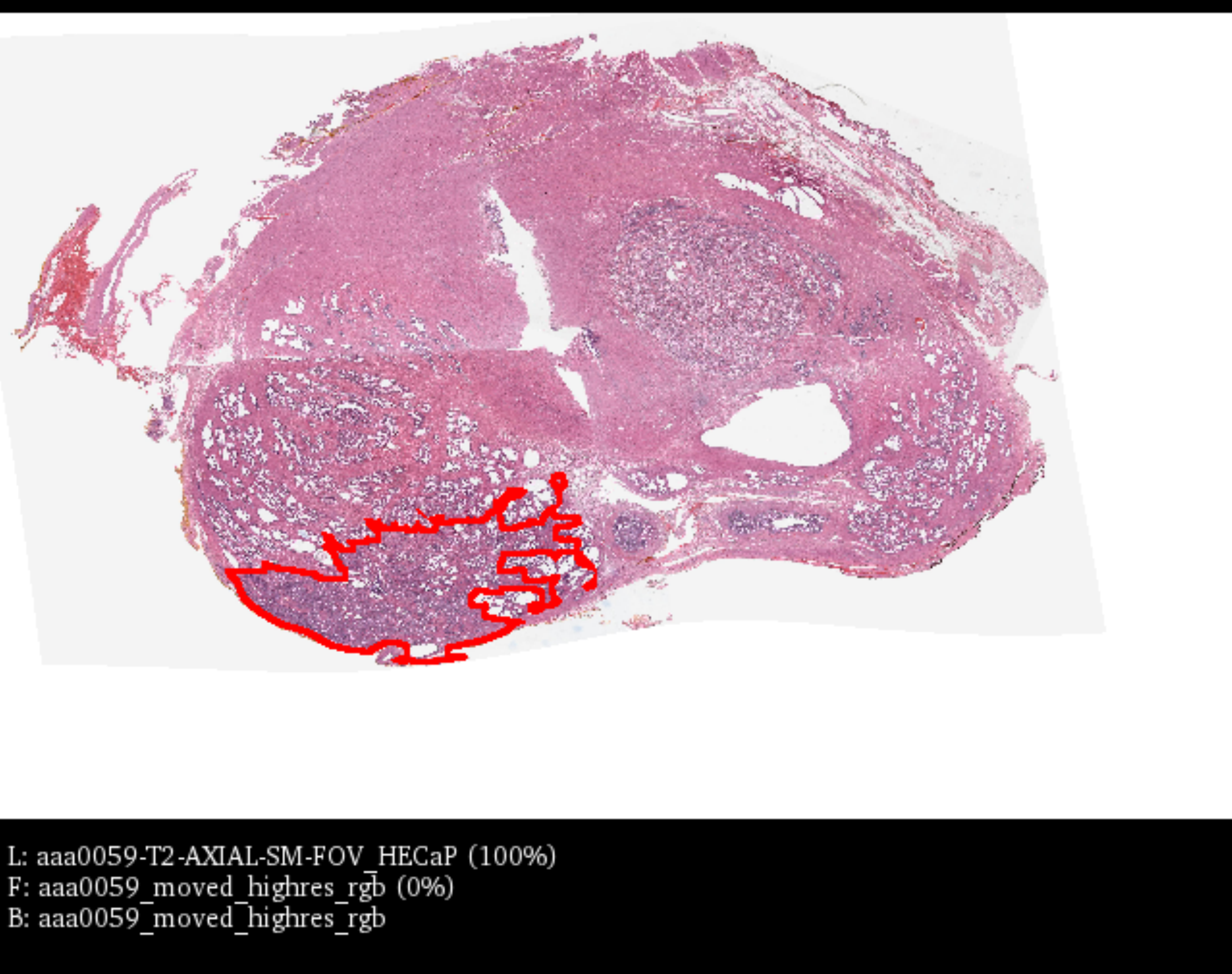}
\end{subfigure}
\begin{subfigure}[b]{0.24\textwidth}
\centering
\includegraphics[trim={0.5cm 4.3cm 0.5cm 0.3cm},clip,width=\linewidth]{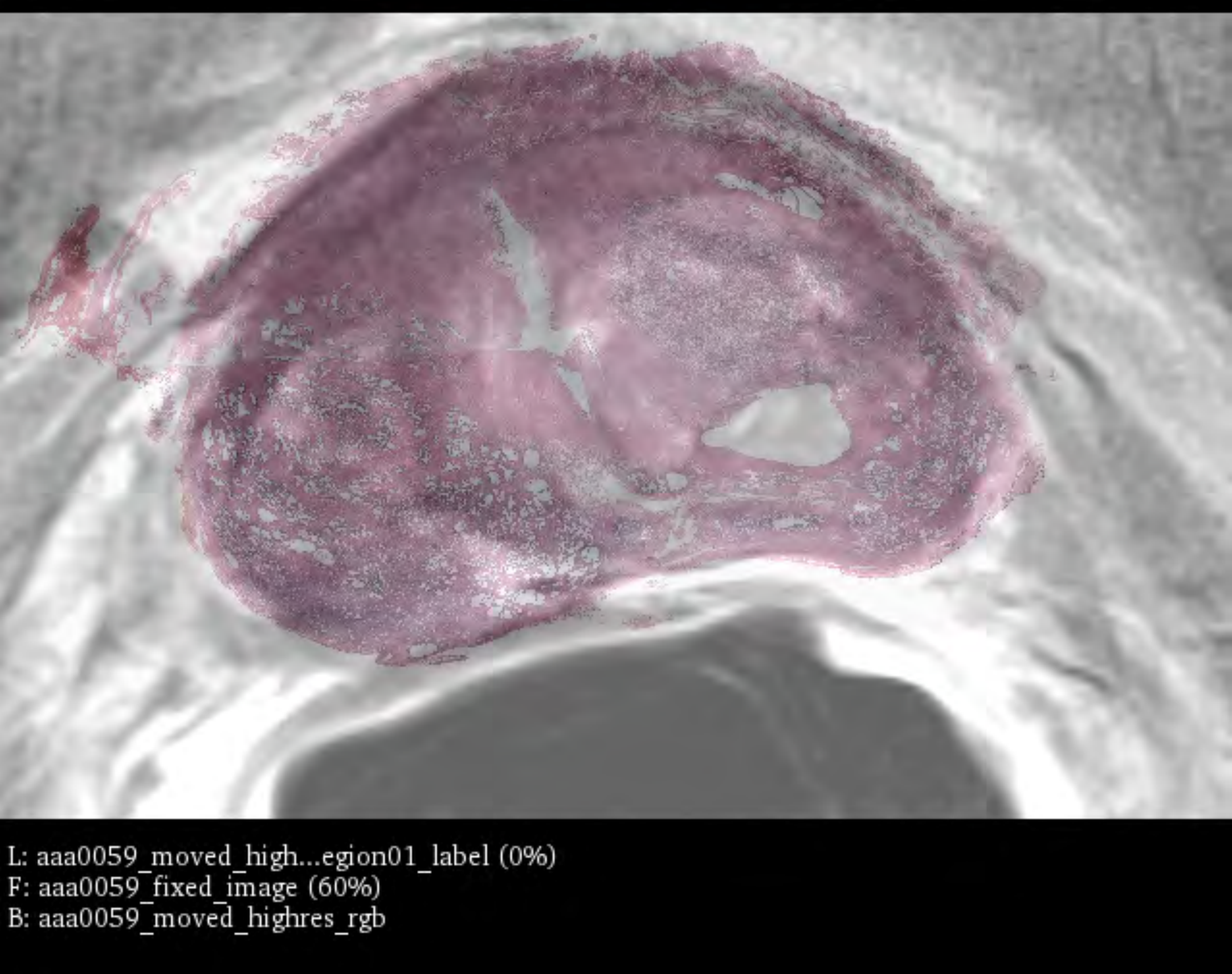}
\end{subfigure}
\begin{subfigure}[b]{0.24\textwidth}
\centering
\includegraphics[trim={0.5cm 4.0cm 0.5cm 0.0cm},clip,width=\linewidth]{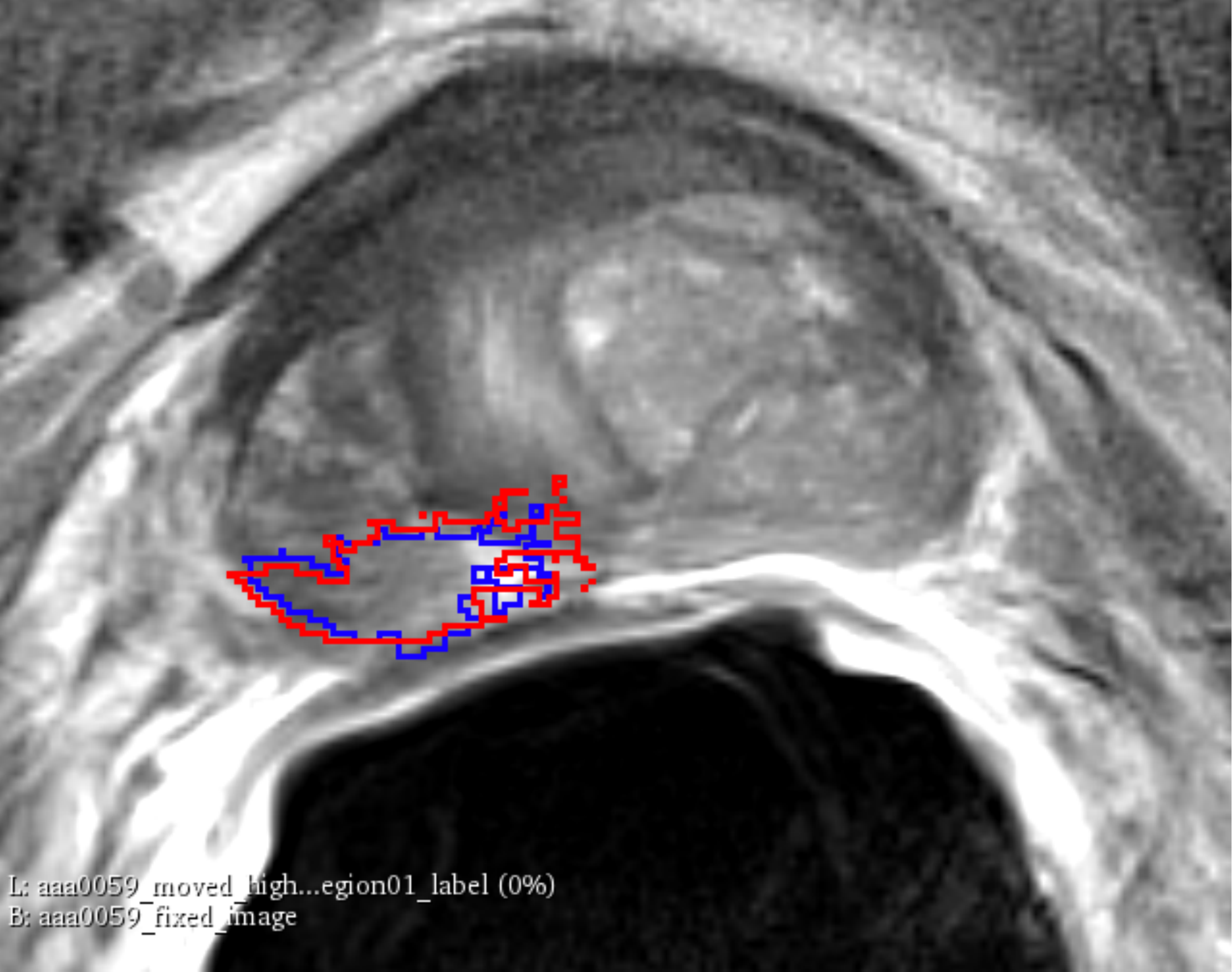}
\end{subfigure}
\begin{subfigure}[b]{0.24\textwidth}
\centering
\includegraphics[trim={3.5cm 5cm 11.5cm 8cm},clip,width=\linewidth]{figures/TCIA/MRI_w_outlines_5.pdf}
\end{subfigure}

\caption{Qualitative results showing the registration for all the histopathology slices from apex to base in subject aaa0059 from Cohort C2. (Column 1) Histopathology slices with cancer outlines (red); (Column 2) Overlay of the registered histopathology and corresponding T2w MRI with histopathology images shown transparent. (Column 3) Corresponding T2w MRI with cancer outlines obtained via RAPSODI (red) or provided by dataset authors (blue); (Column 4) Closeup into the cancer region with outlines shown at the same resolution as the T2w MRI.}
\label{fig:ResultsQualitativeC2}
\end{figure*}

{\color{black}
\figurename~\ref{fig:ResultsQualitativeC2} shows a subject in cohort C2 for which the alignment of the prostate achieved a Dice coefficient of 0.98 and a Hausdorff distance of 2.50 mm on the prostate boundary. As for the results in \figurename{s}~\ref{fig:ResultsQualitativeC1}-\ref{fig:ResultsQualitative2}, the results for this subject are average and not outline. The five histopathology images (\figurename~\ref{fig:ResultsQualitativeC2} Column 1) were registered with the MRI (\figurename~\ref{fig:ResultsQualitativeC2} Column 2), and the cancer outline (red) was mapped onto MRI (\figurename~\ref{fig:ResultsQualitativeC2} Columns 3-4). The public dataset includes the cancer annotation (blue) for this subject, which was obtained using a landmark-based registration \cite{singanamalli_identifying_2016}. The cancer annotations obtained via RAPSODI overlaps well with the labels provided by the dataset authors, with a dice overlap of 0.58 and a Hausdorff Distance of 2.71 mm. The relatively low overlap indicated by the dice coefficient may be accounted by the relatively small size of the tumor, and the misalignment of the regions in the apex slice (\figurename~\ref{fig:ResultsQualitativeC2} Row 1).}

\subsection*{Quantitative Results}

\begin{figure*}[h]
\begin{subfigure}[b]{0.33\textwidth}
\centering
\includegraphics[width=\linewidth]{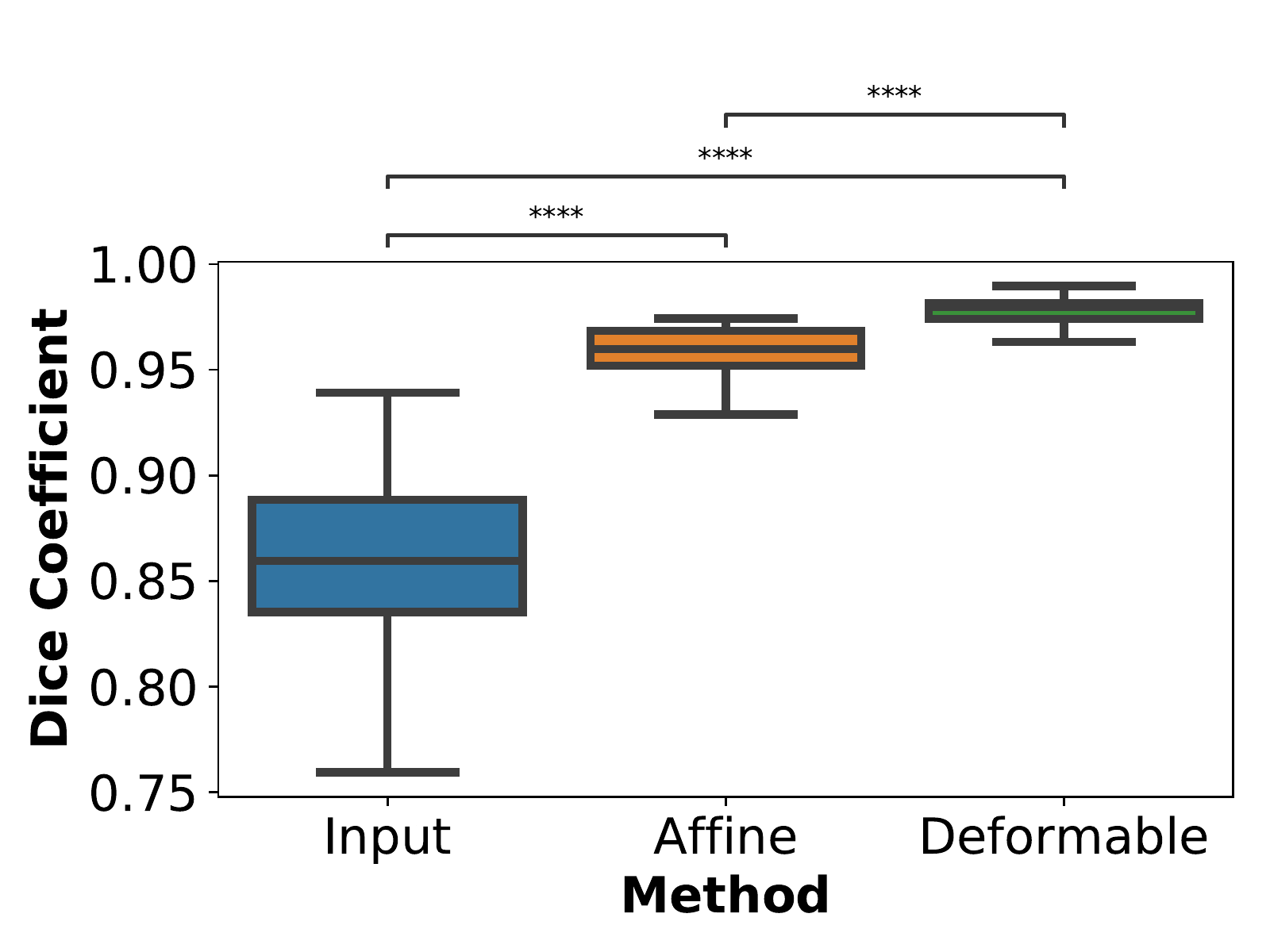}
\caption{}
\end{subfigure}
\begin{subfigure}[b]{0.33\textwidth}
\centering
\includegraphics[width=\linewidth]{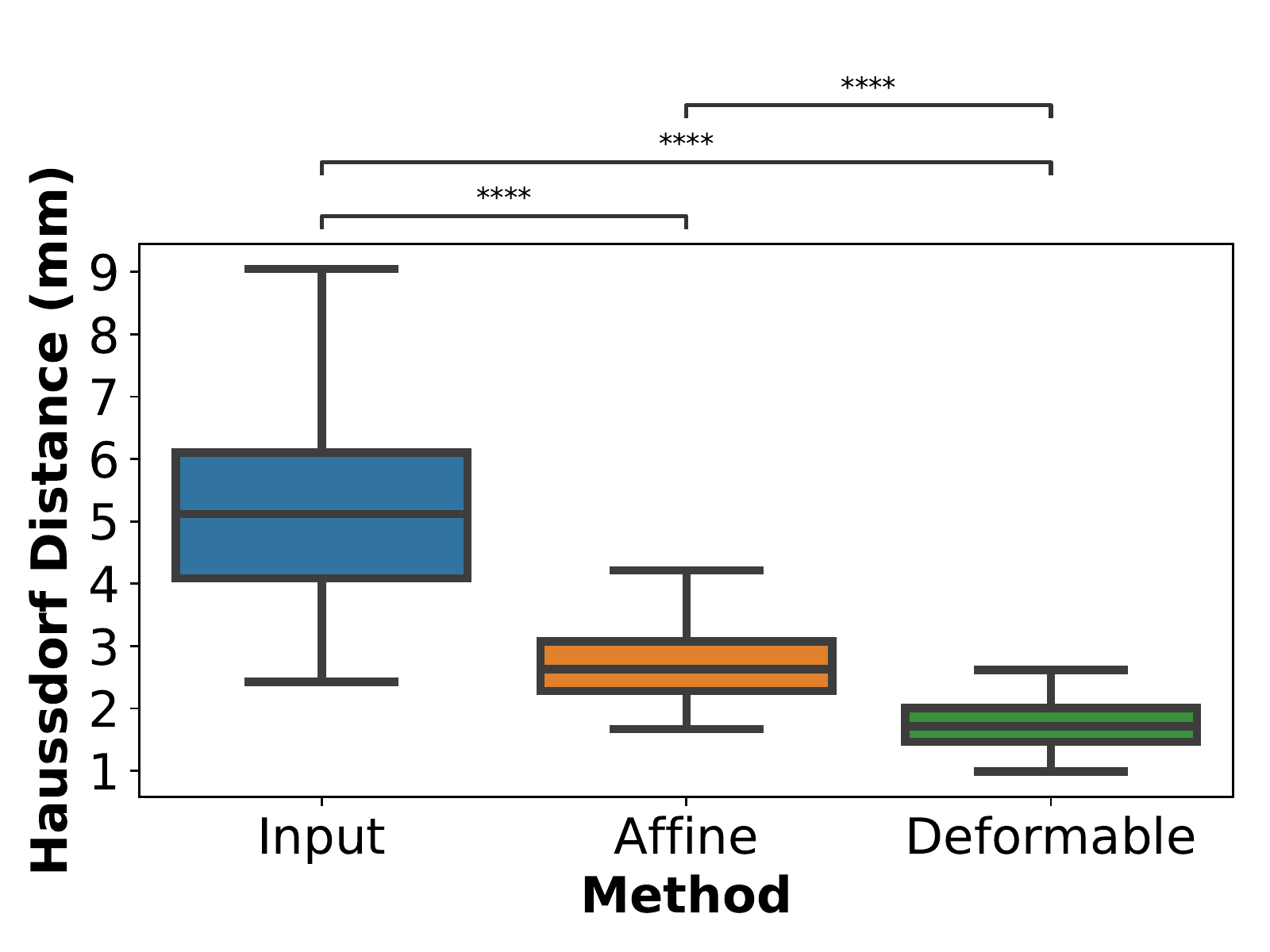}
\caption{}
\end{subfigure}
\begin{subfigure}[b]{0.33\textwidth}
\centering
\includegraphics[width=\linewidth]{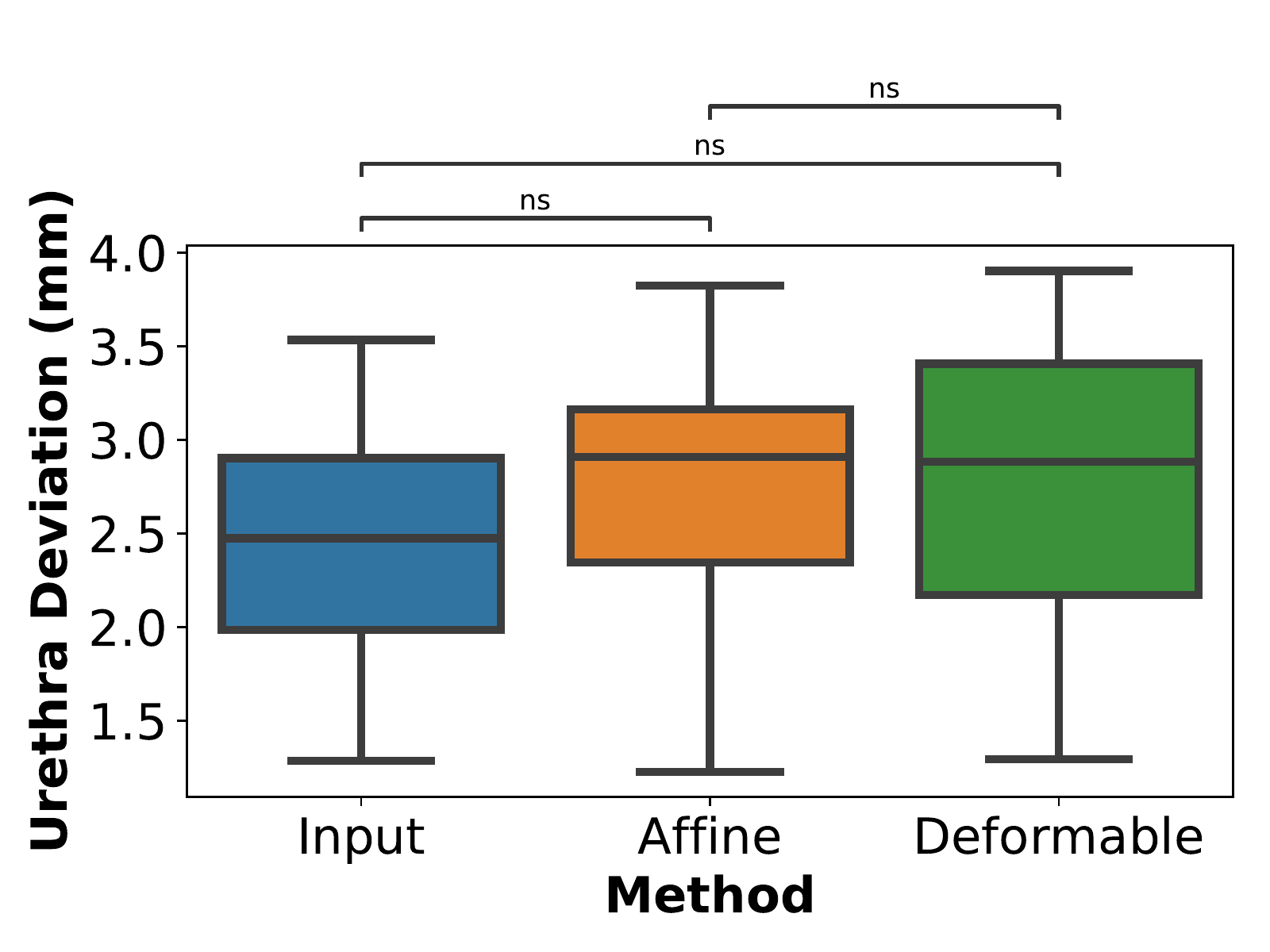}
\caption{}
\end{subfigure}
\begin{subfigure}[b]{0.33\textwidth}
\centering
\includegraphics[width=\linewidth]{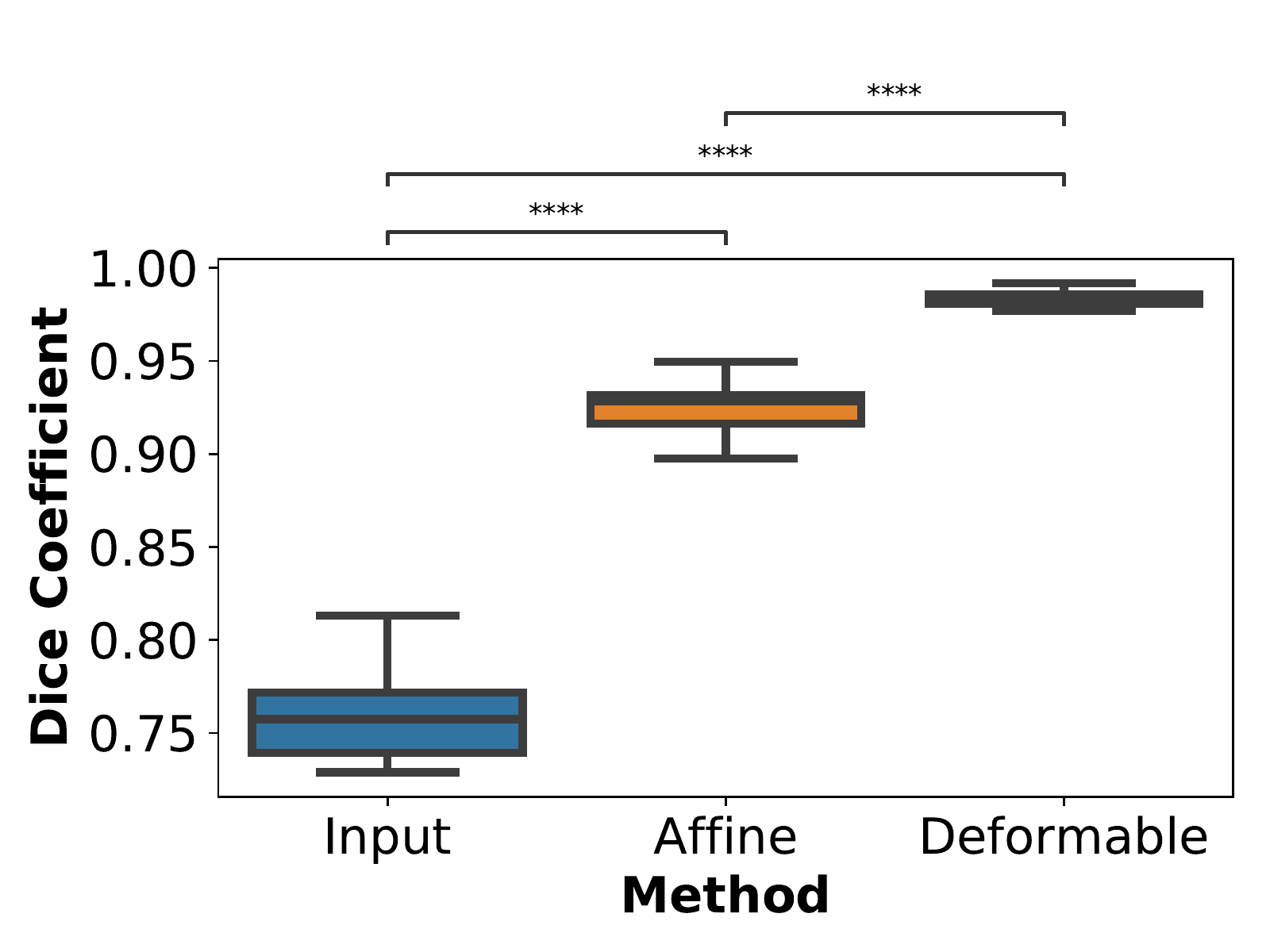}
\caption{}
\end{subfigure}
\begin{subfigure}[b]{0.33\textwidth}
\centering
\includegraphics[width=\linewidth]{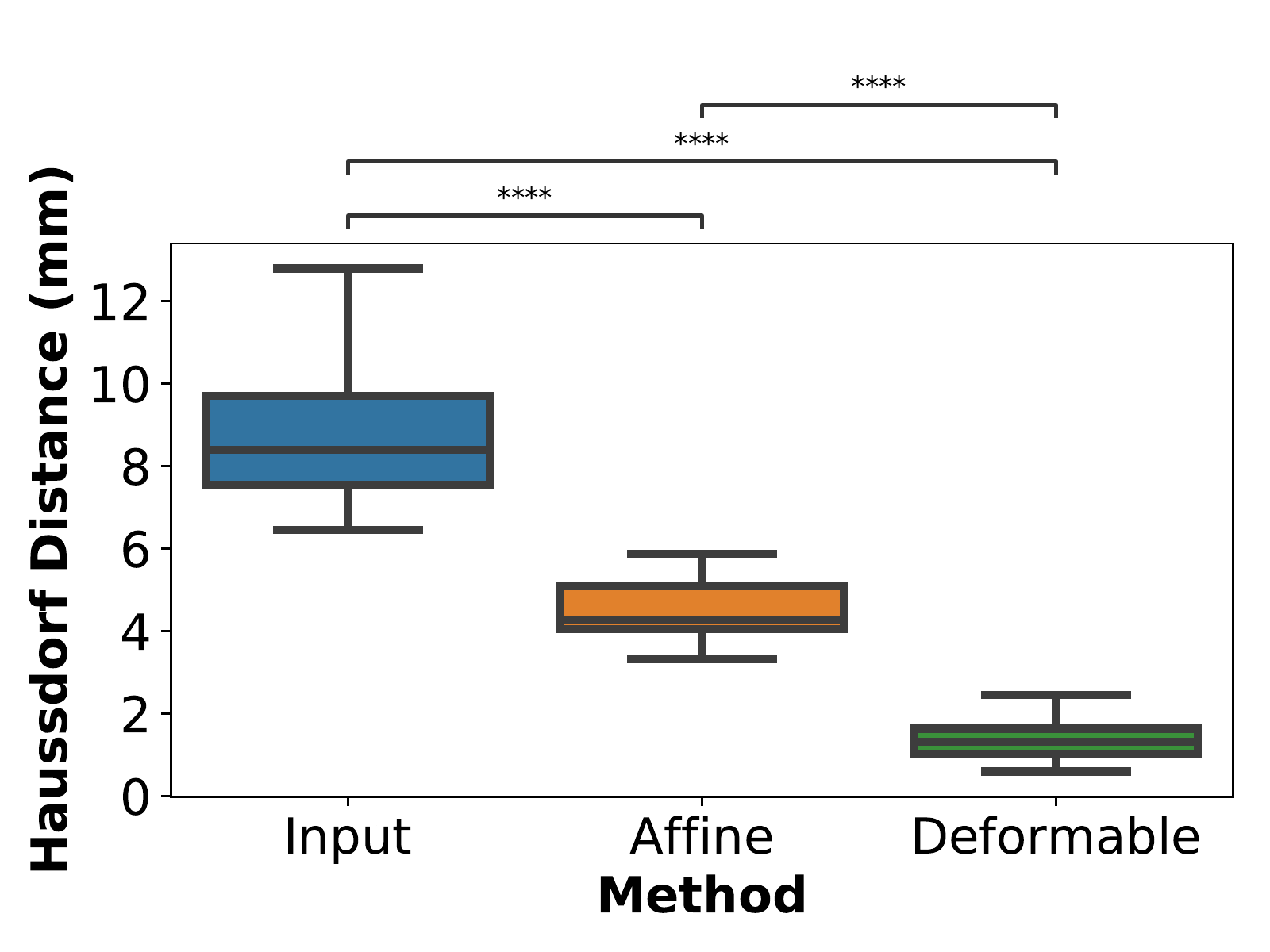}
\caption{}
\end{subfigure}
\begin{subfigure}[b]{0.33\textwidth}
\centering
\includegraphics[width=\linewidth]{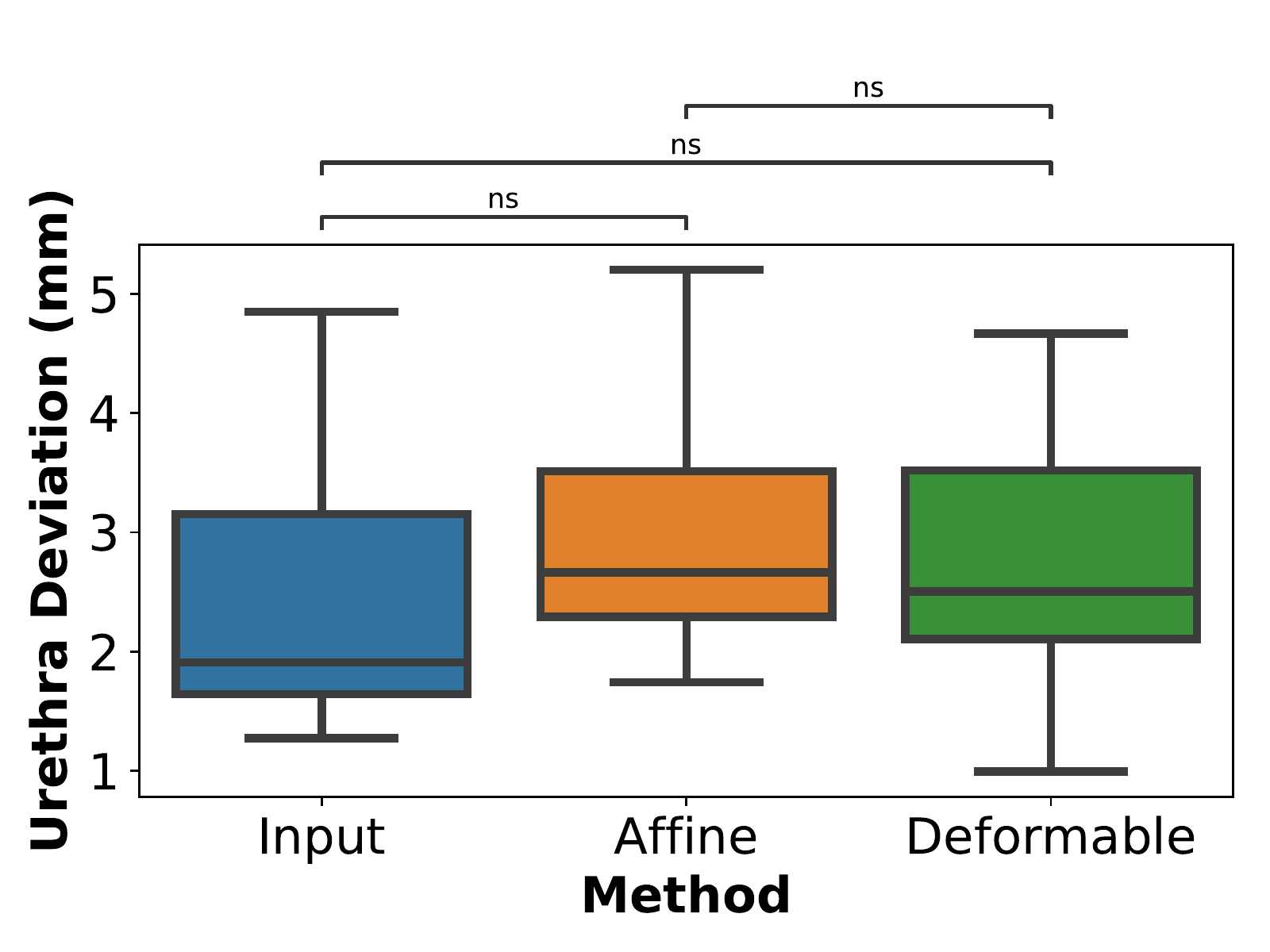}
\caption{}
\end{subfigure}

\caption{Quantitative evaluation of RAPSODI; (a-c) Results for Cohort C1; (d-f) Results for Cohort C2;  (a,d) Dice Coefficients computed for the prostate and (b,e) Hausdorff distance of the prostate border. (c,f) The urethra deviations. p-value for Mann-Whitney test; legend: ns-non significant: 0.05 < p <= 1.00; ****: p <= 0.0001
\label{fig:ResultsQuantitative}
}

\end{figure*}

An improvement in the alignment of the histopathology images and the T2w MRI can be observed across the different steps of our framework (\figurename~\ref{fig:ResultsQuantitative}). Statistically significant differences in Dice coefficients and Hausdorff distances were found between the input and the results of the registration performed using RAPSODI(Mann-Whitney test is statistically significant for $\alpha<0.05$). {\color{black} These statistically significant differences were observed both between the input and the affine registration in step 2 as well as between the affine registration and the deformable registration in step 2, suggesting that both affine and deformable registrations are required to facilitate an accurate alignment.} The urethra (\figurename{s}~\ref{fig:ResultsQuantitative}c,f) and the landmark deviations failed to show a clear trend, but we observe a moderate decrease of landmark deviation between the input, 2.94$\pm$0.54 mm, and after applying RAPSODI, 2.88$\pm$0.7 mm, however, these differences were not statistically significant..

{\color{black} The comparison of results from cohorts C1 and C2 indicated that RAPSODI produces consistent results, with Dice coefficients on the prostate border of 0.98, and Hausdorff distances averaging 1.36-1.78 mm (\tablename~\ref{tab:ResultsQuantitative}). The subjects in cohort C2 have MRIs acquired using an endorectal coil which causes larger deformations of the prostate. Thereby, the input data and affine registration results show worse alignment in cohort C2 compared to cohort C1. However, similar metrics are evaluated after the deformable registration in RAPSODI, suggesting that our approach generalizes even for larger deformations, as those induced by an endorectal coil.} 

\begin{table}[t]
\begin{tabular}{|m{0.50in}|m{0.7in}|m{1.2in}|m{1.0in}|m{1.2in}|m{0.8in}|}
\hline
Cohort & Dice Prostate & Haussdorff Distance (mm) & Urethra Deviation (mm) & Landmarks Deviation (mm) &  Dice Cancer \\ \hline
C1 & 0.98$\pm$0.01 & 1.79$\pm$0.45 & 2.74$\pm$0.82 & 2.88$\pm$0.70 & - \\ \hline
C2 & 0.98$\pm$0.01 & 1.37$\pm$0.47 & 3.14$\pm$1.71& - &0.53$\pm$0.18 \\ \hline
All & 0.98$\pm$0.01 & 1.71$\pm$0.48 & 2.91$\pm$1.25 & 2.88$\pm$0.70 & 0.53$\pm$0.18 \\ \hline
\end{tabular}
\caption{Quantitative results for the two cohorts and aggregated or all subjects in our study. }
\label{tab:ResultsQuantitative}
\end{table}

{\color{black} Additional evaluation was possible in cohort C2, since the authors of the dataset \cite{madabhushi_fused_2016} have provided the mapped cancer obtained via landmark-based registration \cite{singanamalli_identifying_2016}. Thereby, we compared the mapped cancer from RAPSODI with those provided by the dataset authors, and we observed a dice coefficient of $0.53\pm 0.18$ and deviation computed on the center of mass of $2.71\pm 1.31$ mm. The relatively reduced alignment of the cancer labels may be attributed to the general misalignment error, which is within 3 mm inside the prostate and 2 mm on the prostate border. This misalignment can have a significant effect on the value of the overlap evaluated via Dice coefficient for regions of small size, such as the cancer.  

Due to the use of stitched histopathology images, and of endorectal coil MRI, larger deformations needed to be recovered when aligning the histopathology images to MRI in the patients in cohort C2. The pseudo-whole mounts can have stitching artifacts that are absent in the whole-mount histopathology images. For example, the stitched pseudo-whole mount images are elongated in the anterior-posterior direction, e.g. slice C1234 of patient aaa0054. The affine parameters, i.e. scales, of the registration were relaxed, in order to enable the recovery of such large anisotropic stretching. Such modifications were only required for processing two patients, aaa0054 and aaa0072.
}

In order to identify the optimal set of steps in our registration, we tested multiple combinations of processing steps (data not shown) for the subjects in cohort C1, e.g., skipping step 1 (histopathology reconstruction), or applying step 2 without performing the deformable registration. We found that the RAPSODI approach (Histopathology reconstruction followed by 2D affine and deformable transforms) achieves the highest accuracy, showing the highest Dice coefficients and lowest Hausdorff distances and landmark deviations compared to approaches where we skipped step 1 or the deformable registration in step 2.

\section*{Discussion}

Here, we introduced the RAPSODI platform that enables the registration of histopathology and MR images in the prostate. {\color{black}RAPSODI first reconstructs the histopathology volume using pair-wise registration starting from the mid-gland slices towards the apex and base slices, respectively, followed by a slice-to-slice alignment between the corresponding histopathology and T2w images. The reconstruction ensures the consistent stacking of the histopathology slices relative to each other, independent of the MRI, which results in a better initialization of the histopathology slices in the registration with the MRI images.}

We first evaluated RAPSODI in a digital phantom and showed that our framework can recover the rotation angles of the histopathology slices resulting from the slide mounting on glass slides when these angles are within 0-15$^\circ$ from the correct solution and with tissue shrinkage up to 10\%. Correcting for large rotation angles can be achieved prior to applying RAPSODI either by using automated approaches, e.g., by aligning the major axis of the data \cite{rusu_framework_2019}, or via manual approaches where the user indicates an angle, as was done in our study. The tissue shrinks during fixation with a factor that is outside our control. The affine transform helps identify the shrinkage factor, yet the accuracy of the registration declines as the initial conditions are further away from the optimal solution. Registration errors are most apparent at the prostate apex, where the prostate size, shape and textures are reduced.

RAPSODI successfully registered histopathology images with corresponding T2w MR images in the 89 subjects (543 slices) achieving a prostate boundary error within 2 mm and an interior error within 3 mm. Through the use of the prostate segmentation during the registration, we emphasize the importance of the prostate border resulting in a better alignment compared to the interior landmarks. Moreover, picking the landmarks used for evaluation can be challenging as we sought to capture 3+ landmarks/slice, and the resolution of the MRI is relatively reduced due to the surface coil acquisition.

We acknowledge the following limitations for our approach. Although the registration approach is fully automated and does not require patient-specific parameterization for general cases without unusual artifacts, similar to existing approaches, some manual interventions are needed to either segment the prostate on both MRI and histopathology images, to identify slice correspondences between the histopathology and T2w MRI or to correct the gross rotation of the histopathology slices. Unlike other approaches, RAPSODI does not require landmark selection, but only uses them to evaluate the accuracy of registration. 

The registration assumes that a slice-to-slice correspondence exists between the histopathology and MR images. While this is improved by using 3D printed molds, slice misalignment is possible due to the shrinking of the prostate during fixation and shifting in the mold during slicing. Such misalignment is occasionally observed at the base and apex of the prostate. The digital phantom allowed us to study the effect of such misalignment and showed that a perfect alignment cannot be obtained in this situation (we observed a $\sim$4 pixels error) yet the induced shrinkage and rotations are well recovered.

The registration runtime for our approach is 6-8 minutes which is limiting for a Graphical User Interface execution, yet it is acceptable when running the approach in batch mode. We investigated a computationally efficient method that skips Step 1 - reconstructing the histopathology, and runs the registration at lower resolution. The fast approach run in 3.1 minutes  and achieved a Dice coefficient of 0.97$\pm$0.01, a Hausdorff distance of 2.23$\pm$0.66 mm, urethra deviation 2.81$\pm$0.73mm and landmark deviation of 2.92$\pm$0.7 mm in a 66 patient subset from Cohort C1. Although the results are slightly less accurate, the fast approach is 2-3 times faster, and is more suited to running in a Graphical User Interface.

Although our study only includes 89 patients from two institutions, to date it represents the largest study of this magnitude with data from multiple institutions, with MRI acquired either using surface or endorectal coils, and the only study to evaluate the approach in a digital histopathology-MRI prostate phantom. Compared to previous approaches outlined in \tablename~\ref{tab:prior_work}, our quantitative results place us close to the method by Kalavagunta et. al. \cite{kalavagunta_registration_2015} in terms of Dice similarity coefficient. The latter approach relies on heavily annotated datasets that include the border of the transitional zone and the peripheral zone as well as other landmarks. Such landmarks are used to drive the registration at the interior of the prostate resulting in better landmark alignment, yet the approach is labor-intensive and requires careful examination of the data to identify matching landmarks in the pathology images and MRI, which is not trivial.

{\color{black} RAPSODI aims at registering the histopathology and MRI images with the sole goal of mapping the extent and grade of cancer from histopathology images onto T2 weighted MRI, thus creating careful and objective spatial labels on pre-operative MRI. Such mapping may help develop advanced image analysis tools to reliably predict prostate cancer and its aggressiveness on MRI, help improve current MRI interpretation schemes as well as help validate novel MRI protocols or other imaging techniques. Better imaging accompanied by better interpretation schemes can have great impact in reducing overdiagnosis of low-grade cancers, the underdiagnosis of aggressive cancers, and infectious complications of biopsy.}

\section*{Conclusion}
Our radiology-pathology registration framework, RAPSODI, allowed the alignment of histopathology slices and pre-surgical MRI, enabling the accurate mapping of the labels from histopathology onto MRI. The reconstruction of the 3D histopathology specimen followed by 2D registration of corresponding histopathology and T2w MRI slices ensured a robust alignment that provides accurate prostate cancer labels for MRI.

\section*{Acknowledgments}
We thank the Department of Radiology at Stanford University, for their support for this work. 

\bibliographystyle{unsrt}
\bibliography{draft} 

\begin{thebibliography}{10}

\bibitem{siegel_cancer_2019}
Rebecca~L. Siegel, Kimberly~D. Miller, and Ahmedin Jemal.
\newblock Cancer statistics, 2019.
\newblock {\em CA: A Cancer Journal for Clinicians}, 69(1):7--34, 2019.

\bibitem{futterer_can_2015}
Jurgen~J. Fütterer, Alberto Briganti, Pieter De~Visschere, Mark Emberton,
  Gianluca Giannarini, Alex Kirkham, Samir~S. Taneja, Harriet Thoeny, Geert
  Villeirs, and Arnauld Villers.
\newblock Can {Clinically} {Significant} {Prostate} {Cancer} {Be} {Detected}
  with {Multiparametric} {Magnetic} {Resonance} {Imaging}? {A} {Systematic}
  {Review} of the {Literature}.
\newblock {\em European Urology}, 68(6):1045--1053, December 2015.

\bibitem{ahmed_diagnostic_2017}
Hashim~U Ahmed, Ahmed El-Shater~Bosaily, Louise~C Brown, Rhian Gabe, Richard
  Kaplan, Mahesh~K Parmar, Yolanda Collaco-Moraes, Katie Ward, Richard~G
  Hindley, Alex Freeman, Alex~P Kirkham, Robert Oldroyd, Chris Parker, and Mark
  Emberton.
\newblock Diagnostic accuracy of multi-parametric {MRI} and {TRUS} biopsy in
  prostate cancer ({PROMIS}): a paired validating confirmatory study.
\newblock {\em The Lancet}, 389(10071):815--822, February 2017.

\bibitem{van_der_leest_head--head_2019}
Marloes van~der Leest, Erik Cornel, Bas Israël, Rianne Hendriks, Anwar~R.
  Padhani, Martijn Hoogenboom, Patrik Zamecnik, Dirk Bakker, Anglita~Yanti
  Setiasti, Jeroen Veltman, Huib van~den Hout, Hans van~der Lelij, Inge van
  Oort, Sjoerd Klaver, Frans Debruyne, Michiel Sedelaar, Gerjon Hannink,
  Maroeska Rovers, Christina Hulsbergen-van~de Kaa, and Jelle~O. Barentsz.
\newblock Head-to-head {Comparison} of {Transrectal} {Ultrasound}-guided
  {Prostate} {Biopsy} {Versus} {Multiparametric} {Prostate} {Resonance}
  {Imaging} with {Subsequent} {Magnetic} {Resonance}-guided {Biopsy} in
  {Biopsy}-naïve {Men} with {Elevated} {Prostate}-specific {Antigen}: {A}
  {Large} {Prospective} {Multicenter} {Clinical} {Study}.
\newblock {\em European Urology}, 75(4):570--578, 2019.

\bibitem{sonn_prostate_2017}
Geoffrey~A. Sonn, Richard~E. Fan, Pejman Ghanouni, Nancy~N. Wang, James~D.
  Brooks, Andreas~M. Loening, Bruce~L. Daniel, Katherine~J. To'o, Alan~E.
  Thong, and John~T. Leppert.
\newblock Prostate {Magnetic} {Resonance} {Imaging} {Interpretation} {Varies}
  {Substantially} {Across} {Radiologists}.
\newblock {\em European Urology Focus}, December 2017.

\bibitem{weinreb_pi-rads_2016}
Jeffrey~C. Weinreb, Jelle~O. Barentsz, Peter~L. Choyke, Francois Cornud,
  Masoom~A. Haider, Katarzyna~J. Macura, Daniel Margolis, Mitchell~D. Schnall,
  Faina Shtern, Clare~M. Tempany, Harriet~C. Thoeny, and Sadna Verma.
\newblock {PI}-{RADS} {Prostate} {Imaging} - {Reporting} and {Data} {System}:
  2015, {Version} 2.
\newblock {\em European Urology}, 69(1):16--40, January 2016.

\bibitem{barentsz_synopsis_2016}
Jelle~O. Barentsz, Jeffrey~C. Weinreb, Sadhna Verma, Harriet~C. Thoeny,
  Clare~M. Tempany, Faina Shtern, Anwar~R. Padhani, Daniel Margolis,
  Katarzyna~J. Macura, Masoom~A. Haider, Francois Cornud, and Peter~L. Choyke.
\newblock Synopsis of the {PI}-{RADS} v2 {Guidelines} for {Multiparametric}
  {Prostate} {Magnetic} {Resonance} {Imaging} and {Recommendations} for {Use}.
\newblock {\em European urology}, 69(1):41--49, January 2016.

\bibitem{cooperberg_trends_2015}
Matthew~R. Cooperberg and Peter~R. Carroll.
\newblock Trends in {Management} for {Patients} {With} {Localized} {Prostate}
  {Cancer}, 1990-2013.
\newblock {\em JAMA}, 314(1):80--82, July 2015.

\bibitem{park_registration_2008}
Hyunjin Park, Morand~R. Piert, Asra Khan, Rajal Shah, Hero Hussain, Javed
  Siddiqui, Thomas~L. Chenevert, and Charles~R. Meyer.
\newblock Registration methodology for histological sections and in vivo
  imaging of human prostate.
\newblock {\em Academic Radiology}, 15(8):1027--1039, August 2008.

\bibitem{chappelow_elastic_2011}
Jonathan Chappelow, B.~Nicolas Bloch, Neil Rofsky, Elizabeth Genega, Robert
  Lenkinski, William DeWolf, and Anant Madabhushi.
\newblock Elastic registration of multimodal prostate {MRI} and histology via
  multiattribute combined mutual information.
\newblock {\em Medical Physics}, 38(4):2005--2018, April 2011.

\bibitem{ward_prostate:_2012}
Aaron~D. Ward, Cathie Crukley, Charles~A. McKenzie, Jacques Montreuil, Eli
  Gibson, Cesare Romagnoli, Jose~A. Gomez, Madeleine Moussa, Joseph Chin, Glenn
  Bauman, and Aaron Fenster.
\newblock Prostate: registration of digital histopathologic images to in vivo
  {MR} images acquired by using endorectal receive coil.
\newblock {\em Radiology}, 263(3):856--864, June 2012.

\bibitem{kalavagunta_registration_2015}
Chaitanya Kalavagunta, Xiangmin Zhou, Stephen~C. Schmechel, and Gregory~J.
  Metzger.
\newblock Registration of in vivo prostate {MRI} and pseudo-whole mount
  histology using {Local} {Affine} {Transformations} guided by {Internal}
  {Structures} ({LATIS}).
\newblock {\em Journal of Magnetic Resonance Imaging}, 41(4):1104--1114, 2015.

\bibitem{reynolds_development_2015}
H.~M. Reynolds, S.~Williams, A.~Zhang, R.~Chakravorty, D.~Rawlinson, C.~S. Ong,
  M.~Esteva, C.~Mitchell, B.~Parameswaran, M.~Finnegan, G.~Liney, and
  A.~Haworth.
\newblock Development of a registration framework to validate {MRI} with
  histology for prostate focal therapy.
\newblock {\em Medical Physics}, 42(12):7078--7089, December 2015.

\bibitem{li_co-registration_2017}
Lin Li, Shivani Pahwa, Gregory Penzias, Mirabela Rusu, Jay Gollamudi, Satish
  Viswanath, and Anant Madabhushi.
\newblock Co-{Registration} of ex vivo {Surgical} {Histopathology} and in vivo
  {T}2 weighted {MRI} of the {Prostate} via multi-scale spectral embedding
  representation.
\newblock {\em Scientific Reports}, 7(1):8717, August 2017.

\bibitem{losnegard_intensity-based_2018}
Are Losnegård, Lars Reisæter, Ole~J. Halvorsen, Christian Beisland, Aurea
  Castilho, Ludvig~P. Muren, Jarle Rørvik, and Arvid Lundervold.
\newblock Intensity-based volumetric registration of magnetic resonance images
  and whole-mount sections of the prostate.
\newblock {\em Computerized Medical Imaging and Graphics: The Official Journal
  of the Computerized Medical Imaging Society}, 63:24--30, January 2018.

\bibitem{wu_system_2019}
Holden~H. Wu, Alan Priester, Pooria Khoshnoodi, Zhaohuan Zhang, Sepideh
  Shakeri, Sohrab Afshari~Mirak, Nazanin~H. Asvadi, Preeti Ahuja, Kyunghyun
  Sung, Shyam Natarajan, Anthony Sisk, Robert Reiter, Steven Raman, and Dieter
  Enzmann.
\newblock A system using patient-specific 3d-printed molds to spatially align
  in vivo {MRI} with ex vivo {MRI} and whole-mount histopathology for prostate
  cancer research.
\newblock {\em Journal of magnetic resonance imaging: JMRI}, 49(1):270--279,
  January 2019.

\bibitem{rusu_framework_2019}
Mirabela Rusu, Christian Kunder, Richard Fan, Pejman~Ghanouni M.d, Robert West,
  Geoffrey Sonn, and James~Brooks M.d.
\newblock Framework for the co-registration of {MRI} and histology images in
  prostate cancer patients with radical prostatectomy.
\newblock In {\em Medical {Imaging} 2019: {Image} {Processing}}, volume 10949,
  page 109491P. International Society for Optics and Photonics, March 2019.

\bibitem{priester_system_2014}
Alan Priester, Shyam Natarajan, Jesse~D Le, James Garritano, Bryan Radosavcev,
  Warren Grundfest, Daniel~JA Margolis, Leonard~S Marks, and Jiaoti Huang.
\newblock A system for evaluating magnetic resonance imaging of prostate cancer
  using patient-specific 3d printed molds.
\newblock {\em American Journal of Clinical and Experimental Urology},
  2(2):127--135, July 2014.

\bibitem{priester_registration_2019}
A.~Priester, H.~Wu, P.~Khoshnoodi, D.~Schneider, Z.~Zhang, N.~H. Asvadi,
  A.~Sisk, S.~Raman, R.~Reiter, W.~Grundfest, L.~S. Marks, and S.~Natarajan.
\newblock Registration {Accuracy} of {Patient}-{Specific},
  {Three}-{Dimensional}-{Printed} {Prostate} {Molds} for {Correlating}
  {Pathology} {With} {Magnetic} {Resonance} {Imaging}.
\newblock {\em IEEE Transactions on Biomedical Engineering}, 66(1):14--22,
  January 2019.

\bibitem{costa_improved_2017}
Daniel~N. Costa, Yonatan Chatzinoff, Niccolo~M. Passoni, Payal Kapur, Claus~G.
  Roehrborn, Yin Xi, Neil~M. Rofsky, Jose Torrealba, Franto Francis, Cecil
  Futch, Phyllis Hagens, Hollis Notgrass, Susana Otero-Muinelo, Ivan Pedrosa,
  and Rajiv Chopra.
\newblock Improved {Magnetic} {Resonance} {Imaging}-{Pathology} {Correlation}
  {With} {Imaging}-{Derived}, 3d-{Printed}, {Patient}-{Specific}
  {Whole}-{Mount} {Molds} of the {Prostate}.
\newblock {\em Investigative Radiology}, 52(9):507--513, 2017.

\bibitem{penzias_identifying_2018}
Gregory Penzias, Asha Singanamalli, Robin Elliott, Jay Gollamudi, Natalie Shih,
  Michael Feldman, Phillip~D. Stricker, Warick Delprado, Sarita Tiwari, Maret
  Böhm, Anne-Maree Haynes, Lee Ponsky, Pingfu Fu, Pallavi Tiwari, Satish
  Viswanath, and Anant Madabhushi.
\newblock Identifying the morphologic basis for radiomic features in
  distinguishing different {Gleason} grades of prostate cancer on {MRI}:
  {Preliminary} findings.
\newblock {\em PloS One}, 13(8):e0200730, 2018.

\bibitem{hurrell_optimized_2018}
Sarah~L. Hurrell, Sean~D. McGarry, Amy Kaczmarowski, Kenneth~A. Iczkowski,
  Kenneth Jacobsohn, Mark~D. Hohenwalter, William~A. Hall, William~A. See,
  Anjishnu Banerjee, David~K. Charles, Marja~T. Nevalainen, Alexander~C.
  Mackinnon, and Peter~S. LaViolette.
\newblock Optimized b-value selection for the discrimination of prostate cancer
  grades, including the cribriform pattern, using diffusion weighted imaging.
\newblock {\em Journal of Medical Imaging (Bellingham, Wash.)}, 5(1):011004,
  2018.

\bibitem{sumathipala_prostate_2018}
Yohan Sumathipala, Nathan Lay, Baris Turkbey, Clayton Smith, Peter~L. Choyke,
  and Ronald~M. Summers.
\newblock Prostate cancer detection from multi-institution multiparametric
  {MRIs} using deep convolutional neural networks.
\newblock {\em Journal of Medical Imaging (Bellingham, Wash.)}, 5(4):044507,
  October 2018.

\bibitem{cao_joint_2019}
Ruiming Cao, Amirhossein~Mohammadian Bajgiran, Sohrab~Afshari Mirak, Sepideh
  Shakeri, Xinran Zhong, Dieter Enzmann, Steven Raman, and Kyunghyun Sung.
\newblock Joint {Prostate} {Cancer} {Detection} and {Gleason} {Score}
  {Prediction} in mp-{MRI} via {FocalNet}.
\newblock {\em IEEE transactions on medical imaging}, February 2019.

\bibitem{reynolds_voxel-wise_2019}
Hayley~M. Reynolds, Scott Williams, Price Jackson, Catherine Mitchell,
  Michael~S. Hofman, Rodney~J. Hicks, Declan~G. Murphy, and Annette Haworth.
\newblock Voxel-wise correlation of positron emission tomography/computed
  tomography with multiparametric magnetic resonance imaging and histology of
  the prostate using a sophisticated registration framework.
\newblock {\em BJU International}, 123(6):1020--1030, June 2019.

\bibitem{turkbey_multiparametric_2011}
Baris Turkbey, Haresh Mani, Vijay Shah, Ardeshir~R. Rastinehad, Marcelino
  Bernardo, Thomas Pohida, Yuxi Pang, Dagane Daar, Compton Benjamin, Yolanda~L.
  McKinney, Hari Trivedi, Celene Chua, Gennady Bratslavsky, Joanna~H. Shih,
  W.~Marston Linehan, Maria~J. Merino, Peter~L. Choyke, and Peter~A. Pinto.
\newblock Multiparametric 3t prostate magnetic resonance imaging to detect
  cancer: histopathological correlation using prostatectomy specimens processed
  in customized magnetic resonance imaging based molds.
\newblock {\em The Journal of Urology}, 186(5):1818--1824, November 2011.

\bibitem{fedorov_3d_2012}
Andriy Fedorov, Reinhard Beichel, Jayashree Kalpathy-Cramer, Julien Finet,
  Jean-Christophe Fillion-Robin, Sonia Pujol, Christian Bauer, Dominique
  Jennings, Fiona Fennessy, Milan Sonka, John Buatti, Stephen Aylward, James~V.
  Miller, Steve Pieper, and Ron Kikinis.
\newblock 3d {Slicer} as an image computing platform for the {Quantitative}
  {Imaging} {Network}.
\newblock {\em Magnetic Resonance Imaging}, 30(9):1323--1341, November 2012.

\bibitem{madabhushi_fused_2016}
Anant Madabhushi and Michael Feldman.
\newblock Fused {Radiology}-{Pathology} {Prostate} {Dataset}.
\newblock {\em The Cancer Imaging Archive}, 2016.

\bibitem{priester_magnetic_2017}
Alan Priester, Shyam Natarajan, Pooria Khoshnoodi, Daniel~J. Margolis,
  Steven~S. Raman, Robert~E. Reiter, Jiaoti Huang, Warren Grundfest, and
  Leonard~S. Marks.
\newblock Magnetic {Resonance} {Imaging} {Underestimation} of {Prostate}
  {Cancer} {Geometry}: {Use} of {Patient} {Specific} {Molds} to {Correlate}
  {Images} with {Whole} {Mount} {Pathology}.
\newblock {\em The Journal of Urology}, 197(2):320--326, February 2017.

\bibitem{rusu_framework_2015}
Mirabela Rusu, Thea Golden, Haibo Wang, Andrew Gow, and Anant Madabhushi.
\newblock Framework for 3d histologic reconstruction and fusion with in vivo
  {MRI}: {Preliminary} results of characterizing pulmonary inflammation in a
  mouse model.
\newblock {\em Medical Physics}, 42(8):4822--4832, August 2015.

\bibitem{rusu_co-registration_2017}
Mirabela Rusu, Prabhakar Rajiah, Robert Gilkeson, Michael Yang, Christopher
  Donatelli, Rajat Thawani, Frank~J. Jacono, Philip Linden, and Anant
  Madabhushi.
\newblock Co-registration of pre-operative {CT} with ex vivo surgically excised
  ground glass nodules to define spatial extent of invasive adenocarcinoma on
  in vivo imaging: a proof-of-concept study.
\newblock {\em European Radiology}, 27(10):4209--4217, October 2017.

\bibitem{rusu_rad-path_2019}
Mirabela Rusu.
\newblock Rad-{Path} {Fusion} in the {Lung}, April 2019.
\newblock original-date: 2019-04-21.

\bibitem{singanamalli_identifying_2016}
Asha Singanamalli, Mirabela Rusu, Rachel~E. Sparks, Natalie N.~C. Shih, Amy
  Ziober, Li-Ping Wang, John Tomaszewski, Mark Rosen, Michael Feldman, and
  Anant Madabhushi.
\newblock Identifying in vivo {DCE} {MRI} markers associated with microvessel
  architecture and gleason grades of prostate cancer.
\newblock {\em Journal of magnetic resonance imaging: JMRI}, 43(1):149--158,
  January 2016.

\bibitem{rueckert_nonrigid_1999}
D.~Rueckert, L.~I. Sonoda, C.~Hayes, D.~L. Hill, M.~O. Leach, and D.~J. Hawkes.
\newblock Nonrigid registration using free-form deformations: application to
  breast {MR} images.
\newblock {\em IEEE transactions on medical imaging}, 18(8):712--721, August
  1999.

\bibitem{johnson_itk_2013}
Hans~J. Johnson, M.~McCormick, L.~Ibáñez, and The Insight~Software
  Consortium.
\newblock {\em The {ITK} {Software} {Guide}}.
\newblock Kitware, Inc., third edition, 2013.

\end{thebibliography}




\end{document}